\documentclass[11pt, a4paper, twoside]{report}
\usepackage[utf8]{inputenc}

\usepackage[margin=1in]{geometry}
\usepackage{parskip}

\usepackage[ngerman, main=english]{babel}
\usepackage[table,xcdraw,dvipsnames]{xcolor}
\usepackage{float}

\usepackage{fancyhdr}
\fancypagestyle{fancy}{
    \fancyhf{}
    \fancyhead[LE]{}%Boundary identification in chemical reaction subnetworks using Gaussian linear regression}
    \fancyhead[RO]{\nouppercase \leftmark}
    \fancyfoot[LE, RO]{\thepage}
}

\fancypagestyle{plain}{
    \fancyhf{}
    \fancyhead[LE]{}%Boundary identification in chemical reaction subnetworks using Gaussian linear regression}
    \fancyfoot[LE, RO]{\thepage}
}

\usepackage[autostyle]{csquotes}
\usepackage{booktabs}

\usepackage{enumitem}
\usepackage{physics}
\usepackage{siunitx}
\usepackage{graphicx}
\usepackage{float}
\usepackage[labelfont=bf]{caption}

\patchcmd{\abstract}{\titlepage}{\thispagestyle{plain}}{}{}
\patchcmd{\endabstract}{\endtitlepage}{}{}{}

\usepackage{amsmath, amssymb, amsfonts, amsthm, mathtools, latexsym}
\usepackage{bm}
\usepackage{thmtools}
\usepackage{epigraph}

\usepackage[most]{tcolorbox}

\usepackage[colorlinks, linktocpage, linkcolor=Blue, citecolor=Blue, urlcolor=Blue]{hyperref} % dark blue to stick with PRL standard
\usepackage[capitalise]{cleveref}

\usepackage{todonotes}

\newcommand{\dw}{\ensuremath{\,\text{d}w}}
\newcommand{\dx}{\ensuremath{\,\text{d}x}}
\newcommand{\wmin}{\ensuremath{w_\text{min}}}
\newcommand{\aand}{\ensuremath{\hspace{0.5cm}\text{and}\hspace{0.5cm}}}

\newtcolorbox{math_box}[2][breakable]{breakable=true, %break at=8cm, pad at break=1mm, 
    colback=blue!3!white, colframe=black!3!white, fonttitle=\bfseries, colbacktitle=blue!20!white,
    coltitle=white!1!black,
    title={#2}}

\newtcolorbox{example_box}[2][breakable]{
    breakable=true, colback=green!3!white, colframe=black!5!white, fonttitle=\bfseries, colbacktitle=green!20!white,
    coltitle=white!1!black,
    title={#2}}

\title{On the statistical physics of wealth distribution}
\author{Joel Wagner}
\date{2022}

\begin{document}

\pagestyle{empty}
\begin{titlepage}
    % university logo:
    \includegraphics[width=.48\textwidth]{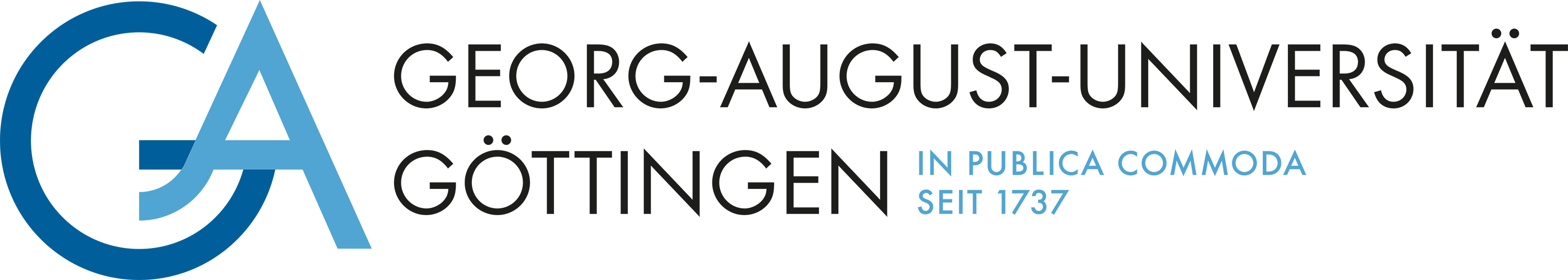}
    % second logo
    \hfill
    \includegraphics[width=.48\textwidth]{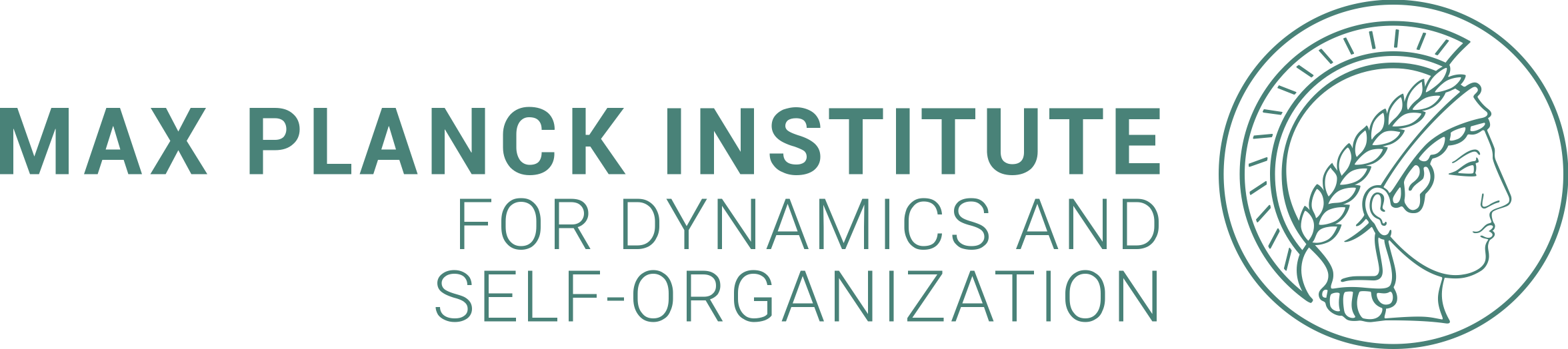}
    \begin{center}
    \includegraphics[width=.248\textwidth]{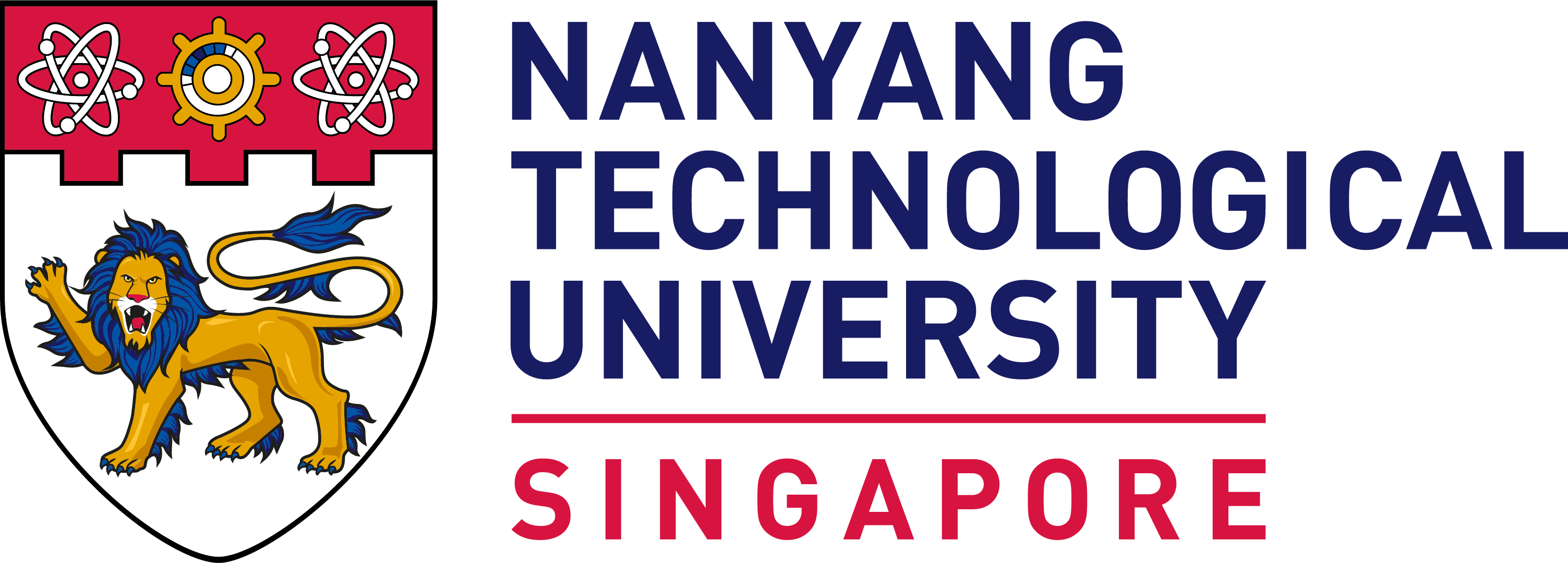}
        \vspace*{1cm}

        \textbf{\LARGE Master's Thesis }

        \vspace{1.5cm}

        \rule{\textwidth}{1pt}

        \vspace{2mm}

        \textbf{\huge {On the Statistical Physics of Wealth Distribution}}
        %\textbf{\huge {First line\\[1ex]second line\\[2.5ex]third line}}

        \vspace{2mm}

        \rule{\textwidth}{1pt}

        \vspace{1cm}

        \rule{\textwidth}{1pt}

        \vspace{2mm}

        %\textbf{\huge {Erste Zeile\\[1ex] zweite Zeile\\[2.5ex]dritte Zeile}}
        \textbf{\huge {Über die Statistische Physik von Wohlstandsverteilung}}

        \vspace{2mm}

        \rule{\textwidth}{1pt}

        \vspace{2cm}

        \large
        Master's Thesis \\[1.3ex]
        by\\[1.3ex]
        \textbf{Joel Wagner}\\[1.3ex]

        \vspace{2cm}

        %Nanyang Technical University\\[1.3ex]

        \vspace{2mm}

        % put group logo here
        %\includegraphics[height=1.2cm]{figures/Group_Logo_new_HP.png}

        \vfill

        Thesis submitted to the Faculty of Physics, Georg-August-Universit\"at G\"ottingen
        \vspace{5mm}

        \begin{tabular}{ll}
            Thesis period: & 12.08.2024 - 12.02.2025 \\
            %Submission date: & 12.01.2025
            %07.11.2022 --- 13.02.2023\\[1.3ex]
            First referee: & Prof. Viola Priesemann \\%Prof.~Dr.~Cheong~Siew~Ann\\[1.3ex]
            Second referee: & Prof. Cheong Siew Ann\\%Prof.~Dr.~Viola~Priesemann
        \end{tabular}
    \end{center}
\end{titlepage}

\section*{Preface}

This master’s thesis, titled 'On the Statistical Physics of Wealth Distribution', explores the study of wealth distribution within the field of econophysics. The work is structured into two main parts: a comprehensive background and theory section, followed by new results. 
Chapter 2 provides a detailed overview of the most important findings from asset exchange models and their steady-state distributions, serving as a useful resource for readers seeking a solid foundation in this research area. Chapter 3 introduces a novel spectral-theoretic analysis of the Markov transition matrices underlying these models—an approach that, to the best of our knowledge, has not been applied to asset exchange models before. The second part of the results section presents an analysis of Ethereum transaction data, which, again to the best of our knowledge, has not been studied in the context of wealth exchange models previously. By making this thesis available on arXiv, we aim not only to offer a literature overview but also to inspire further research in this interdisciplinary field.

\section*{Disclaimer}

While this thesis was proofread and evaluated as part of the academic examination process, it has not undergone formal peer review. As such, the findings and interpretations presented here should be considered preliminary and subject to further validation by the scientific community. Readers are encouraged to approach the results with appropriate critical scrutiny.

\newpage 
\tableofcontents
\newpage 

\pagestyle{fancy}
\pagenumbering{arabic}
\setcounter{page}{1}

%%% BEGIN OF THESIS BODY %%%%%%%%%%%%%%%%%%%%%%%%%%%%%%%%%%%%%%%%%%%%%%%%%%%%%%%

%Abstract
\begin{abstract}

    Asset exchange models (AEMs) offer a physics-inspired framework for understanding wealth formation within societies. These models describe the dynamics of wealth distribution through pairwise exchanges of money between individuals. Depending on the specific rules governing these exchanges, various steady-state wealth distributions can emerge, ranging from exponential equilibrium distributions to heavy-tailed power-law distributions.
    Despite their theoretical appeal, asset exchange models lack empirical validation. Quantitative comparisons between model mechanisms and real-world transaction data have been scarce due to the challenges in acquiring comprehensive data on real-world money exchanges. In this study, we address this gap by analysing the spectral properties of Markov transition matrices derived from both asset exchange models and blockchain transaction data from Ethereum.
    Our approach allows to quantitatively compare the exchange dynamics in the model with those observed empirically.
    Specifically, we examine the extent to which exchange processes are in thermodynamic equilibrium, namely if they satisfy detailed balance, and derive the steady state wealth distribution from the transition matrices. We show that the spectrum of transition matrices describing equilibrium (detailed-balance) systems consist of only real-valued eigenvalues, and identify changes in the price of Ethereum with a downward shift of the whole spectrum.
    Additionally, we explore the impact of external factors (e.g., taxes) on wealth distribution and show that when richer individuals gain additional advantages, the initial distribution of wealth can dictate the further evolution of the system.
    Our results introduce a novel quantitative framework for assessing asset exchange models against real-world data and reveal the conditions under which wealth formation exhibits path dependence. By doing so, they contribute to the broader field of economic modelling and enhance our understanding of wealth formation processes.

\end{abstract}
% Introduction
\chapter{Introduction}
\label{sec:Introduction}

Wealth inequality has posed a substantial challenge to human civilisations throughout history. The equitable distribution of wealth is not only necessary for ensuring adequate living conditions for the impoverished, but is also a prerequisite for societal stability \cite{motesharrei2014modeling,neckerman2007inequality}. Regrettably, despite overall economic growth, the latter half of the 20th century witnessed a surge in inequality in many parts of the world \cite{neckerman2007inequality, albers2022wealth, dreher2008has,iglesias_inequality_2020}.
In an effort to understand the underlying causes of this trend, researchers have explored various approaches to model the distribution of wealth, including models inspired by physics. Among these, asset exchange models (AEMs) have provided important insights into how wealth distributions form from the pairwise exchange of money between individuals \cite{yakovenko_colloquium_2009,ispolatov_wealth_1998}. For instance, AEMs have demonstrated that even in 'fair' economies, where no individual has a systematic advantage, money exchange can still lead to a high degree of inequality \cite{boghosian_h_2015,boghosian_fokkerplanck_2014,cardoso_equal_2022}.
Despite these contributions, wealth formation remains a highly complex process and is challenging to model in its full extent. Asset exchange models, while insightful, are simplifications and lack robust empirical support, which impedes their usage. This highlights the need for continued research for a better understanding of the causes of economic inequality.

In asset exchange models, economic agents exchange money based on rules that dictate the monetary amounts that the agents exchange. Such an exchange process bears similarity to the exchange of energy among particles, which has been extensively studied in statistical physics. Indeed, asset exchange models can reach a Boltzmann steady-state distribution of wealth, similar to how energy distributes among particles in fluids. However, this outcome only arises if the exchange rule adheres to \textit{detailed balance}, a condition that is unlikely to hold for real-world wealth formation \cite{dragulescu_statistical_2000}. Moreover, a Boltzmann distribution has a Gini index of $G=0.5$, which is substantially lower than the wealth Gini index observed in real economies. When detailed balance is not enforced, the wealth distribution changes from the Boltzmann distribution to a non-equilibrium steady-state distribution. In some cases, it may even result in wealth condensation, where a single individual accumulates all available wealth. These various outcomes highlight the sensitivity of asset exchange models to the imposed exchange rules, which thus far have been chosen somewhat arbitrarily. Therefore, improving the accuracy of asset exchange models necessitates a deeper understanding of realistic mechanisms governing monetary transactions between individuals.

The analysis of asset exchange models currently relies mostly on numerical simulations to determine the steady-state distribution of wealth. For some exchange rules, an analytical derivation via the Master equation is possible \cite{ispolatov_wealth_1998}. In this thesis, we propose an alternative method to study asset exchange, which reveals information on the structure of the exchange process and ultimately allows for a quantitative comparison to data. In particular, we treat the exchange of money as transitions between different states of the wealth distribution, utilising the fact that asset exchange models are Markovian. By simulating asset exchange models and tracking all transactions, one can construct probabilistic Markov transition matrices. Studying the spectral properties of these matrices provides insights into the underlying mechanisms of the model. For example, asset exchange models that satisfy detailed balance correspond to \textit{reversible} Markov processes, characterised by diagonalisable transition matrices with real-valued eigenvalues. Further, the eigenvector associated with the unique eigenvalue $\lambda=1$ represents the steady-state distribution of the system, offering a way to infer the equilibrium distribution of wealth based on the transition probabilities between states. This method is particularly useful as it can readily be used for analysing transition matrices derived from empirical data, thereby bridging the gap between theoretical models and real-world observations.

While data on real-world currency exchanges is limited, blockchain technology enables to analyse extensive datasets on digital currency transactions. For instance, Ethereum allows the reconstruction of individual wallet balances, representing 'crypto wealth' of users, with which one can compute transition matrices. In this thesis, we analyse such matrices by collecting data from millions of individual Ethereum transactions. We show that this approach reveals a highly unequal steady-state distribution resulting from the exchange process, and compare the spectral properties of the Ethereum matrices with those of asset exchange models. All in all, this work represents a novel effort to quantitatively compare the structure of asset exchange models with empirical data.

This thesis is organised into three chapters. Chapter~\ref{ch:theoretical_background} begins with an overview of wealth distribution analysis in physics in Sec.~\ref{sec:empiricaldata}, introducing the topic with Vilfredo Pareto's observation of the 80/20 rule. It covers key concepts such as the Gini index and offers a comprehensive review of asset exchange models in Sec.~\ref{sec:wealth_modelling_overview}, including their numerical and analytical solutions. This section aims to equip the reader with a solid understanding of the various equilibrium wealth distributions that can emerge from asset exchange models.\\
Chapter~\ref{ch:results} presents the results of this study, starting with an introduction to essential theoretical concepts related to Markov processes in Sec.~\ref{sec:markovchains}, which are necessary for the subsequent analysis. These theoretical concepts will be highlighted in blue and presented in separate boxes to distinguish them from the main text. We will use asset exchange models to illustrate these theoretical concepts before applying the analysis to Ethereum transaction data in Sec.~\ref{sec:crypto}. The theory on Markov chains further provides a way to study under which conditions the steady state distribution of asset exchange models depends on the initial wealth distribution, which will be presented in Sec.~\ref{sec:bistability}. Chapter~\ref{ch:conclusion_outlook} concludes the thesis by summarising and discussing the results, and exploring alternative approaches to studying wealth distribution in physics. Supplementary figures and additional theoretical background are provided in the appendix.

\chapter{Wealth Distribution Models in Physics}
\label{ch:theoretical_background}

\section{History, notation and empirical data}
\label{sec:empiricaldata}

\subsection{The Pareto distribution}

The study of wealth inequality is said to date back to Vilfredo Pareto, who at the beginning of the 20th century observed, that 80\% of the land in Italy was owned by only 20\% of the landowners \cite{pareto1964cours,chakrabarti2013econophysics}. This observation has become known as the ‘80-20 rule’ or 'Pareto principle'. With respect to landownership, it has proven remarkably consistent across various countries and historical periods. Moreover, the universality of this principle extends beyond realms directly related to landownership, even in seemingly disparate fields, such as business or software development \cite{sanders1987pareto}. In this report, we specifically focus on the distribution of wealth, which is closely related to land ownership, and where the 80-20 rule also finds notable application.

In order to formalise the concept underlying the 80-20 rule, we introduce the following notation. Let $P(w)$ be the population probability density function according to which wealth $w\geq0$ is continuously distributed among a population. It is defined such that, when randomly picking an individual out of that population, the probability that his or her wealth lies within $w$ and $w+\dw$ is $P(w)\dw$. Hence, $\int_a^b P(w)\dw$ denotes the fraction of the population with wealth between $a$ and $b$. The zeroth and first moments $m_0$ and $m_1$ of $P(w)$ are given by 

\begin{equation*}
    m_0 = \int_0^\infty P(w)\dw = 1  \hspace{0.5cm} \text{and} \hspace{0.5cm} m_1 = \int_0^\infty w\,P(w)\dw = \bar w\,,
\end{equation*}

where $\bar w$ denotes the average wealth of the population. Pareto's insight into the 80-20 rule stemmed from plotting the fraction of individuals whose wealth (in the context of land ownership) exceeded $w$ as a function of $w$ \cite{boghosian_kinetics_2014}. This 'Pareto function', often denoted by $A(w)$, is given by

\begin{equation*}
    A(w) = \int_w^\infty P(w')\dw'\,.
\end{equation*}

It is related to the population probability density function (PDF) $P(w)$ via $P(w) = -\frac{\text{d}A(w)}{\text{d}w}$ and to the cumulative population probability density function (CDF) $C(w)$ via $C(w) = 1-A(w)$, which follows from normalisation.

The '80-20' rule that Pareto observed does not apply to any probability distribution $P(w)$. Instead, is a characteristic of the Pareto distribution, which Pareto found to be a good fit to the data on land ownership. The Pareto distribution is defined such that the fraction of individuals with wealth greater than $w$ is given by a power-law distribution with lower cutoff $\wmin$ \cite{boghosian_kinetics_2014}:

\begin{equation*}
    A(w) = \begin{cases} 
      1 & \text{if } w < \wmin \\
      \left(\frac{\wmin}{w}\right)^\alpha & w\geq \wmin\,.
      \end{cases}
\end{equation*}

The exponent $\alpha>0$ is called the Pareto exponent and defines how unequal the distribution is; lower values of $\alpha$ indicate a more unequal distribution with increased weight in the tail of $P(w)$. The corresponding PDF $P(w)$ and CDF $C(w)$ of the Pareto distribution are 

\begin{equation}
    P(w) = \begin{cases} 
      0 & \text{if } w < \wmin \\
      \frac{\alpha \wmin^\alpha}{w^{\alpha + 1}} & w\geq \wmin\,
      \end{cases} \hspace{0.5cm} \text{and}  \hspace{0.5cm} C(w) = \begin{cases} 
      0 & \text{if } w < \wmin \\
      1-  \left(\frac{\wmin}{w}\right)^\alpha &w\geq \wmin\,.
      \end{cases}\label{eq:pareto_pdf_cdf}
\end{equation}

The Pareto distribution is a type of power-law distribution, as the probability density $P(w)$ decays to the power of the exponent $\alpha$ plus one, $P(w) \sim w^{-(\alpha+1)}$. 
%Power-law distributions are found in many areas of nature and science, from describing the probability distributions of different magnitude earthquakes to the frequency of words appearing in texts. They often emerge as the natural result of mechanisms related to multiplicative processes, or systems self-organising towards a critical state. In section Sec.~\ref{} we will come back to this topic and see how wealth accumulation processes can give rise to power-law distributions. 
Importantly, only a specific value of the exponent $\alpha$ fulfils the 80-20 rule. Denoting by $w_{\rm top}$ the 20\% fraction of the richest population which holds 80\% of the total wealth, two conditions must be satisfied:

\begin{equation*}
    \int_{w_{\rm top}}^\infty w'P(w')\dw' = 0.8\bar w\, \hspace{0.5cm} \text{and} \hspace{0.5cm} \int_{w_{\rm top}}^\infty P(w')\dw' = A(w_{\rm top}) = 0.2\,.
\end{equation*}

The first condition states that the wealth held by the top fraction of the population equals 80\% of the average wealth $\bar w$, and the second condition states that the top fraction makes up 20\% of the population.
Inserting the Pareto distribution $P(w)$ (Eq.~\ref{eq:pareto_pdf_cdf}) and computing the integrals, one finds that the Pareto exponent must take the value $\alpha \approx 1.16$. This can be generalised such that a fraction $f$ of the richest population holds a fraction $1-f$ of the total wealth, which leads to the requirement $\alpha = \log(f)/\log(\frac{f}{1-f})$ \cite{boghosian_kinetics_2014}. For a perfectly equal distribution of wealth, 50\% of the richest population would hold 50\% of the total wealth. It would thus correspond to $f=0.5$, which leads to a diverging Pareto exponent $\alpha$. Hence, large values of $\alpha$ make the Pareto distribution more equal, small values more unequal. 
 
\begin{figure}
    \centering
    \includegraphics[width=6in]{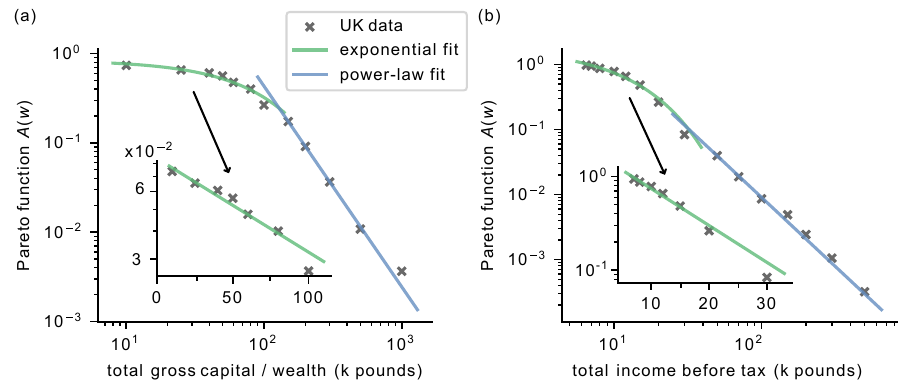}
    \caption{\textbf{Wealth and income distributions in the United Kingdom follow an exponential- power-law distribution.} (a) The cumulative fraction of individuals $A(w)$ with wealth exceeding $w$ falls onto a straight line in a log-log plot for the high wealth range. It is well fitted by a power-law distribution (blue, last 5 data points used for the fit). In contrast, data from the low to intermediate wealth range falls onto a straight line in a semi-logarithmic plot and is well fitted by an exponential distribution (green, first 7 data points used for the fit). The data reports the amount of wealth, defined as the total gross capital value of \SI{16415} individuals in 2002 \cite{ukgovernment2002distribution}. (b) For data on income, the cumulative fraction of individuals $A(w)$ follows a similar power-law - exponential trend. The data reports the pre-tax income of \SI{31281} individuals in 2005 \cite{ukgovernment2005distribution}. The first 7 data points were used for the exponential fit and the last 7 were used for the power-law fit.} \label{fig:wealth_data_UK}
\end{figure}

While the Pareto distribution often provides a good fit for high values of wealth, i.e., the tail of the wealth distribution, it fails to do so for small to intermediate wealth, namely the bulk of the distribution. Instead, wealth in the intermediate range is often better fitted by an exponential distribution \cite{yakovenko_colloquium_2009,dragulescu_exponential_2001,iglesias_entropy_2012}. Wealth data from the United Kingdom indicates that the transition between exponential and power-law behaviour occurs at a value of approximately \SI{100000}{Pounds} of net wealth (Fig.~\ref{fig:wealth_data_UK}a).

The ubiquity of this exponential - power-law characteristic of wealth distributions is believed to stem from fundamental differences in how wealth is accumulated on both scales \cite{nirei_two_2007,silva406385thermal}. While individuals on the lower end of the distribution accumulate wealth predominately through wages, wealth of individuals on the upper end of the distribution is often accumulated through assets. The former is an additive process, while the latter is a multiplicative process, which lead to the qualitatively different distributions \cite{silva406385thermal}.

While wealth is a theoretically well-defined concept, practical assessments can be challenging. It is defined as the aggregate value of assets and liabilities, and thus encompasses diverse variables, including cash, land, property and loans \cite{yakovenko_colloquium_2009}. Reporting practices of those variables vary, with some factors frequently disclosed to authorities and made public, while others may not be or may be subject to privacy regulations. Although some countries impose inheritance taxes that necessitate the listing of assets for deceased individuals, this data only reflects the distribution of wealth among the deceased. Such distributions may differ from those of the living population, due to, e.g., economic growth or age disparities \cite{yakovenko_colloquium_2009}. Consequently, comprehensive data on wealth distributions is often relying on surveys or entirely lacking.

In contrast, data on income distributions is more straightforward to obtain. Tax records are oftentimes made public, from which it is possible to approximate the underlying income distribution \cite{dragulescu_evidence_2001,yakovenko_colloquium_2009}. Notably, the distribution of income often exhibits qualitative similarities to that of wealth, featuring an exponential lower part and a power-law upper part (Fig.~\ref{fig:wealth_data_UK}b) \cite{neda_scaling_2020}. Despite these resemblances, models explaining the two phenomena differ significantly \cite{yakovenko_colloquium_2009,neda_scaling_2020}. This report will focus exclusively on the modelling of wealth distributions.

\subsection{The Gini index as a measure of inequality}

Evidently, the Pareto distribution exhibits significant inequality.
The power-law behaviour causes a heavy concentration of probability in the tails, differing from that of an exponential distribution. Notably, the probability of extreme values, signifying the presence of exceptionally rich individuals, is not exponentially suppressed; rather, it occurs with considerably higher probability. 
In fact, for a set of $N$ individuals with wealth $w_1,...w_N$ that is power-law distributed with Pareto exponent $\alpha$, the sum of the series of random variables grows with $N$ in the same way as the maximum of that series grows with $N$ (Sec.~\ref{sec:theory:thin_vs_fat_tails}): $\sum_i^N w_i \sim N^{\frac{1}{\alpha}}$ and $\text{max}\,(w_1,...w_N) \sim N^{\frac{1}{\alpha}}$. This implies that a single agent within the distribution has the potential to dominate the entire wealth distribution, attaining a level of wealth comparable to the combined wealth of all other individuals \cite{taleb2010black}.
While the exponent $\alpha$ measures the extend of this inequality, other inequality measures are required if the wealth distribution is not purely a power-law or obeys an entirely different probability distribution. For that purpose, the most common indicator for inequality used in economics is the Gini index \cite{chakrabarti2013econophysics}. It is defined in continuous and discrete cases as

\begin{equation}
    G = \frac{1}{2\bar w} \int_0^\infty \int_0^\infty \dw \dw'\, P(w)P(w') \,|w-w'| \hspace{0.5cm} \text{and} \hspace{0.5cm} G = \frac{1}{2N^2\bar w} \sum_i^N\sum_j^N|w_i-w_j|
\end{equation}

and ranges between the values of $G=0$ and $G=1$ \cite{cardoso_equal_2022}. $G=0$ corresponds to maximum equality and arises when all terms $|w_i-w_j|$ are zero, i.e., when all individuals possess the same amount of wealth. $G=1$ corresponds to maximum inequality and arises when one agent is in possession of all the wealth. In that case, the only contribution of $|w_i-w_j|$ equals the total wealth $N\bar w$ whenever $i$ or $j$ is the richest agent. It is counted $2N$ times (the factor 2 comes from double counting), such that the terms $2N^2\bar w$ cancel out to yield $G=1$.

An alternative, but equivalent, definition of the Gini index involves the use of the Lorenz curve \cite{boghosian_h_2015}. The Lorenz curve is constructed by plotting the cumulative wealth percentage of the population $y(w)$ against the cumulative fraction of the population holding that wealth, $x(w)$. The two are defined as

\begin{equation}
    y(w) = \frac{1}{\bar w}\int_0^w \dw'\,P(w')\,w' \hspace{0.5cm} \text{and} \hspace{0.5cm} x(w) =  \int_0^w \dw' \,P(w')\,.\label{eq:lorenzcurve}
\end{equation}

\begin{figure}[!ht]
    \centering \includegraphics[width=6in]{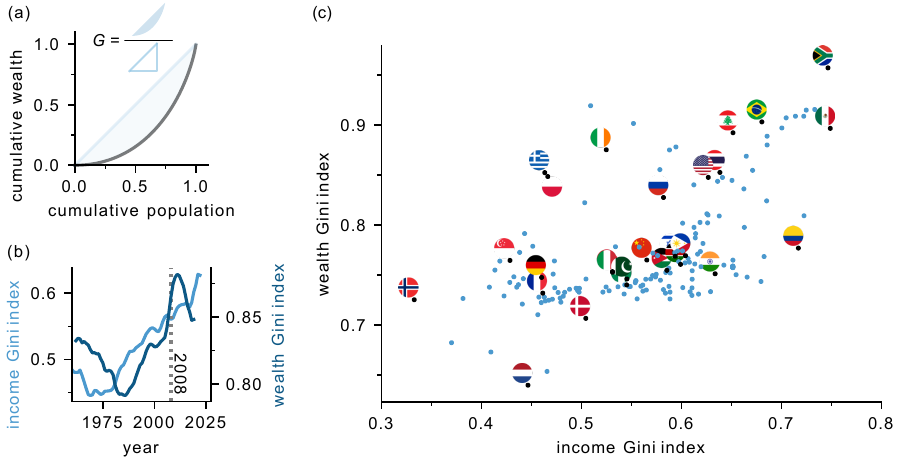}
    \caption{\textbf{The Gini index reveals the presence of severe inequality in most countries.} (a) The Gini index $G$ can be computed using the Lorenz curve, as the ratio of area between the Lorenz curve (grey) and the diagonal line (blue) to the total area below the diagonal line. The Lorenz curve is depicted for an exponential wealth distribution $P(w) \sim e^{-w/T}$. (b) Both the Gini indices for income and wealth have been increasing in the US since the 1980s. The financial crisis of 2008 caused a drop in the wealth Gini index. (c) The relation between income and wealth Gini index shows a linear relation (Pearson correlation coefficient $r=0.55$). Data in (b) and (c) taken from the World Inequality Database \cite{wid}.}   \label{fig:gini_lorenz_curve_data}
\end{figure}

For a perfectly equal distribution the two curves increase identically with $w$, e.g., the poorer half of the population holds half of the total wealth. In such a scenario, the Lorenz curve would be given by a diagonal line. In contrast, for a non perfectly equal distribution, the Lorenz curve deviates from a straight line and lies strictly below it \cite{boghosian_h_2015}. This discrepancy indicates that the less wealthy fraction of the population possesses a smaller share of the total wealth than they would in an equitable setting. The Gini index can be calculated as the fraction of the area between the Lorenz curve and the diagonal line relative to the area beneath the diagonal line using Eq.~\ref{eq:lorenzcurve} (Fig.~\ref{fig:gini_lorenz_curve_data}a) \cite{chakrabarti2013econophysics}:

\begin{equation}
    G = \frac{\int_0^1 \dx\, (x-y)}{\int_0^1 \dx\, x } = 2 \int_0^1 \dx\, (x-y)\,.\label{eq:gini_using_lorenz_curve}
\end{equation}

The Gini index reveals prevalent and substantial inequality across the populations of nearly all countries. While the Gini index for income is typically lower than the Gini index for wealth, both metrics surpass $G>0.4$ for income and $G>0.7$ for wealth in the majority of nations. The disparity between income and wealth Gini indices can be partly attributed to the additional impact of property when assessing wealth, considering that it can easily be inherited. When accounting for consumption in the income distribution, namely computing it for 'what is left at the end of the month', the Gini index increases under the assumption that individuals on the upper end of the wealth distribution can save larger shares of their income than individuals on the lower end. Overall, the relation between the Gini indices for income and wealth are closely related (Pearson correlation coefficient $r=0.55$, Fig.~\ref{fig:gini_lorenz_curve_data}c). Besides the relation between wealth and income Gini index, both indicators are negatively correlated with the democracy index ($r=-0.43$ for income and $r=-0.18$ for wealth) and the Human Development Index (HDI) ($r=-0.48$ for income and $r=-0.18$ for wealth) (Sec.~\ref{sec:appendix-giniindex_vsdemocracy_vshdi}).

In spite of overall economic growth and a reduction in absolute poverty, the latter half of the 20th century witnessed a surge in inequality across most developed nations \cite{neckerman2007inequality,dreher2008has}. Notably, the United States has experienced a continual increase in both wealth and income Gini indices since the 1980s (Fig.~\ref{fig:gini_lorenz_curve_data}b). Economic crises, such as the 2008 financial downturn, temporarily reversed this trend, causing a decline in the wealth Gini index (Fig.~\ref{fig:gini_lorenz_curve_data}b). However, this came at a substantial societal cost, such as increasing unemployment.

Traditionally, the exploration of wealth and its distribution falls under the domain of economics, with metrics such as the Gini index serving as macroscopic indicators \cite{boghosian_kinetics_2014}. Other macroscopic quantities in economics include the gross domestic product (GDP), the unemployment rate, or the inflation rate. These economic indicators are quantitative descriptions of the economic system as a whole. They are analogous to macroscopic quantities in physics such as pressure or temperature, that describe the state of a gas or liquid. In principle, the derivation of such macroscopic quantities should be feasible from microscopic interactions. For instance, in physics, particles in a gas exchanging energy describe a microscopic process. The Boltzmann distribution characterises the exponential energy distribution of these particles. It serves as the foundation for deducing macroscopic variables, including temperature and pressure, relating the micro- to the macroscopic description. In an ideal scenario, a similar connection should exist in economics, where the collective actions of economic agents, such as wealth exchange, should explain macroscopic quantities such as the Gini index \cite{boghosian_kinetics_2014}. However, a strong separation between the fields of micro- and macroeconomics persists \cite{boghosian_kinetics_2014}.

Inspired by the success of statistical physics, physicists working in the field of econophysics have aspired to bridge the gap between micro- and macroeconomics for over two decades. The official inception of the field of econophysics is often attributed to the 1995 conference on \textit{Dynamics of Complex Systems} in Kolkata, where the term 'Econophysics' was first mentioned by Eugene Stanley \cite{yakovenko_colloquium_2009}. While 1995 is thus the official birth date of wealth distribution modelling in econophysics, first models were developed as early as 1986 \cite{yakovenko_colloquium_2009,angle1986surplus}. Furthermore, ideas sprang as early as 1960, the year in which Benoît Mandelbrot first brought up the idea of using models from statistical physics to model economic interactions \cite{mandelbrot1960pareto}:

\begin{quote}
“There is a great temptation to consider the exchanges of money which occur in economic interaction as analogous to the exchanges of energy which occur in physical shocks between molecules. In the loosest possible terms, both kinds of interactions 'should' lead to 'similar' states of equilibrium. That is, one 'should' be able to explain the law of income distribution by a model similar to that used in statistical thermodynamics...”
\begin{flushright}
--- Benoît Mandelbrot, 1960
\end{flushright}
\end{quote}

%Overall, physicists' endeavour to address unresolved questions within the realm of wealth distribution thus appears promising for several reasons. 
Since the inception of econophysics, various models to study the distribution of wealth have been analysed using tools and concepts from statistical physics. In the forthcoming chapter, we will provide an overview of a specific category of models, known as \textit{asset exchange models}, whose invention was motivated by Benoît Mandelbrot's idea.

\newpage

\section{Asset exchange models}
\label{sec:wealth_modelling_overview}

\begin{figure}[!ht]
    \centering
    \includegraphics[width=6in]{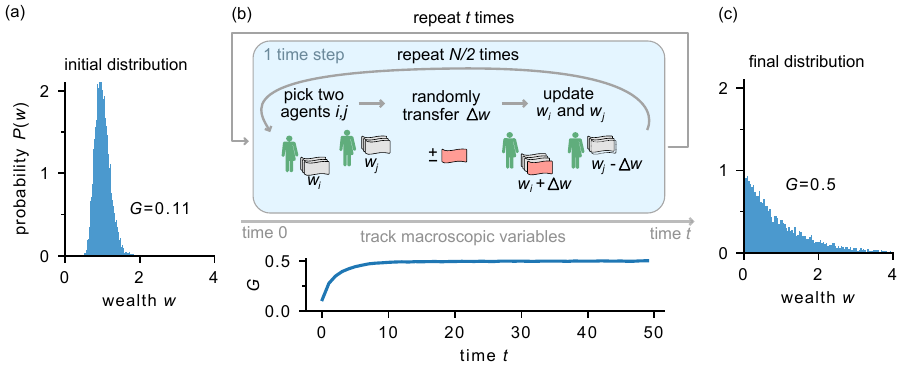}
    \caption{\textbf{Schematic overview of an asset exchange model (AEM).} (a) AEMs are initialised by distributing wealth $W$ among $N$ agents, for example according to a lognormal distribution with a Gini index $G=0.11$. (b) One time step of an AEM involves picking two agents at random and exchanging a wealth amount $\Delta w$, the value of which depends on the interaction rule. This process is repeated $N/2$ times, such that, on average, each agent has participated in one transaction. This process is again repeated $t$ times, and macroscopic variables such as the Gini index can be tracked over time. (c) The model is evolved until the steady state distribution of wealth has been formed, in this case an exponential distribution with $G=0.5$.}
    \label{fig:aem_scheme}
\end{figure}

Asset exchange models generally involve a population of $N$ economic agents, each possessing an amount of wealth $w_i$. These agents engage in sequential pairwise economic transactions, where a random transfer of wealth $\Delta w$ occurs between two agents. In a simulated scenario, the process involves the random selection of two agents $i$ and $j$, a coin toss, and, depending on the outcome, the transfer of $\Delta w$ from $i$ to $j$, or vice versa. This entire sequence repeats until a stable steady state distribution of wealth is reached (Fig.~\ref{fig:aem_scheme}). As every exchange conserves the total amount of wealth, asset exchange models are meant to model a closed economy in which wealth changes hands between agents solely through monetary exchange.

% Boghosian2014fokker-planck: more details on the justification. Perhaps move this paragraph into the discussion. 
There are generally two different interpretations for the transaction scheme represented by asset exchange models. The first is that the two agents engaging in the transaction are exchanging some tangible good in return for the monetary amount $\Delta w$. While the model only tracks the monetary aspect, it implies an underlying flow of goods in the opposite direction for each transaction \cite{yakovenko_colloquium_2009}. The second interpretation is that a specific good is traded for a monetary amount, but with the transaction being imperfectly fair – for example, the payer may slightly overpay \cite{hayes_computing_2002,boghosian_fokkerplanck_2014}. Consequently, the effective wealth of the two agents undergoes a change $\Delta w$, reflecting the profitability of the transaction.

In the subsequent sections, we will provide an overview of different asset exchange models and their resulting steady-state distributions. We will start with a model for $\Delta w$ that draws analogy to the kinetic energy exchange of molecules and see how it recovers the Boltzmann equilibrium distribution.

\subsection{Dragulescu model \& the Boltzmann distribution}
\label{sec:dragulescumodel}

As a first example, consider an asset exchange model where the amount of wealth exchanged is a fraction $(1-\beta)$ of the average of the wealth of the two agents $i$ and $j$, $\Delta w = (1-\beta)\frac{w_i+w_j}{2}$ \cite{dragulescu_statistical_2000,chakrabarti2013econophysics,chakraborti_distributions_2002}. We will refer to this model as the \textit{Dragulescu model} \cite{dragulescu_statistical_2000}. The parameter $0\leq \beta \leq 1$ is fixed and can be viewed as a measure for risk aversion among the agents. After a transaction the wealth of the agents changes to 

\begin{equation*}
    w_i \to w_i' = w_i + \eta (1-\beta)\frac{w_i+w_j}{2} \aand w_j \to w_j' = w_j - \eta (1-\beta) \frac{w_i+w_j}{2}\,,\label{eq:dragulescu_delta_w}
\end{equation*}

where $\eta$ is a random variable taking the values $\eta \in \{-1,1\}$ with equal probability, $\mathbf{E}(\eta) = 0$.

To prevent negative wealth, the exchange is executed only if the paying agent possesses sufficient wealth to cover the transaction. This exchange mechanism draws analogy to energy exchange during collisions between particles, where the transferred energy $\Delta e$ is a fraction of the total available energy, which is the sum of the two individual energies, $e_i+e_j$. 
From statistical physics it is known that the energy distribution of particles in an ideal gas follow the Boltzmann distribution. Due to the apparent similarities between energy exchange and wealth exchange under this exchange rule, it can be anticipated that wealth also distributes according to the Boltzmann distribution in this asset exchange model \cite{yakovenko_colloquium_2009,dragulescu_statistical_2000}.

One can derive this outcome by dividing the system into two subsystems. The total wealth of the two systems is then given by $w=w_1+w_2$ and the probability of the whole system being in that state is given by $P(w)=P(w_1)P(w_2)$. Hence, $P(w_1+w_2)=P(w_1)P(w_2)$, which is only solved by an exponential function and leads to the Boltzmann distribution of wealth \cite{yakovenko_colloquium_2009}

\begin{equation*}
    P(w) = \frac{1}{\bar w} e^{-w/\bar w} \,.
\end{equation*}

\begin{figure}
    \centering
    \includegraphics[width=6in]{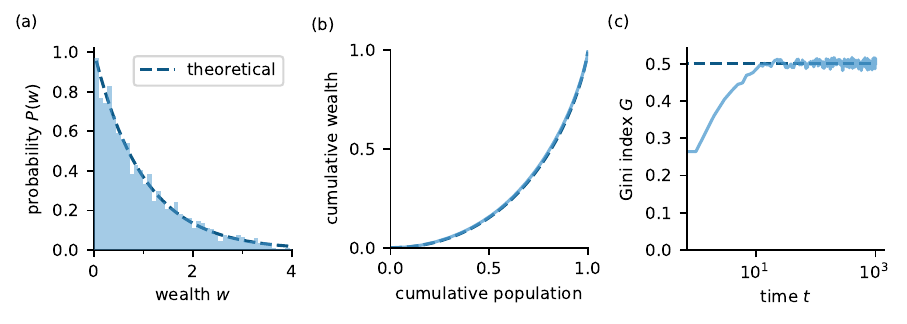}
    \caption{\textbf{An asset exchange model obeying detailed balance leads to the Boltzmann distribution} (a) Model simulation for the exchange $\Delta w = (1-\beta)\frac{w_i+w_j}{2}$ with $\beta=0.5$ lead to an exponential distribution of wealth, $P(w)=\frac{1}{\bar w}e^{-w/\bar w}$. Theoretical results are given by the dark blue dashed line. (b) The Lorenz curve for the exponential distribution is given by $y(x) = x + (1-x)\log (1-x)$. (c) The Gini index approaches the theoretical equilibrium value of an exponential distribution, $G=1/2$.}
    \label{dragulescu_detailed-balance}
\end{figure}

Simulations of the Dragulescu model with $N=2000$ agents confirm the convergence of the wealth distribution towards the Boltzmann distribution (Fig.~\ref{dragulescu_detailed-balance}a).

The Gini index of the exponential distribution can be computed using the Lorenz curve (Eq.~\ref{eq:lorenzcurve}). The cumulative population density $x(w)$ and the cumulative wealth density $y(w)$ are given by 

\begin{equation*}
    x(w) = \int_0^w \frac{1}{\bar w}e^{-w'/\bar w}\dw' = 1-e^{-w/\bar w}\,\hspace{0.5cm}\text{and}\hspace{0.5cm} y(w) = \frac{1}{\bar w}\int_0^w w'\frac{1}{\bar w}e^{-w'/\bar w}\dw' =  x(w) -\frac{w}{\bar w}e^{-w/\bar w}\,.
\end{equation*}

Replacing $\frac{w}{\bar w}$ by $-\log(1-x)$ and $e^{-w/\bar w}$ by $1-x$ yields the analytical Lorenz curve $y(x) = x + (1-x)\log (1-x)$ for the Boltzmann distribution, which is, notably, independent of the average wealth $\bar w$ \cite{dragulescu_exponential_2001}. Using $y(x)$, the Gini index can be computed as the ratio of the area between the Lorenz curve and the diagonal line to the area beneath the diagonal (Eq.~\ref{eq:gini_using_lorenz_curve}),

\begin{equation*}
    G =  2 \int_0^1 \dx \left(x-\left(x + (1-x)\log (1-x)\right)\right) = 0.5\,,
\end{equation*}
which is also confirmed numerically in simulations (Fig.~\ref{dragulescu_detailed-balance}b,c). The fact that the Lorenz curve is independent of the mean wealth $\bar w$ implies that the Gini index is invariant against an increase of mean wealth, that could be interpreted as inflation.

The preceding derivation made no explicit assumption about the specific form of $\Delta w$. It solely relied on the conservation of wealth and deduced the Boltzmann distribution through the maximisation of entropy. One might thus anticipate that any wealth-conserving exchange rule naturally results in a Boltzmann steady-state distribution through entropy maximisation. However, this assumption is incorrect. The Boltzmann distribution arises only for \textit{additive} exchange rules, which $\Delta w \sim w_i+w_j$ is an example of \cite{yakovenko_colloquium_2009,ispolatov_wealth_1998}. It implies that wealth can be transferred to and from agents at the same rate in both directions, which is linked to the concept of detailed balance, described in the following section.

\subsection{Detailed balance in asset exchange models}
\label{sec:detailedbalance_aems}

In order to understand the conditions under which the Boltzmann distribution may or may not represent the equilibrium wealth distribution of asset exchange models, we consider an alternative derivation of the steady state distribution, using the Boltzmann equation. The Boltzmann equation was originally formulated by Ludwig Boltzmann to describe a dilute gas, where only two-body molecular collisions are significant \cite{krapivsky_kinetic_2010,zwanzig2001nonequilibrium}. Due to the two-natured interactions in asset exchange models, it also finds application in solving those \cite{dragulescu_statistical_2000}. Generally, Boltzmann wrote the change of particle distribution density $P(r,v,t)$, describing the state of a gas given the positions $r$ and velocities $v$, as 

\begin{equation}
    \left(\frac{\partial}{\partial t} + v\frac{\partial}{\partial r} + a\frac{\partial }{\partial v}\right) P(r,v,t)= \left(\frac{\partial P}{\partial t}\right)_{\text{collision}}\,.\label{eq:boltzmanneq_general}
\end{equation}

The left hand side of Eq.~\ref{eq:boltzmanneq_general} describes the change in $P$ by particle motion in the absence of collisions. It includes a term to account for intrinsic particle motion due to advection, as well as a term for particle motion due to an external force that causes acceleration $a$ \cite{krapivsky_kinetic_2010}. The right hand side of Eq.~\ref{eq:boltzmanneq_general} describes how collisions influence the particle density and its form depends on the nature of the interaction \cite{zwanzig2001nonequilibrium}. It generally involves an integral over the momentum space as only particles with the same position $r$ can interact. Besides, it generally includes a gain and a loss term, representing the different ways in which particles with position $r$ and velocity $v$ are gained or lost. For example, it could be given by

\begin{equation}
    \left(\frac{\partial P}{\partial t}\right)_{\text{collision}} = \int \sigma [-P(r,v;r,w) + P(r,v';r,w')]\, \delta(v+w-v'-w')\delta(v^2+w^2-v'^2-w'^2)\,\text{d}v'\,\text{d}w\,\text{d}w'\,, \label{eq:boltzmann_collision_term}
\end{equation}

where $\sigma$ represents the cross section of the interaction and the $\delta$ functions ensure conservation of momentum and energy, respectively \cite{krapivsky_kinetic_2010}. The loss term $-P(r,v;r,w)$ accounts for particles with velocity $v$ to be lost due to collisions with particles of velocity $w$. Similarly, the gain term $+ P(r,v';r,w')$ produces particles of velocity $v$ by a collision with particles of velocities $v'$ and $w'$. Importantly, in this form Eq.~\ref{eq:boltzmann_collision_term} is not a closed equation because the single particle distribution $P(r,v)$ is expressed via the two particle distribution $P(r,v;r,w)$. One could proceed by writing an equation for the two particle distribution, but it would involve the three particle distribution, and so on, leading to an infinite hierarchy of equations, known as the BBGKY hierarchy \cite{boghosian_kinetics_2014,krapivsky_kinetic_2010}. Boltzmann resolved this issue at the very beginning by approximating the two particle distribution as the product of one particle distributions:

\begin{equation}
    P(r,v;r,w) = P(r,v)P(r,w)\,.
\end{equation}

This is known as the \textit{molecular chaos approximation} or \textit{Stosszahlansatz}, as originally called by Boltzmann \cite{krapivsky_kinetic_2010}. It is valid when one can neglect inter-particle correlations, and thus represents the mean-field approximation of the process. With the molecular chaos approximation, one can write the change of particle density as a closed equation. Unfortunately, due to the complexity of the problem, an analytical solution of the Boltzmann equation is lacking. It involves a partial integro differential equation of usually at least 6 variables --- 3 dimensions for both position and velocity. Moreover, for a realistic system, one often needs to account for vibrational or rotational degrees of freedom, adding to further complexity \cite{krapivsky_kinetic_2010}. 

In asset exchange models the only variable is wealth $w$ and collisions are the only way in which the wealth distribution $P(w)$ changes. This leads to a significantly simplified version of the Boltzmann equation. For the exchange of wealth $\Delta w$ the Boltzmann equation is given by \cite{dragulescu_statistical_2000}

\begin{align}
    \frac{\partial P}{\partial t} = \int \int \dw'\text{d}\Delta w\, \{ & - P( w\pm \Delta w, w'\mp \Delta w|w,w') P(w,w') \notag \\ 
    &+ P(w,w'|w\pm \Delta w, w'\mp \Delta w )P(w\pm \Delta w,w'\mp \Delta w )\} \,. \label{eq:boltzmann_eq_dragulescu}
\end{align}

Here $P( w\pm \Delta w, w'\mp \Delta w|w,w')$ represents the rate at which wealth $\Delta w$ is exchanged between two agents of wealth $w$ and $w'$, leasing to a decrease in $P(w)$. Similarly, $P(w,w'|w\pm \Delta w, w'\mp \Delta w )$  represents the rate at which agents with wealth $w$ emerge through interactions of agents with wealth $w\pm \Delta w$ and $w' \mp \Delta w$, leading to an increase in $P(w)$. Eq.~\ref{eq:boltzmann_eq_dragulescu} is not a closed form equation as it describes the change of the one agent distribution $P(w)$ by the two agent distribution $P(w,w')$. Analogously to Boltzmann's molecular chaos approximation, one can approximate $P(w,w')=P(w)P(w')$. This assumption, also called the \textit{random agent approximation}, is rather strong \cite{boghosian_kinetics_2014}. It implies that two agents entering a transaction are uncorrelated. In reality this is often violated, e.g., as soon as choosing to frequent the same grocery store instead of choosing one at random \cite{boghosian_kinetics_2014}. 

For the purpose of solving asset exchange models, where agents are drawn randomly, one can neglect the correlations. Solving for the steady state distribution of Eq.~\ref{eq:boltzmann_eq_dragulescu}, $\frac{\partial P}{\partial t}$ must be equal to zero. If the integrand itself vanishes for any $w'$ and $\Delta w$, the process obeys \textit{detailed balance}. It leads to the condition

\begin{equation*}
    P( w\pm \Delta w, w'\mp \Delta w|w,w') P(w)P(w') = P(w,w'|w\pm \Delta w, w'\mp \Delta w )P(w\pm \Delta w)P(w'\mp \Delta w)\,.
\end{equation*}

If the forward rate $P( w\pm \Delta w, w'\mp \Delta w|w,w')$ is the same as the backward rate $ P(w,w'|w\pm \Delta w, w'\mp \Delta w )$, an exponential distribution provides the solution to $P(w)P(w')=P(w\pm \Delta w)P(w'\mp \Delta w)$ and the Boltzmann distribution for wealth exchange $P(w)=\frac{1}{\bar w}e^{-w/ \bar w}$ is recovered. 

A process for which the forward and the backward rates $P(\cdot | \cdot )$ are the same is said to obey \textit{time reversal symmetry} \cite{zwanzig2001nonequilibrium,krapivsky_kinetic_2010} or to be \textit{microscopically reversible}. It implies that every process has the same probability of occurring in reverse order. Thus, in principle, by reverting time, one could restore the initial state of the system. If violated, the process is also said to break time reversal symmetry.

For asset exchange models, it implies that  only transfer rules that are microscopically reversible lead to a Boltzmann distribution \cite{yakovenko_colloquium_2009}. In the case of the wealth exchange considered in this section, $\Delta w = (1-\beta)\frac{w_i+w_j}{2}$, it is evident that this process is reversible. If, after a first transaction, wealth has changed to $w_i' = w_i + (1-\beta)\frac{w_i+w_j}{2}$ and $w_j = w_j - (1-\beta)\frac{w_i+w_j}{2}$, and the same two agents meet again, $\Delta w$ will not have changed compared to the previous exchange. Thus, in the second transaction $\Delta w$ can be transferred back to the agent who previously lost $\Delta w$ and the initial state of the system is restored. Hence, time reversal symmetry is obeyed and the system leads to the Boltzmann distribution. In Sec.~\ref{sec:markovchains} we will study this in more detail, where we view asset exchange models as Markov processes, and show that those obeying time reversal symmetry correspond to reversible Markov chains. If time symmetry reversal is broken, the steady state distribution does not follow the Boltzmann distribution, and instead, depends on subtleties of the exchange rules. In the following section we will give an example of such an asset exchange model.

% Maths Box on detailed balance. Cite alberty2004principle and gnesotto2018broken

\subsection{Ispolatov model \& broken time reversal symmetry}
\label{sec:ispolatov-broken}

An asset exchange model breaking time reversal symmetry is given by the exchange rule where the amount of wealth exchanged $\Delta w$ is a fraction of the paying agent's wealth. We will refer to this model as \textit{Ispolatov model} \cite{ispolatov_wealth_1998}. The exchange is given by 

\begin{equation}
    w_i \to w_i' = w_i - (1-\beta)w_i \aand w_j \to w_j' = w_j + (1-\beta)w_i\,,\label{eq:delta_w_espolatov_losers_rule}
\end{equation}
where $i$ is the paying agent and is determined by a coin flip \cite{ispolatov_wealth_1998}. In the following we will also refer to the paying agent as \textit{losing} agent, as an asset exchange resembles a gamble on an amount $\Delta w$ with a winner and a loser. Note that in this exchange rule the winning agent is determined before the amount that is being transferred, as opposed to the previous rule where $\Delta w$ was determined first.

The exchange rule Eq.~\ref{eq:delta_w_espolatov_losers_rule} breaks time reversal symmetry, as can be seen when trying to reverse the process in time. After an initial transaction $\Delta w = (1-\beta)w_i$, where $i$ pays and $j$ receives, the amount transferred in the next interaction, when $j$ is the payer, would be $\Delta w' = (1-\beta)w_j' = (1-\beta)(w_j + (1-\beta)w_i)$. Hence, after the second transaction agent $i$ would have $w_i'' = w_i + (1-\beta)(w_j + (1-\beta)w_i) \neq w_i$. The system does not return to its original state and the equilibrium wealth distribution does not take the form of the Boltzmann distribution \cite{ispolatov_wealth_1998,dragulescu_statistical_2000}.

To derive the non-Boltzmann equilibrium distribution, one can write down the Master equation of the transaction rule, which governs how the agent distribution $P$ evolves in time \cite{ispolatov_wealth_1998}:

\begin{equation}
    \frac{\partial P(w,t)}{\partial t } = \frac{1}{2}\int \int \dw'\dw'' P(w')P(w'') [-\delta(w-w') - \delta(w-w'') + \delta(w''\beta -w) + \delta(w''+(1-\beta)w'-w)]\label{eq:master_eq_ispolatov}
\end{equation}

The first two terms $-\delta(w-w')$ and $-\delta(w-w'')$ account for interactions of agents with wealth $w$ and are therefore negative contributions, as after the exchange they no longer have wealth $w$. The third term $\delta(w''\beta -w)$ arises if an agent with $w''$ interacts with an agent $w'$ and loses $\Delta w = (1-\beta)w''$. His or her wealth changes accordingly to $w = w'' - (1-\beta)w'' = \beta w''$. The third term $ \delta(w''+(1-\beta)w'-w)$ describes an agent with $w''$ interacting with an agent with $w'$ and making a profitable transaction of $(1-\beta)w'$. His or her wealth thus becomes $w=w''+(1-\beta)w'$.

Integrating Eq.~\ref{eq:master_eq_ispolatov} once, the expression leads to \cite{ispolatov_wealth_1998}

\begin{equation}
    \frac{\partial P(w,t)}{\partial t } = -P(w) + \frac{1}{2\beta}P\left(\frac{w}{\beta}\right) + \frac{1}{2(1-\beta)}\int_0^w \dw''\, P(w'')P\left(\frac{w-w''}{1-\beta}\right)\,.\label{eq:ispolatov_master_integrated}
\end{equation}

In order to see whether the Boltzmann distribution is a solution to the steady state $\frac{\partial P}{\partial t} = 0$, one can insert a test solution $P(w) = \frac{1}{\bar w}e^{-w/\bar w}$, which reveals that the equation only holds for $\beta=\frac{1}{2}$ and the Boltzmann distribution is thus only a special case of the steady state distribution \cite{ispolatov_wealth_1998}. 

Generally, the steady state distribution takes on a more complex form, which is not fully analytically derivable. However, the fact that the system settles into a steady state can be seen by analysing the moments of $P(w)$. Only if all moments of $P(w)$ become constant in time, will $P(w)$ be in a steady state satisfying $\frac{\partial P(w,t)}{\partial t} = 0$. The time evolution of the moments is given by 

\begin{equation}
    \dot m_n = \frac{dm_n}{dt} =  \int_0^\infty \frac{\partial P(w,t)}{\partial t}\,w^n\dw\,. \label{eq:moment_timederivative}
\end{equation}

The zeroth and the first moment are conserved, $\dot m_0 = \dot m_1 = 0$, as they should, in order to conserve agents and wealth. Inserting Eq.~\ref{eq:ispolatov_master_integrated} into Eq.~\ref{eq:moment_timederivative}, the second moment obeys 

\begin{align*}
    \dot m_2(t) &= -\int_0^\infty w^2\,P(w)\dw + \frac{1}{2\beta}\int_0^\infty w^2\,P\left(\frac{w}{\beta}\right)\dw + \frac{1}{2(1-\beta)}\int_0^\infty  \left(\int_0^w\dw'P\left(w'\right)P\left(\frac{w-w'}{1-\beta}\right)\right)w^2 \dw\\
    &= -m_2 + \frac{\beta}{2}m_2 + \frac{1}{2(1-\beta)}\int_0^\infty \left(\int_0^\infty w^2 P\left(w'\right)P\left(\frac{w-w'}{1-\beta}\right)\dw \right) \dw' \\
    &= -\beta(1-\beta)m_2 + (1-\beta)m_1^2\,,
\end{align*}

where for the last term, one can swap the order of integration to first integrate over $w$ and then substitute $u = \frac{w-w'}{1-\beta}$. Using $m_1 = \bar w$, the solution to the differential equation for $m_2$ is 

\begin{equation*}
    m_2(t) = \frac{\bar w}{\beta} + \left(m_2\left(0\right) -\frac{\bar w}{\beta }\right)e^{-\beta(1-\beta)t}\,.
\end{equation*}

Hence, the second moment converges exponentially to a constant $\frac{\bar w}{\beta}$. Similarly, higher moments also converge exponentially to constant values, which signifies that the wealth distribution stabilises in its steady state \cite{ispolatov_wealth_1998}. This is to be expected for the considered exchange, because rich agents have far more to lose in an exchange than poor agents, which stabilises the wealth distribution. In the next section Sec.~\ref{sec:yardsale-wealthcondensatoin} we will show an example where this is not the case.

Generally, one can distinguish between additive and multiplicative exchange rules. The first exchange rule considered, $\Delta w \sim w_i+w_j$, described an additive exchange, where the amount of wealth transferred $\Delta w$ does not change when two agents meet repeatedly. Contrarily, $\Delta w \sim w_i$ with $i$ the paying agent, is multiplicative in nature as $\Delta w$ is a fraction of one of the agent's wealth. This violates time reversal symmetry, which is why multiplicative exchange rules generally do not result in Boltzmann steady state distributions \cite{yakovenko_colloquium_2009,ispolatov_wealth_1998}. % Perhaps add something with Maxwell's demon. 

Unlike in physics, there is no fundamental reason why economic transactions should obey time-reversal symmetry \cite{dragulescu_statistical_2000}. For the exchange $\Delta w \sim w_i$ with $i$ the paying agent, the wealth distribution reaches a stationary state as rich agents have far more wealth to lose in a transaction than poor agents. This refrains them from ever accumulating more wealth and thus leads to a stationary distribution. However, in a realistic setting, no profit oriented agent would engage in transactions with poorer agents for such an exchange, as the expected loss would far exceed the expected gain. Hence, another model was introduced in which the wealth amount at stake is a fraction of the poorer agent, $\Delta w \sim \text{min}(w_i,w_j)$. As will be explained in the next section, this subtle difference will give rise to the emergence of an important phenomenon in asset exchange models --- wealth condensation.

\subsection{Yard-Sale model \& wealth condensation}
\label{sec:yardsale-wealthcondensatoin}

The two preceding exchange rules for $\Delta w$ are arguably unrealistic, as they allow poor agents to generate large profits when interacting with richer agents. A model that addresses this issue is known as the \textit{Yard-Sale model}, characterised by the exchange rule \cite{chakraborti_distributions_2002}

\begin{equation*}
    \Delta w = (1-\beta)\text{min}(w_i,w_j)\,,
\end{equation*}

where $\beta$ again reflects risk aversion. 
This exchange was termed the Yard-Sale model as it resembles the concept of a yard sale or garage sale, where individuals bring items they no longer need or want and offer them for sale \cite{hayes_computing_2002}. The fact that the wealth exchanged $\Delta w$ is always a fraction of the poorer agent's wealth, restricts poorer agents from making large profits when interacting with richer agents. 

The master equation of the Yard-Sale model is given by \cite{boghosian_kinetics_2014}

\begin{align*}
    \frac{\partial P(w,t)}{\partial t} &= \frac{1}{2}\int_0^\infty \int_0^\infty \dw'\dw'' P(w')P(w'')\, [-\delta\left(w'-w\right) - \delta\left(w''-w\right) \\
    &+ \theta\left(w-\left(2- \beta\right)w'\right)\delta\left(w''-w+(1-\beta) w'\right) + \theta\left(\left(2- \beta\right)w' - w\right)\delta\left(w''\left(2- \beta\right)-w\right) \\
    &+ \theta\left(w-\beta w'\right)\delta\left(w''-w-(1-\beta) w'\right) + \theta\left(\beta w'-w\right)\delta\left(\beta w'' - w\right)  ]\,,
\end{align*}

where $\theta$ denotes the Heaviside step function. The two negative contributions in the first line arise again through interactions of agents with wealth $w$, who subsequently no longer possess $w$. The two positive contributions in the second row arise from winning interactions of agents with $w''>w'$ and $w''<w'$, respectively. In the former, the agent wins $(1-\beta)w'$ and has subsequently $w=w''+(1-\beta)w' > (2-\beta)w'$, and in the latter the agent wins $(1-\beta)w''$ and has subsequently $w=(2-\beta)w'' < (1-\beta)w'$. Similarly, the two contributions in the last line account for losing interactions for $w''$ with $w''>w'$ and $w''<w'$, respectively \cite{boghosian_kinetics_2014}. 

Integrating over the delta functions and utilising agent conservation $\int_0^a P(w)\dw + \int_a^\infty P(w)\dw = 1$, one obtains 

\begin{align}
    \frac{\partial P(w,t)}{\partial t} &= -P(w) + \frac{1}{2(2-\beta)}P\left(\frac{w}{2-\beta}\right)\left(1-\int_0^{\frac{w}{2-\beta}}P(w')\dw'\right) + \frac{1}{2\beta}P\left(\frac{w}{\beta}\right)\left(1-\int_0^{\frac{w}{\beta}}P(w')\dw'\right) \notag \\
    & + \frac{1}{2} \int_0^{\frac{w}{2-\beta}}P(w')P(w-(1-\beta)w')\dw' + \frac{1}{2} \int_0^{\frac{w}{\beta}}P(w')P(w+(1-\beta)w')\dw'\,.\label{eq:yard-sale-master-eq-after-int}
\end{align}

A first insight into the behaviour of the system can again be obtained by analysing the behaviour of the moments $m_n(t)$. The zeroth and first moment are conserved, as required \cite{boghosian_oligarchy_2017}. However, the behaviour of the second moment indicates that the Yard-Sale model behaves qualitatively different than the previous exchange rules encountered. In order to compute $\dot m_2$, one can see from Eq.~\ref{eq:yard-sale-master-eq-after-int} that the second and third term in the first row cancel out, as one integrates all the way to $w =\infty$, so that the integrals over $w'$ become unity. The remaining terms are 

\begin{align*}
    \dot m_2 &= -m_2 + \frac{1}{2}\int \left(\int_0^{\frac{w}{2-\beta}}P(w')P(w-(1-\beta)w') \dw'\right)w^2\dw  \\
    &\hspace{1.35cm} +\frac{1}{2}\int \left( \int_0^{\frac{w}{\beta}}P(w')P(w+(1-\beta)w')\dw' \right)w^2\dw\,,
\end{align*}

where one can again swap the order of integration to obtain $\dot m_2= (1-\beta)^2 m_2$. Thus, the second moment diverges exponentially as $m_2(t) \propto e^{(1-\beta)^2t}$. This indicates that the wealth distribution never settles into a steady state. 

\begin{figure}
    \centering
    \includegraphics[width=6in]{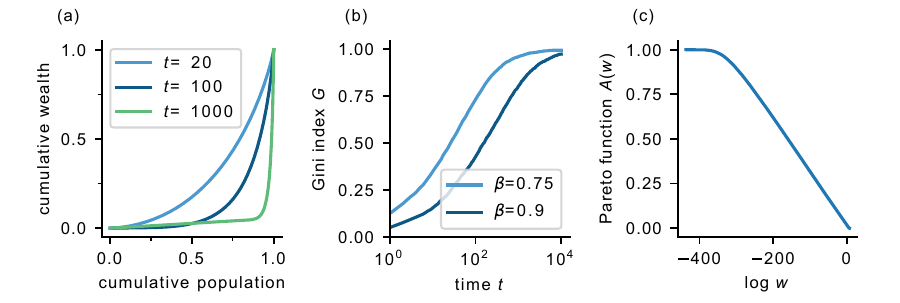} 
    \caption{\textbf{The wealth distribution in the Yard-Sale model approaches a condensed state.} (a) With time, an increasingly small fraction of the population holds an increasing portion of the wealth, as shown by the Lorenz curves. (b) The Gini index $G$ approaches 1, indicating wealth condensation as the outcome of the Yard-Sale model. Higher risk aversion $\beta$ merely prolongs the time it takes to reach that state. (c) The Pareto distribution of the logarithm of wealth, $A(\log w) = \int_{\log w}^\infty P(w')\dw'$ at time $t=\SI{10000}{}$ shows a linear decrease of the form $A(w) = b(t) - a(t)\log w$, indicating a Pareto distribution with Pareto exponent $\alpha = 0$. Parameters used: $\beta=0.75$, $N=\SI{5e5}{}$ and $W=\SI{5e7}{}$.}
    \label{fig:yard-sale}
\end{figure}

The solution of the Yard-Sale model is indeed \textit{wealth condensation}, a state in which a vanishingly small fraction of agents accumulates all available wealth, and leaves the rest of the population at zero wealth \cite{cardoso_equal_2022,boghosian_h_2015,boghosian_kinetics_2014}. The time it takes to reach that state takes infinitely long, which is why the moments of $P(w)$ never converge. When simulating the exchange with discrete agents, one agent eventually accumulates all available wealth, while all others are left with zero wealth. The Lorenz curve and Gini index show increasing inequality with time, with the latter steadily increasing  and approaching the value of $G=1$ (Fig.~\ref{fig:yard-sale}a,b).

While the condensed steady state of the Yard-Sale model is not an adequate representation of empirical wealth distribution, one may ask if it ever resembles such distributions on its way to the condensed state. Namely, if it at some point features a power-law tail with Pareto exponent $\alpha \approx 1.5$ \cite{chakrabarti2013econophysics}. Simulations show that this is not the case, as the Pareto function $A(w)$ closely follows $A(w) \approx b(t) - a(t)\log w$ (Fig.~\ref{fig:yard-sale}c). Differentiating both sides leads to $P(w) \approx \frac{a(t)}{w}$ and hence a Pareto exponent of $\alpha \approx 0$. While the distribution of the Yard-Sale model thus goes through a series of power-laws, its Pareto exponent is considerably smaller than in empirical wealth data. Extensions of the Yard-Sale model are required to make it an adequate fit to data, featuring additional affects such as wealth tax and increased rich advantages in the asset exchange, as will be explained later \cite{boghosian_economically_nodate}.

\subsubsection{Understanding the causes of wealth condensation}

% give more examples, like in the presentation, on why wealth condensation happens. Also explain like in boghosian2020economically, why it happens (exponential decay of wealth)

The fact that the Yard-Sale model leads to a condensed state seems counter-intuitive at first, as the winning agent is selected with even odds. However, the fact that $\Delta w$ always constitutes a fraction of the poorer agent's wealth enables the wealthier agent to endure longer sequences of losses. Thus, once an agent becomes poor, which eventually happens because of pure randomness, escaping this situation becomes exceedingly challenging. With time, this happens to all agents except the richest, resulting in a state of wealth condensation.

\begin{figure}
    \centering
    \includegraphics[width=6in]{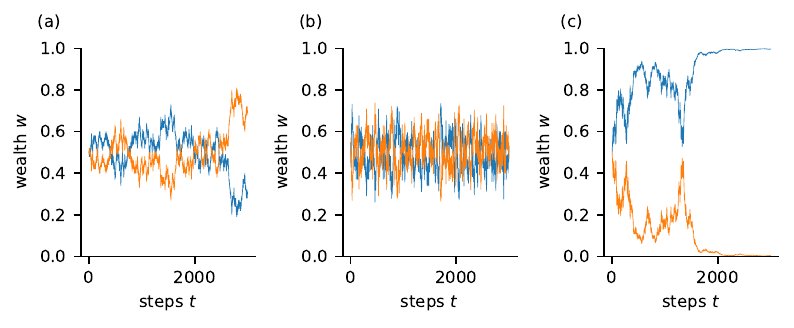}
    \caption{\textbf{A simplified economy with only two agents reveals the different nature of the transaction rules.} For a model with $N=2$ agents and total wealth $W=1$, these same two agents meet at every time step and exchange wealth $\Delta w$. (a) For the rule $\Delta w =(1-\beta)\frac{w_1+w_2}{2} = \frac{1-\beta}{2}$ the step size is constant, leading to a normal random walk confined in the region $0\leq w \leq 1$. (b) For the rule $\Delta w = (1-\beta)w_i$, where $i$ is the paying agent, the step length is always a fraction of the wealth of the downward moving agent. Thus, the richer agent has more to lose, which stabilises the wealth of both agents around the equitable state $w_1=w_2 = 1/2$. (c) For the rule $\Delta w = (1-\beta)\text{min}(w_1,w_2)$ the step length decreases as the agents move apart and one of the two comes closer to the line $w=0$. This makes it increasingly hard to recover wealth, leading to wealth condensation in the Yard-Sale model.}
    \label{fig:zenowalk_randomwalk}
\end{figure}

To comprehend why this model behaves so differently than the previous ones, one can consider the asset exchange in a simplified economy of just two agents, $N=2$, a total amount of wealth $W=1$, and initial conditions $w_1 = w_2 = \frac{1}{2}$ (Fig.~\ref{fig:zenowalk_randomwalk}a-c). The asset exchange dynamics is then equivalent to a random walk with the step length depending on the agent's wealth. 

For the first exchange rule, $\Delta w = (1-\beta)\frac{w_i+w_j}{2}$, the step length remains constant at $1-\beta$, as the total wealth $w_1+w_2 = 1$ is conserved. Thus, the dynamics is a normal random walk confined to the region $0\leq w \leq 1$ (Fig.~\ref{fig:zenowalk_randomwalk}a). For the second exchange $\Delta w = (1-\beta)w_i$ (with $i$ the losing agent), the step length is given by a fraction of the agent moving downwards. The poorer agent, namely the one closer to the bottom line $w=0$, makes smaller steps when going down but larger steps when going up. This drives the system to the equal state around $w=1/2$ (Fig.~\ref{fig:zenowalk_randomwalk}b).

In contrast, in the Yard-Sale model the dynamics is equivalent to a random walk with the step length being proportional to $\text{min}(w_1,w_2)$. This equals the shortest distance to any of the two boundaries $w=0$ or $w=1$. As either agent approaches $w=0$, the step length diminishes, making it challenging for that agent to recover their wealth (Fig.~\ref{fig:zenowalk_randomwalk}c). In a model with more agents this increasing wealth disparity eventually leads to a condensed state.

\subsubsection{Proof of wealth condensation in the Yard-Sale model}

There are various ways to prove that an asset exchange mode following Yard-Sale dynamics leads to wealth condensation \cite{boghosian_h_2015,cardoso_equal_2022}. One way is to write the temporal change of the Gini index as 

\begin{equation}
    \frac{dG}{dt} = \int_0^\infty \dw\, \frac{\delta G}{\delta P(w)}\frac{\partial P(w,t)}{\partial t}\,, \label{eq:dGdt_frechet}
\end{equation}

where $\frac{\delta G}{\delta P(w)}$ is the Fréchet derivative of $G[P]$, which is the analogue of the gradient in function space \cite{frigyik2008introduction}. Eq.~\ref{eq:dGdt_frechet} can thus be seen as an infinite dimensional version of the ordinary chain rule $\frac{df(r)}{dt} = \frac{\partial f(r)}{\partial r}\frac{dr}{dt}$.

Computing $\frac{\delta G}{\delta P(w)}$ and inserting Eq.~\ref{eq:yard-sale-master-eq-after-int}, one can show that $\frac{dG}{dt} \geq 0$ and that $G(t)$ asymptotically approaches the value of $G=1$. This proof of wealth condensation closely resembles Boltzmann's $H$-Theorem, an early attempt to prove the second law of thermodynamics using the Boltzmann equation (Sec.~\ref{sec:theory:boltzmann-htheorem} for details) \cite{krapivsky_kinetic_2010,kardar2007statistical,zwanzig2001nonequilibrium}.

While inserting the master equation of the Yard-Sale model (Eq.~\ref{eq:yard-sale-master-eq-after-int}) into the change of Gini index Eq.~\ref{eq:dGdt_frechet} provides the correct solution, the computation is tedious. A simpler way is to assume that risk aversion is large ($\Delta w$ is small), and wealth thus only changes in small steps. This allows to derive a Fokker-Planck equation for $P(w)$ which takes a simpler form than the master equation and thus allows to prove wealth condensation more easily \cite{boghosian_fokkerplanck_2014,boghosian_h_2015}.

\subsubsection{Fokker-Planck equation of the Yard-Sale model}

In order to derive the Fokker-Planck equation for the Yard-Sale model, one has to take the \textit{small transaction limit} of the asset exchange dynamics. For example, if we interpret $\Delta w$ as the profitability of a transaction under imperfect fairness, it is expected that $\Delta w \ll \bar w$. 

When wealth $w$ fluctuates in small, stochastic steps $\Delta w$, its probability distribution $P(w)$ adheres to the Fokker-Planck equation (Sec.~\ref{sec:theory:fokkerplanck}) \cite{boghosian_fokkerplanck_2014,risken1996fokker}

\begin{equation}
    \frac{\partial P(w,t)}{\partial t} = -\frac{\partial }{\partial w}[\langle \Delta w\rangle P(w)] + \frac{\partial^2}{\partial w^2}\left[\frac{\langle \Delta w\rangle ^2}{2}P(w)\right]\,.\label{eq:fokkerplanck-general}
\end{equation}

The first term is called the drift term and the second term is called the diffusion term \cite{risken1996fokker}. 
Taking the average of the transaction size $\langle \Delta w\rangle$ over the distribution $P(w)$ results in $\langle \Delta w\rangle=0$, namely there is no deterministic drift affecting the wealth (details in Sec.~\ref{sec:theory:fokkerplanck}). Instead, all its change is due to the stochastic diffusion term resulting from $\langle \Delta w\rangle ^2$, which is nonzero, and leads to the Fokker-Planck equation of the Yard-Sale model, 

\begin{equation}
     \frac{\partial P(w,t)}{\partial t} =  (1-\beta)^2 \frac{\partial^2}{\partial w^2}\left[\left(\frac{w^2}{2}\int_w^\infty \dw'\, P(w')+\int_0^w \dw'\,P(w')w'^2\right)P(w)\right]\,.\label{eq:fokkerplanck-yardsale}
\end{equation}

This equation is a nonlinear integrodifferential equation and represents a state dependent diffusion process that leads to wealth condensation. While Eq.~\ref{eq:fokkerplanck-yardsale} does not admit a closed-form solution, the Fokker-Planck description is useful to prove wealth condensation by inserting it into the change of Gini index Eq.~\ref{eq:dGdt_frechet} (Sec.~\ref{sec:theory:boltzmann-htheorem} for details).

Overall, wealth condensation in the Yard-Sale model arises because in every transaction the poorer agent has a fraction of his or her wealth at stake. In similar fashion, a model with $\Delta w \propto \frac{w_iw_j}{w_i+w_j}$ also leads to wealth condensation \cite{iglesias_entropy_2012,cardoso_wealth_2021}. However, a condensed state does not reflect real-world wealth distribution either, prompting extensions to the model for a better fit to empirical data. Two popular extensions are effects of tax mechanisms and adjusting the probability of having profitable interactions.

\subsubsection{Additions and modifications to the Yard-Sale model}
\label{sec:taxmechanisms}

\begin{figure}
    \centering
    \includegraphics{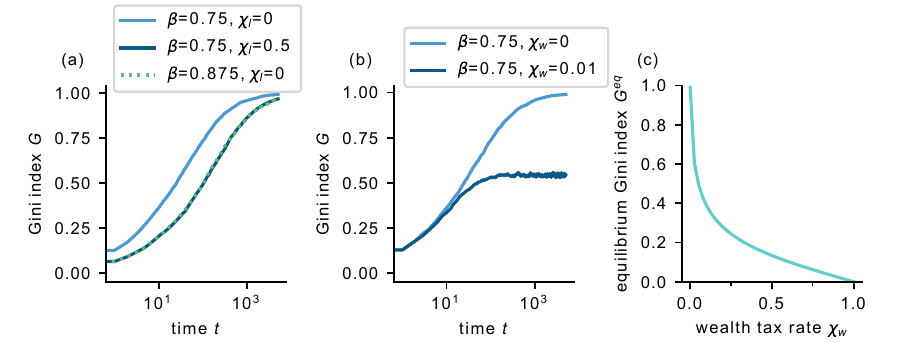}
    \caption{\textbf{Effects of income and wealth tax in the Yard-Sale model.} (a) An income tax is applied as $w_i \to w_i - \chi_l (w_i(t)-w_i(t-1))$ after one asset exchange step, i.e., $N/2$ transactions of value $\Delta w$. It can be absorbed into a higher value of risk aversion $\beta$ and thus merely slows down wealth condensation. For example, risk aversion $\beta=0.75$ and income tax rate $\chi_l=0.5$ features the same dynamics as risk aversion $\beta=0.875$ (dark blue line and dotted green line). (b) In contrast, a wealth tax applied as $w_i \to w_i + \chi_w(\bar w - w_i)$ prevents wealth condensation. (c) The equilibrium Gini index $G^{\rm eq}$ decreases steeply for small wealth tax rates and reaches $G^{\rm eq}$ for full redistribution with $\chi_w=1$.}
    \label{fig:yard-sale_tax}
\end{figure}

Real economies often encompass tax mechanisms, which can be instituted by governments to finance their operations, but also to potentially mitigate wealth inequality. However, the extent to which tax mechanisms contribute to the reduction of inequality depends on their implementation --- various forms of taxation exist, including income tax, wealth tax, and inheritance tax. In the following we will consider income and wealth taxes, as these can be implemented straightforwardly, without having to introduce modifications such as finitely living agents.

\textbf{Income tax}

To establish the framework for an income tax, the concept of income has to be defined first. Considering that asset exchange models do not feature wages, we define income, denoted as $l(t)$, as the alteration in an agent's wealth between consecutive time steps $t-1$ and $t$ \cite{banzhaf_effects_2021}:

\begin{equation*}
    l_i(t) = w_i(t) - w_i(t-1)\,.
\end{equation*}

One time step consists of $N/2$ transactions. Income thus results from the asset exchange, and can be positive or negative. An income tax mechanism with income tax rate $\chi_l$ can be defined as 

\begin{equation*}
    w_i(t) \to w_i'(t) = w_i(t) - \chi_l l_i(t) = \chi_l w_i(t-1) + (1-\chi_l)w_i(t)\,, 
\end{equation*}

which acts on all agents after one asset exchange time step. It becomes apparent that this income tax formulation does not prevent wealth condensation. For an individual agent $i$ generating profits, namely $l_i(t) > 0$, the income tax translates merely into a reduction in the magnitude of their earnings by a factor of $1-\chi_l$. Conversely, losses are similarly mitigated. Hence, the influence of the income tax rate $\chi_l$ can be assimilated into the agent's risk aversion parameter, $\beta$, only resulting in a deceleration of wealth condensation (Fig.~\ref{fig:yard-sale_tax}a). 

An alternative definition of income tax could involve constraining the definition of income to positive values, thereby imposing tax solely on agents with $l(t) > 0$, followed by uniform redistribution of the accumulated tax revenue. However, also this adjusted tax mechanism does not prevent wealth condensation \cite{banzhaf_effects_2021}. Instead, preventing wealth condensation in the Yard-Sale model necessitates a wealth tax, which directly redistributes wealth from the upper part of the distribution to the lower parts.

\textbf{Wealth tax}

Consider an implementation of a wealth tax as follows. After each asset exchange step, a fraction $\chi_w$ of every agent's wealth is collected and subsequently redistributed among the population \cite{banzhaf_effects_2021,boghosian_kinetics_2014}. Furthermore, we assume that it is redistributed equally among the population, thus every agent receiving a fraction $\frac{1}{N}$ of the total collected tax amount. This redistribution alters the wealth of agent $i$ according to

\begin{equation}
    w_i \to w_i' = w_i + \chi_w(\bar w-w_i)\,, \label{eq:wealthtax}
\end{equation}

where $\chi_w \bar w = \frac{1}{N}\sum_i^N \chi_w w_i$ is the equally redistributed portion of wealth. Agents with wealth above the average $\bar w$ experience a decrease in wealth, while those with wealth below the average experience an increase. If only this taxation step were implemented, without any asset exchange, all agents' wealth would converge to the mean $\bar w$. Thus, taxation steers the system towards complete equality $P(w)=\delta(w-\bar w)$, whereas the Yard-Sale exchange dynamics steer the system towards maximal inequality. 

In combination, it is evident that the wealth tax prevents wealth condensation for any positive tax rate $\chi_w > 0$. Agents with wealth close to zero benefit from the taxation step by gaining an amount $\chi_w \bar w$ without incurring any costs themselves, thereby maintaining a nonzero wealth level. Consequently, the Gini index stabilises at a value $0<G<1$ (Fig.~\ref{fig:yard-sale_tax}a) and its equilibrium value depends on the relative strength between tax and asset exchange. Varying $\chi_w$, $G^{\rm eq}$ experiences a sharp decline for small tax rates $\chi_w$, levelling off thereafter and reaching total equality $G^{\rm eq}=0$ at $\chi_w=1$ (Fig.~\ref{fig:yard-sale_tax}c). 

All in all, extending the Yard-Sale model by tax mechanisms shows that a wealth tax is able to prevent wealth condensation, which a simple income tax acting on the asset exchange revenues does not achieve. Next up we will consider modified winning probabilities that can give the poorer agents an advantage in the asset exchange and investigate whether this prevents wealth condensation.

\textbf{Uneven winning probabilities}

\begin{figure}[!ht]
    \centering
    \includegraphics[width=6in]{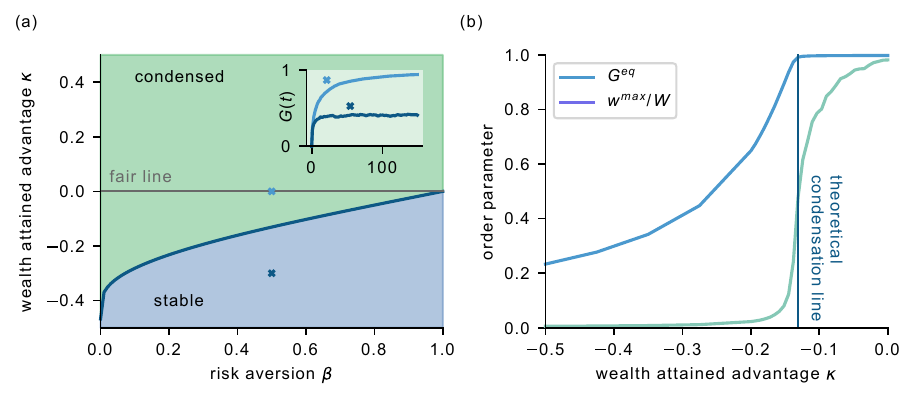} 
    \caption{\textbf{Adjusting winning probabilities in a Yard-Sale asset exchange leads to wealth condensation through a phase transition.} (a) For any value of risk aversion $\beta$ there exists a critical value of wealth attained advantage $\kappa^*$ which leads to wealth condensation in the Yard-Sale model. The theoretical condensation surface is given by Eq.~\ref{eq:critical-waa}. High wealth attained advantage $\kappa>\kappa^*$ leads to wealth condensation (condensed phase, green) and the Gini index approaches $G=1$ (inset, light blue). Wealth attained advantage below $\kappa^*$ leads to a non-trivial wealth distribution (stable phase, blue) and a Gini index $G<1$ (inset, dark blue). (b) The transition from the stable to the condensed phase resembles a second order phase transition with the fraction of wealth owned by the richest agent $w^{\rm max}/W$ as order parameter. The Gini equilibrium Gini index $G^{\rm eq}$ approaches its maximum value $G^{\rm eq}=1$ at the transition. Here for $\beta=0.5$.}
    \label{fig:yard-sale-adjusted-winning-prob-waa}
\end{figure}

In the previous asset exchange models we considered, both agents had an equal chance of winning the exchange, characterised by a winning probability of $p=1/2$. Let this now be generalised such that the richer agent wins with probability $p=1/2 + \kappa$. Parameter $\kappa$ is called the \textit{wealth attained advantage} (WAA), describing the advantage of having a favourable transactions due to higher wealth \cite{boghosian_economically_nodate}. 
There are various reasons to believe why $\kappa$ should be positive. Such an advantage could stem from the capacity of richer individuals to leverage their financial resources, such as through hiring legal or financial professionals or securing more favourable loan terms. 

In our subsequent analysis, we focus on scenarios where the WAA $\kappa$ remains constant, ranging from $\kappa=-1/2$ to $\kappa = 1/2$. In the former, the poorer agent wins in every exchange, while in the latter, the richer agent always wins. Alternatively, one could adopt a dynamic approach by making the WAA $\kappa$ dependent on the wealth difference between the participating agents. For example, $\kappa \propto |w_i-w_j|$ would give the richer agent a higher winning probability proportional to their wealth difference to the other agent \cite{boghosian_oligarchy_2017}.  
 
One might expect that as soon as $\kappa<0$, poor agents should be able to restore their wealth, which would prevent condensation. However, for any value of risk aversion $\beta$, there exists a critical wealth attained advantage $\kappa^*(\beta)<0$ for which wealth condensation occurs despite favouring the poor agent \cite{moukarzel_wealth_2007}: 

\begin{equation*}
    \kappa^*(\beta) = \frac{\log(\frac{1}{2-\beta})}{\log(\frac{\beta}{2-\beta})}-\frac{1}{2}\,.\label{eq:critical-waa}
\end{equation*}

Whenever the wealth attained advantage $\kappa$ is larger than $\kappa^*$, the result of the model is wealth condensation. The critical WAA $\kappa^*$ approaches the value of a 'fair' economy $\kappa^*=0$ only in the limit $\beta \to 1$, namely when the transaction size $\Delta w$ becomes zero (Fig.~\ref{fig:yard-sale-adjusted-winning-prob-waa}a). On the other hand, with increasing transaction sizes (decreasing $\beta$) the critical WAA approaches $\kappa = -0.5$, namely the state in which poorer agents are guaranteed to win every exchange. Hence, for any given advantage of the poor indicated by $\kappa < 0$, there exists a critical risk aversion $\beta^*$ for which wealth condensation occurs despite favouring poorer agents. Any value of $\kappa$ that is small enough to prevent wealth condensation will lead to non-trivial steady state distributions with a Gini index smaller than one (Fig.~\ref{fig:yard-sale-adjusted-winning-prob-waa}a, inset).

The transition between the stable phase of the model with a non-trivial wealth distribution and the condensed phase resembles a second order phase transition with the fraction of the wealthiest agent's wealth as order parameter (Fig.~\ref{fig:yard-sale-adjusted-winning-prob-waa}b). It suggests that, in principle, one could restore a much more equal state of the system by achieving that poorer agents have a higher winning probability in an exchange. Unfortunately, as pointed out earlier, this is unlikely due to various advantages richer individuals enjoy \cite{boghosian_economically_nodate}. 

In summary, modifications made to the Yard-Sale model indicate that only a redistribution mechanism of wealth can feasibly prevent wealth condensation, even though the practical implementation of such a strategy would most likely encounter numerous challenges. While the statistical physics approach to modelling wealth distribution has provided valuable insights, it may fall short of answering such questions of political nature. As we conclude this overview, we will instead briefly spotlight attempts to extend asset exchange models to incorporate power-law tails to provide a better fit to empirical wealth distributions.

\subsection{Emergence of power-law distributions}
\label{sec:emergence-of-powerlaws}

The previous asset exchange models we considered revealed the emergence of three qualitatively different steady-state distributions of wealth: an equilibrium Boltzmann distribution, non-equilibrium steady state distributions and a condensed state. While the Boltzmann distribution provides an adequate fit to the bulk of most empirical wealth distributions, none of the above models features a power-law tail, which is an essential part of empirical distributions. There are various approaches to introduce power-law tails to asset exchange models \cite{chakrabarti2013econophysics,chatterjee_pareto_2004,nirei_two_2007,boghosian_fokkerplanck_2014}. In the following we will outline one popular approach, which is given by introducing a saving factor $\lambda$ into the asset exchange \cite{chakrabarti2013econophysics,chatterjee_pareto_2004}. 

In this approach, before engaging in transactions, agents keep a fraction $\lambda$ of their wealth and trade with the remainder. If $i$ is the winning and $j$ the losing agent, this can be introduced by the exchange rule

\begin{equation}
    \Delta w = (1-\lambda)[ \epsilon (w_i + w_j) - w_i]\,,
\end{equation}

where $\epsilon$ is a random fraction $\epsilon \in [0,1]$. If $\epsilon =0$, agent $i$ loses a fraction $1-\lambda $ of their wealth, being only left with $\lambda w_i$, namely the fraction they saved. Similarly, if $\epsilon = 1$, agent $j$ loses all the wealth they did not save and are left with $\lambda w_j$. 

In the limit $\lambda =0$ this asset exchange is similar to the Dragulescu model (Sec.~\ref{sec:dragulescumodel}) and features a Boltzmann distribution as its steady-state distribution. The most probable value of wealth (the mode of the wealth distribution) is thus $w=0$. In contrast, in the limit $\lambda = 1$, agents save everything of their wealth and the most probable value is the mean wealth $w=\bar w$. For intermediate saving factors $\lambda$, going from $\lambda=0$ to $\lambda = 1$, the most probable value thus shifts from $w=0$ to $w=\bar w$.

The key step that introduces power-law tails into this model is to consider heterogeneity among the agents, each possessing a different saving factor $\lambda_i$, which is assumed to be constant in time. In particular, we consider the case where $\lambda_i$ is distributed uniformly and independently in the interval $[0,1]$. The exchange rule is then given by 

\begin{equation*}
    \Delta w = \epsilon (1-\lambda)w_i - (1-\epsilon)(1-\lambda)w_j\,,\label{eq:chakrabarti-saving}
\end{equation*}

where $\epsilon$ is again a random fraction, $\epsilon \in [0,1]$. The resulting steady-state distribution is found to follow a power-law $P(w) \sim \frac{1}{w^{1+\alpha}}$ with Pareto exponent $\alpha = 1$ (Fig.~\ref{fig:chakrabarti-powerlaw}). 

Apart from this model that achieves power-law tails through heterogeneous saving propensity among agents, another notable approach is given by an extended Yard-Sale model which shows remarkable agreement with empirical data \cite{boghosian_oligarchy_2017,boghosian_economically_nodate}. 

First, a wealth redistribution mechanism is imposed to prevent wealth condensation. Next, wealth attained advantage is introduced, giving richer agents a higher winning probabilities in the exchange. This model itself can already describe empirical wealth distributions to an accuracy of about $2\%$ \cite{boghosian_oligarchy_2017}. Importantly, most of the remaining discrepancy can be attributed to the presence individuals with negative wealth in real world economies. About $10\%$ of households are believed to have liabilities in excess of assets, and hence 'negative wealth' \cite{boghosian_economically_nodate}. By design, the Yard-Sale model restricts wealth to be positive as agents cannot lose more than a fraction of what they have. Hence, a further modification was added to shift the whole wealth distribution downward, resulting in negative wealth agents. The resulting model was shown to provide a fit to European and American wealth data to an accuracy of below $1\%$ \cite{boghosian_economically_nodate}.

All in all, this shows how asset exchange models can be extended to provide more accurate fits to empirical data. In the next section we will revert to the basic mechanisms of asset exchange models to gain a deeper understanding of qualitative differences among them.

\chapter{Results}
\label{ch:results}

% Paragraph 1: Issue 1 with AEM, arbitrary outcomes. How to fix it
The review of asset exchange models in the previous chapter has identified several limitations. A significant issue is that the outcomes of these models are highly sensitive to the details of the exchange rules. A deeper understanding of empirical exchange mechanisms could enhance asset exchange models by providing more realistic assumptions and potentially narrowing the range of possible outcomes. However, data collection poses substantial challenges. It requires not only data on transaction sizes $\Delta w$ between individuals but also data on their respective wealth $w_i$ and $w_j$. As wealth data is difficult to obtain even on an aggregated level, obtaining individual wealth data would be even more so challenging and would have to rely on survey data, where large amounts of data are challenging to acquire.

One promising research direction involves crypto/ blockchain technologies, such as the transfer of Bitcoin or Ethereum. The decentralised nature of those digital currencies opens new avenues for research as the entire history of transactions is publicly accessible. Thus, it is possible to reconstruct individual 'wallet balances', which indicate how rich in cryptocurrencies a certain user is. For any given cryptocurrency transaction, one can determine the associated wallet balances, thus obtaining a dataset on transaction sizes $\Delta w$ and 'wealth' $w_i$ and $w_j$. Using this information, one can investigate several questions. For example, one could construct a transaction network and study its statistical properties. Or, one could examine how typical transaction sizes $\Delta w$ relate to $w_i$ and $w_j$, indicating how different groups of the wealth distribution interact among each other and with other groups. While those present interesting questions, we will focus this thesis on constructing probabilistic Markov transition matrices from the data and analysing their spectral properties. This spectral analysis will, for example, reveal to what extent a system obeys detailed balance and what steady-state distribution of wealth can be expected based on transition probabilities. While the extent to which cryptocurrency transactions mirror real world money exchange remains uncertain, exploring this avenue also yields interesting theoretical results. %Among those is the finding that %Among those is the answer to the question under what conditions the steady state distribution of wealth depends on the initial distribution.

In the first section of the following chapter (Sec.~\ref{sec:markovchains}) we will start by introducing required background knowledge on Markov chains and use asset exchange models as illustrative examples to elucidate the theory. The theoretical background will be presented in five distinct boxes separate from the main text. Initially, we will consider simplified versions of asset exchange models with only two agents and discretised wealth, as transition probabilities can be analytically determined for that case. Extending this analysis to models with more agents and continuous wealth, we will present results from the spectral analysis of transition matrices computed for the asset exchange models discussed in the previous chapter. This will provide a solid foundation for the subsequent analysis of crytocurrency exchange in Sec.~\ref{sec:crypto}, where we will focus on the analysis of transition matrices for Ethereum transactions. Finally, in Sec.~\ref{sec:bistability}, we will explore the conditions under which the steady-state wealth distribution depends on the initial wealth distribution, offering potential new avenues for future research.

\vspace{1cm}

\section{Markov Chains}
\label{sec:markovchains}

\subsection{Theory and 2-agent asset exchange models}
\label{sec:2agentcase}

%\subsubsection{Transition matrices and steady states }

\begin{figure}
    \centering
    \includegraphics{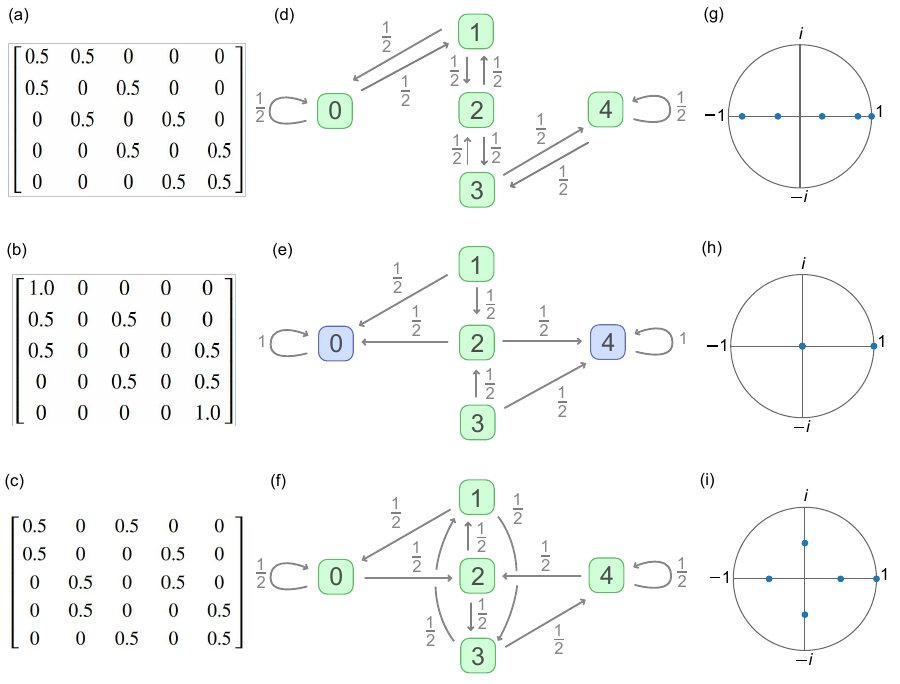}
    \caption{\textbf{Transition matrices of asset exchange models represented as graphs.} (a)-(c) The matrices $\boldsymbol{P}_a$, $\boldsymbol{P}_b$ and $\boldsymbol{P}_c$ (top to bottom) considered as illustrative examples throughout this section. (d)-(f) Transition matrices can be represented as graphs by representing the states as nodes and the transition probabilities as weighted directed links. Absorbing states of $\boldsymbol{P}_b$ are indicated in blue. (g)-(i) Eigenvalues of the transition matrices fall within the unit circle. (g) The eigenvalues are real for system $\boldsymbol{P}_a$ as it describes a reversible Markov process satisfying detailed balance. (h) The eigenvalues of $\boldsymbol{P}_b$ are 0 and 1, which are degenerate, with multiplicity 3 and 2, respectively. (i) The eigenvalues of $\boldsymbol{P}_c$ are complex as the underlying Markov process does not satisfy detailed balance.}
    \label{fig:aem-graphrepresentation}
\end{figure}

%A Markov process is a stochastic process without memory. The probability of being in a future state at time $t+1$ only depends on the current state at time $t$, not on previous states (Box.~\ref{sec:theory:markovchains1}, Eq.~\ref{}). In the following maths box we provide an introduction to Markov chains and introduce the transition matrix. 

\begin{math_box}{\subsection{Markov chains 1: The transition matrix} \label{sec:theory:markovchains1}}

% literature: levin2009markov, jiang19xxmathematical, spinney2012fluctuations, seabrook2023tutorial

% seabrook: 
% definition, evolution through transition matrix, classification of states, ergodicity, reversibility, absorbing chains, graph, eigenvalues, perron-frobenius, reversible chains and their eigenvalues / symmetric Laplacian

% maths markov box 1: transition matrix, evolution in time, stationary distribution 
% maths markov box 2: classification of states, ergodicity, absorbing states
% maths markov box 3: eigenvalues, transient and persistent structures, Perron-Frobenius theorem 

% left to be put: 
% reversibility, symmetric Laplacian, entropy production

This is the first mathematical background box about Markov chains. Here we define what a Markov process is, introduce the Markov transition matrix, and show how it is used to evolve probability distributions in time.  \\

In the following we will focus on discrete Markov processes on a finite size state space $|\mathbf{S}| = N$ with $N$ distinct states in which the system $X$ can be found. When dealing with discrete states, a Markov process is commonly referred to as a Markov chain.  \\

The defining characteristic of a Markov chain is the Markov property, which states that \cite{seabrook2023tutorial,wilmer2009markov}:

\begin{equation}
    \text{Pr}(X_{t+1}=x| X_1=x_1,X_2=x_2,...,X_t=x_t) = \text{Pr}(X_{t+1}=x|X_t=x_t)\,. \label{eq:markovproperty}
\end{equation}

This property implies that the probability of the system being in state $x$ at time $t+1$ depends solely on the state at the previous time step $t$, and not on the entire history of states. If these transition probabilities remain constant in time, they can be represented by a $N\times N$ matrix $\boldsymbol{P}$,

\begin{equation}
    \boldsymbol{P} = \begin{bmatrix}
        P_{11} & P_{12} & \ldots & P_{1N}  \\
        P_{21} & P_{22} & \ldots & P_{2N}  \\
        \vdots &\vdots & \ddots &  \vdots \\
        P_{N1} & P_{N2} & \ldots & P_{NN}  \\
    \end{bmatrix}\,.
\end{equation}\\

The entries $P_{ij}$ denote the probability of transitioning from state $i$ to state $j$, $P_{ij} = \text{Pr}(X_{t+1} = x_j|X_t =x_i)$. With that definition, the rows of $\boldsymbol{P}$ have to sum to one in order to conserve probability: $\sum_j P_{ij} = 1 \,\forall i$. Note that one could equally define $\boldsymbol{P}$ as its transpose, in which case the columns sum to one. Here we will stick to the former definition, as it is more commonly used \cite{seabrook2023tutorial,wilmer2009markov}. \\

% Paragraph: Representation of graph 
The transition matrix of a Markov chain can also be visualised via a directed weighted graph (Fig.~\ref{fig:aem-graphrepresentation}). Nodes correspond to the different states and directed links from $i$ to $j$ correspond to the transition probabilities from $i$ to $j$.  \\

Many aspects of a Markov chain can be understood by analysing its transition matrix $\boldsymbol{P}$. For example, the temporal evolution of a Markov chain is given by matrix multiplication with $\boldsymbol{P}$. Consider a row vector $\boldsymbol{\mu}^T(t)= (\mu_1(t),\ldots \mu_N(t))^T$, describing the distribution of the chain at time $t$. Its entries sum to one, $\sum_i \mu_i(t) = 1$. For example, for a system with 3 states $1,2,3$ and a 50/50 chance of finding the system in either state 1 or 2, $\boldsymbol{\mu}(t)$ would be given by $\boldsymbol{\mu}(t) = (\frac{1}{2},\frac{1}{2},0)^T$. \\

The distribution can be evolved to time $t+1$ via application of $\boldsymbol{P}$. The probability of being in a state $j$ at time $t+1$, denoted by $\mu_j(t+1)$, depends on the probabilities of being in all other states $i$ at time $t$, $\mu_i(t)$, and the transition probabilities from $i$ to $j$, $P_{ij}$ \cite{seabrook2023tutorial,spinney2012fluctuation}. Specifically, it is given by  

\begin{equation}
    \mu_j(t+1) = \sum_i \mu_i(t)P_{ij}\,.
\end{equation}

The whole probability vector $\boldsymbol{\mu}(t+1)^T$ can therefore be evolved by $\boldsymbol{\mu}(t+1)^T = \boldsymbol{\mu}(t)^T\boldsymbol{P}$. This operation can be extended to multiple time steps, as each subsequent iteration is simply given by repeated application of $\boldsymbol{P}$ from the right:

\begin{equation}
    \boldsymbol{\mu}(t+k)^T = \boldsymbol{\mu}(t+k-1)^T\boldsymbol{P} = \boldsymbol{\mu}(t)^T\underbrace{\boldsymbol{P}...\boldsymbol{P}}_{k\, \text{times}} =  \boldsymbol{\mu}(t)^T\boldsymbol{P}^k\,.
    \label{eq:timeevolution}
\end{equation}

Thus, given the transition matix $\boldsymbol{P}$ of a Markov chain, any distribution can be projected into the future via matrix multiplication. \\

% Paragraph: Stationary distribution 
A distribution $\pi^T$ that remains invariant under multiplication with $\boldsymbol{P}$, $\pi^T = \pi^T \boldsymbol{P}$, is said to be a stationary distribution of the Markov chain \cite{seabrook2023tutorial}. For finite state spaces, it is guaranteed that at least one such distribution exists. \\

The stationary distribution satisfies the so-called \textit{global balance} equations. For the $j$th element, this condition is expressed as 

\begin{equation}
    \pi_j = \sum_i \pi_i P_{ij}\,. \label{eq:globalbalance}
\end{equation}

The right hand side of Eq.~\ref{eq:globalbalance} represents the total flow of probability into state $j$ from all other states $i$. Since $\pi_j$ remains invariant under this flow, an equal amount of probability must flow out of state $j$ into all other states $i$. \\

Once a Markov chain reaches a stationary distribution with $\pi^T = \pi^T\boldsymbol{P}$, it remains there forever. One can therefore interpret the underlying process to be in a steady state \cite{seabrook2023tutorial}. 

%Definition Flow Matrix: $F^\pi = \Pi P$. Tells us how much probability mass flows from one state to another, with $\Pi = \text{diag}(\pi)$. The entries $(F^\pi)_{ij} = \pi_i P_{ij}$ tell how much probability flows from $s_i$ to $s_j$.  

\end{math_box}

In asset exchange models, the state of the system is represented by a vector that includes the wealth of all agents, $\mathcal{S} = (w_1,w_2,\hdots ,w_N)$. During an asset exchange, two agents undergo a change in their wealth, resulting in a transition of the system to a new state. These exchange dynamics adhere to the Markov property Eq.~\ref{eq:markovproperty}, because the amount of wealth transferred between agents depends only on their current wealth levels. Although asset exchange models have a continuous state space by default, wealth can be discretised to obtain a finite state space, allowing these models to be treated as Markov chains.

The case of $N=2$ agents is particularly straightforward to analyse because the state of one agent fully determines the state of the entire Markov chain. Consider a discrete system where the wealth is measured in discrete units of $\Delta w \in \mathbf{N}$ and a total amount $W\in \mathbf{N}$ with $ W\geq \Delta w$. The wealth of one agent is dependent on the wealth of the other, and the entire state of the Markov chain can be described by $(w_1,W-w_1)$, with $w_1$ the wealth of the first agent. For this two-agent scenario, the Markov transition matrix is simply an $n\times n$ matrix, where $n$ is the number of distinct states. Note that if $\Delta w = 1$, then $n=W+1$. For example, for $\Delta w=1$ and $W=5$ the states are $S = \{ (0,5), (1,4), (2,3), (3,2), (4,1), (5,0)\}$.

To illustrate concepts related to Markov chains, we will examine three example chains in the following section, which resemble the asset exchange models discussed in Chapter~2. Specifically, we will analyse the following three transition matrices (Fig.~\ref{fig:aem-graphrepresentation}):

\begin{equation}
\boldsymbol{P}_a = 
\begin{bmatrix}
\frac{1}{2} & \frac{1}{2} & 0 & 0 & 0 \\
\frac{1}{2} & 0 & \frac{1}{2} & 0 & 0 \\
0 & \frac{1}{2} & 0 & \frac{1}{2} & 0 \\
0 & 0 & \frac{1}{2} & 0 & \frac{1}{2} \\
0 & 0 & 0 & \frac{1}{2} & \frac{1}{2}
\end{bmatrix}\,, \hspace{0.5cm}
\boldsymbol{P}_b = 
\begin{bmatrix}
1 & 0 & 0 & 0 & 0 \\
\frac{1}{2} & 0 & \frac{1}{2} & 0 & 0 \\
\frac{1}{2} & 0 & 0 & 0 & \frac{1}{2} \\
0 & 0 & \frac{1}{2} & 0 & \frac{1}{2} \\
0 & 0 & 0 & 0 & 1
\end{bmatrix}\,, \hspace{0.5cm}
\boldsymbol{P}_c = 
\begin{bmatrix}
\frac{1}{2}& 0 & \frac{1}{2} & 0 & 0 \\
 \frac{1}{2} & 0 & 0 & \frac{1}{2} & 0 \\
0 & \frac{1}{2} & 0 & \frac{1}{2} & 0 \\
0 & \frac{1}{2} & 0 & 0 & \frac{1}{2} \\
0 & 0 & \frac{1}{2} & 0 & \frac{1}{2}
\end{bmatrix}\,.
\label{eq:exampletransitionmatrices}
\end{equation}

The transition matrices Eq.~\ref{eq:exampletransitionmatrices} can be visualised as weighted directed graphs by representing the different states of the Markov chain as nodes and the transition probabilities as weighed edges (Fig.~\ref{fig:aem-graphrepresentation}a-f). Note that the rows of Eq.~\ref{eq:exampletransitionmatrices} sum to one in order to conserve probability.

Transition matrix $\boldsymbol{P}_a$ represents an asset exchange of $\Delta w = 1$ with equal odds. The entries of $\boldsymbol{P}_a$ are either $0.5$ or $0$, depending on whether a certain transition is possible or not. For example, if $w_1=1$ there is a $50\%$ chance of transitioning to either $w_1=0$ or $w_1=2$ through an exchange, but a zero chance of transitioning to $w_1=3$ or $w_1=4$. 
This model bears similarity to a simplified version of the Dragulescu model satisfying detailed balance, encountered in Sec.~\ref{sec:dragulescumodel} ($\Delta w = (1-\beta)\frac{w_i+w_j}{2}$), with two agents and $\beta = 1/2$.

Transition matrix $\boldsymbol{P}_b$ describes a simplified version of the Yard-Sale model (that leads to wealth condensation, Sec.~\ref{sec:yardsale-wealthcondensatoin}), where $\Delta w = \text{min}\,(w_1,W-w_1)$ is exchanged. If $w_1=1$ or $w_1=3$, the exchanged amount $\Delta w$ is $\Delta w=1$, if $w_1=2$ then  $\Delta w=2$, and else $\Delta w = 0$. In the latter case, no more transactions can occur and the system never leaves that state. 

Transition matrix $\boldsymbol{P}_c$ describes a mechanism similar to the Ispolatov model (Sec.~\ref{sec:ispolatov-broken}), where $\Delta w$ is a fraction $(1-\beta)$ of the losing agent's wealth. Here we take this fraction to be $(1-\beta) = 0.5$ and round up in case of uneven numbers, such that $\Delta w \geq 1$ is exchanged in every transaction.

The temporal evolution of the systems described by Eq.~\ref{eq:exampletransitionmatrices} is given by repeated application of the transition matrices according to Eq.~\ref{eq:timeevolution}. If a system is initialised with all the wealth being held by the second agent, $w_1=0$ and $w_2 = 4$, the distribution over the states is given by $\boldsymbol{\mu}=(1,0,0,0,0)$. 

If we apply the matrix $\boldsymbol{P}_a$ to the vector $\boldsymbol{\mu}=(1,0,0,0,0)$ , there is a 50/50 chance of either remaining at $w_1=0$ or of transitioning to $w_1=1$. Hence, applying $\boldsymbol{P}_a$ once leads to $\boldsymbol{\mu} \boldsymbol{P}_a = (\frac{1}{2},\frac{1}{2},0,0,0)$. Repeated application of $\boldsymbol{P}_a$ eventually leads to a uniform distribution $\boldsymbol{\mu} =(\frac{1}{5},\frac{1}{5},\frac{1}{5},\frac{1}{5},\frac{1}{5})$, in which the agent has equal probabilities to be in any state of the system between $w_1=0$ and $w_1=W$. In contrast, applying the transition matrix $\boldsymbol{P}_b$ to the state $\boldsymbol{\mu}=(1,0,0,0,0)$ will leave it unchanged, $\boldsymbol{\mu P}=\boldsymbol{\mu}$. Hence, $\boldsymbol{\pi}=(1,0,0,0,0)$ represents a stationary state of the system described by $\boldsymbol{P}_b$, just like $\boldsymbol{\pi}=(0,0,0,0,1)$ does. Repeated application of $\boldsymbol{P}_c$, just like $\boldsymbol{P}_a$, leads to a uniform distribution $\boldsymbol{\mu} =(\frac{1}{5},\frac{1}{5},\frac{1}{5},\frac{1}{5},\frac{1}{5})$.  

Evidently, the system described by $\boldsymbol{P}_b$ behaves qualitatively differently from those described by $\boldsymbol{P}_a$ and $\boldsymbol{P}_c$. Temporal evolution through $\boldsymbol{P}_b$ leads to one of two states, either $w=0$ or $w=4$, without ever leaving them again. Both $\boldsymbol{\pi}=(1,0,0,0,0)$ and $\boldsymbol{\pi}=(0,0,0,0,1)$ represent stationary distributions satisfying $\boldsymbol{\pi P}=\boldsymbol{\pi} $, whereas only $\boldsymbol{\pi} =(\frac{1}{5},\frac{1}{5},\frac{1}{5},\frac{1}{5},\frac{1}{5})$ does so for $\boldsymbol{P}_a$ and $\boldsymbol{P}_c$. This qualitative difference in behavior highlights the importance of classifying Markov chains based on their properties. The subsequent background box provides such classifications, which will form the basis for the spectral analysis that follows.

%\subsubsection{Types of Markov chains}

\begin{math_box}{\subsection{Markov chains 2: Classification of states} \label{sec:theory:markovchains2}}

%FIGURE: Example of reducible and recurrent, Seabrock p.18 
% panel a) reducible, recurrent, b) reducible, not recurrent, c) irreducible 

%\centering
    \includegraphics[width=6in]{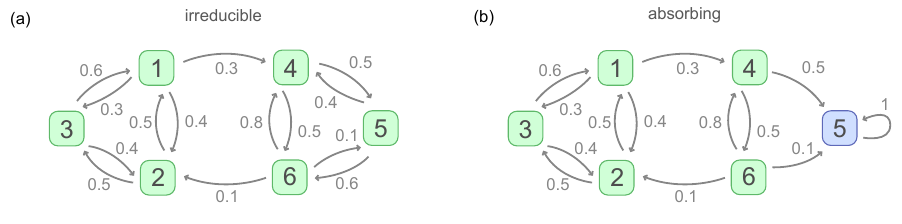}
    \captionof{figure}{\textbf{Classification of Markov chains.} (a) The chain is irreducible because there exists a connecting path between any two pairs of states. (b) The chain is absorbing as there exists a connection path to the absorbing state ($5$) from any other state.}
    \label{fig:markovclassification}
\vspace{0.5cm}
%\raggedright

This is the second background box about Markov chains. Here we categorise different states within a Markov chain, which allows to qualitatively distinguish between different chains \cite{seabrook2023tutorial}. \\

\textbf{Definition (Accessibility):} A state $x_j$ is said to be accessible from state $x_i$, if it is possible to reach $x_j$ from $x_i$ in $k\geq 1$ transitions: $(P^k)_{ij} > 0$. It is denoted by $x_i\to x_j$. \\

\textbf{Definition (Communication):} If two states are mutually accessible, $x_i\to x_j$ and $x_j\to x_i$, $x_i$ and $x_j$ are said to communicate. It is denoted by $x_i\leftrightarrow x_j$. \\

Pairs of states within a Markov chain can be categorised according to whether it is possible to transition from one to the other. This allows to partition the entire state space $\mathcal{S}$ of a Markov chain into communicating classes, each containing states that communicate with each other. A Markov chain can be categorised by how many communicating classes it contains \cite{seabrook2023tutorial}. \\

\textbf{Definition (Irreducibility):} A Markov chain is said to be irreducible if it contains only one communicating class. In that case, there exists a connection path between any two pairs of states (Fig.~\ref{fig:markovclassification}a).\\ 

%Theorem: a Markov chain is irreducible if and only if it has a stationary distribution $\pi$. Furthermore, that distribution has strictly positive elements. For reducible Markov chain the guarantee of uniqueness and positivity of the stationary probabilities no longer holds. 

The concept of irreducibility is important as this condition ensures that the Markov chain has a unique stationary distribution with strictly positive elements. For reducible chains this guarantee of uniqueness and positivity of stationary probabilities no longer holds \cite{seabrook2023tutorial}. \\

Lastly, one important concept of Markov chains is absorption. \\

\textbf{Definition (Absorbing states):} A state $x_i$ is called absorbing if it is possible to transition into it but not out of it, namely $P_{ii} =1$. \\

An absorbing Markov chain is one for which there exists a path to an absorbing state from any other state. It follows that all non-absorbing states of that chain are transient (Fig.~\ref{fig:markovclassification}b). %The presence of absorbing states in a chain imply that the chain cannot be ergodic.  

\end{math_box}

In the chains described by $\boldsymbol{P}_a$ and $\boldsymbol{P}_c$, all states communicate with one another, meaning there exists a path between any pair of states. These chains each contain a single communicating class and are therefore irreducible. In contrast, the chain described by $\boldsymbol{P}_b$ is an absorbing Markov chain because it includes two absorbing states that are accessible from all other states. Consequently, all non-absorbing states are transient, and there is no unique stationary distribution. In this context, wealth condensation in the Yard-Sale model can be viewed as the system reaching an absorbing state, where transaction sizes decrease to zero and the system becomes trapped in a state from which it cannot escape.

The classification of Markov chains into irreducible, recurrent, and absorbing categories clarifies the qualitative differences between the system described by $\boldsymbol{P}_b$ and those described by $\boldsymbol{P}_a$ and $\boldsymbol{P}_c$. However, this classification does not address the differences between $\boldsymbol{P}_a$ and $\boldsymbol{P}_c$, as both chains share the same stationary distribution $\boldsymbol{\pi} = \left(\frac{1}{5}, \frac{1}{5}, \frac{1}{5}, \frac{1}{5}, \frac{1}{5}\right)$. However, in the continuous version of these models with many agents, one model corresponds to the Dragulescu model with a Boltzmann equilibrium steady state distribution, and the other corresponds to the Ispolatov model with a non-equilibrium steady state distribution. In order to understand these differences based on the transition matrices, we have to extend our analysis by examining the eigenvalues of $\boldsymbol{P}_a$ and $\boldsymbol{P}_c$. Such a spectral analysis will enable to classify Markov chains according to whether they satisfy detailed balance or not, and will provide useful tools to analyse transition matrices computed from data.

%\subsubsection{Spectral analysis of Markov chains}

\begin{math_box}{\subsection{Markov chains 3: Eigenvalues} \label{sec:theory:markovchains3}}

This is the third background box about Markov chains. Here we highlight important results from the spectral analysis of Markov transition matrices. \\

Any real matrix $\boldsymbol{A}\in \mathbf{R}^{N\times N}$ can be interpreted as a linear transformation. A vector which, when transformed by $\boldsymbol{A}$, only gets multiplied by a number $\lambda $ is called an eigenvector of $\boldsymbol{A}$ to the eigenvalue $\lambda$. This can be obtained by both left- and right multiplication with $\boldsymbol{A}$. If $\boldsymbol{l}^T \boldsymbol{A} = \lambda \boldsymbol{l}^T$, $\boldsymbol{l}^T$ is called a \textit{left eigenvector}; if $\boldsymbol{A} \boldsymbol{r} = \lambda \boldsymbol{r}$, $\boldsymbol{r}$ is called a \textit{right eigenvector} \cite{seabrook2023tutorial}. For example, the stationary distribution $\boldsymbol{\pi}$ of a Markov chain satisfies $\boldsymbol{\pi P} = \boldsymbol{\pi}$ and is thus a left eigenvector to the eigenvalue $\lambda = 1$.\\ 

Markov transition matrices are real-valued matrices with non-negative entries, the study of which has received much attention in the field of spectral graph theory. Despite having only real entries, transition matrices can have complex eigenvalues. However, they can only occur in complex conjugate pairs. One of the most important results from the spectral analysis of Markov chain is the Perron-Frobenius theorem \cite{seabrook2023tutorial}. \\

\textbf{Theorem (Perron-Frobenius for irreducible Markov chains:)} If $\boldsymbol{P}$ is the transition matrix of an irreducible Markov chain, then the following holds:
\begin{itemize}
    \item $\lambda =1$ is guaranteed to be an eigenvalue
    \item $\lambda$ is an eigenvalue with multiplicity 1, namely, it occurs only once 
    \item eigenvalue $\lambda=1$ has a left eigenvector corresponding to the unique stationary distribution $\boldsymbol{\pi} = (\pi_1,...,\pi_n)$ and a right eigenvector corresponding to $\boldsymbol{\eta} = (1,1,...,1)^T$ (upon appropriate normalisation)
    \item all other eigenvalues have $|\lambda | \leq 1$, hence the spectral radius of $\boldsymbol{P}$ is one
\end{itemize}

\bigskip

One can order the eigenvalues by magnitude and separate $|\lambda_w|=1$ from $|\lambda_w|<1$. The former indicate persistent behaviour of the chain whereas the latter indicates transient behaviour. In the long time limit, only terms (eigenvectors) with $|\lambda|=1$ survive. %Else, the absolute value of $\lambda$ measures the decay rate with which the terms die off. \\

The eigenvalues with $|\lambda|=1$ can be separated into $\lambda=1$, $\lambda=-1$ and $\lambda \in \mathbf{C}, |\lambda|=1$. Eigenvectors with $\lambda=1$ are called persistent structures, an example is the steady state distribution $\boldsymbol{\pi}$. For $\lambda=-1$ the eigenvectors flip their sign under application of $\boldsymbol{P}$, $\boldsymbol{l}^T \boldsymbol{P}= -\boldsymbol{l}^T$, but return to their initial state after two steps. This corresponds to permanent oscillations of probability mass between two states. Similarly, eigenvalues with $\lambda \in \mathbf{C}, |\lambda|=1$ correspond to persistent cycles. \\

For eigenvectors with $|\lambda|<1$ a similar categorisation can be made, just that these terms are transient and eventually die off. When $|\lambda|\approx 1$ the states are said to be metastable because the system spends a lot of time in these states, for example $\lambda \in \mathbf{C}, |\lambda|\approx 1$ indicate cycles that can persist for a long time. \\

An important implication from the Perron-Frobenius theorem is that if a Markov chain only has the eigenvalue $\lambda=1$ on the unit circle, and it occurs only once, then the stationary distribution $\boldsymbol{\pi}$ is the only persistent structure. The chain is guaranteed to end up in this state. \\

If a chain is reducible, namely it has several communicating classes, the Perron-Frobenius Theorem makes weaker statements \cite{seabrook2023tutorial}. Most importantly, the number of linearly independent eigenvectors with $\lambda=1$ equals the number of recurrent communicating classes of the chain. For example, a Markov chain with two recurrent communicating classes has two eigenvalues $\lambda=1$.\\

\end{math_box}

According to the Perron-Frobenius Theorem, the matrices $\boldsymbol{P}_a,\boldsymbol{P}_b$ and $\boldsymbol{P}_c$ always have at least one eigenvalue equal to one, which we will refer to as $\lambda_1$. The corresponding eigenvector $\boldsymbol{v}_1$ corresponds to the stationary state as $\boldsymbol{v_1} \boldsymbol{P} = \lambda_1 \boldsymbol{P} = \boldsymbol{P}$. 

In the systems described by $\boldsymbol{P}_a$ and $\boldsymbol{P}_c$ the eigenvalue $|\lambda_1|=1$ has multiplicity 1, which implies that the chains is guaranteed to end up in its unique stationary distribution. The corresponding eigenvector $\boldsymbol{v}_1$ is constant and is, upon normalisation, given by the stationary distribution $\boldsymbol{\pi} =(\frac{1}{5},\frac{1}{5},\frac{1}{5},\frac{1}{5},\frac{1}{5})$. In contrast, the absorbing chain described by $\boldsymbol{P}_b$ has a degenerate eigenvalue $\lambda_1=1$ with multiplicity 2. These correspond to the absorbing states $w_1=0$ and $w_1=4$ and the eigenvectors are given by $\boldsymbol{\pi} = (1,0,0,0,0)$ and $\boldsymbol{\pi} = (0,0,0,0,1)$, respectively. 

The difference between systems $\boldsymbol{P}_a$ and $\boldsymbol{P}_c$ is that for $\boldsymbol{P}_a$ all eigenvalues are real, whereas for $\boldsymbol{P}_c$ they include complex eigenvalues (Fig.~\ref{fig:aem-graphrepresentation}). In the latter case they are given by $(\frac{1}{2},-\frac{1}{2},\frac{i}{2},-\frac{i}{2})$. This presence of complex eigenvalues with absolute value smaller than one signifies damped oscillatory behaviour as the system approaches its steady state distribution. As will be explained in the subsequent background box, $\boldsymbol{P}_a$ only features real eigenvalues because the associated Markov chain obeys detailed balance.

\begin{math_box}{\subsection{Markov chains 4: Reversibility and detailed balance} \label{sec:theory:markovchains4}}

This is the fourth background box about Markov chains. Here we introduce the concept of detailed balance and how it relates to the spectrum of transition matrices.  \\

\textbf{The concept of detailed balance}\\

A defining characteristic of Markov chains is that the future $X$ depends only on the present $Y$ and not on the past $Z$. Conversely, the past $Z$ depends only on the present $Y$ and not on the future $X$ \cite{seabrook2023tutorial}. An interesting question is how a Markov process would behave if the direction of time were reversed. Since both the past and the future depend solely on the present but not on each other, the time-reversed process would still satisfy the Markov property Eq.~\ref{eq:markovproperty}. Thus, we can define a new chain $\mathcal{\hat X}$ as the time reversal of the original chain $\mathcal{X}$. The open question is how the transition matrix of $\mathcal{\hat X}$, denoted by $\boldsymbol{\hat P}$, relates to that of $\mathcal{X}$, denoted by $\boldsymbol{P}$.\\
  
Using Bayes theorem, one finds that \cite{wilmer2009markov,seabrook2023tutorial}

\begin{equation}
    (\boldsymbol{\hat P})_{ij} = \text{Pr}(X_t = x_j|X_{t+1}=x_i) = \frac{\text{Pr}(X_{t+1} = x_i|X_t=x_j)\text{Pr}(X_t=x_j)}{\text{Pr}(X_{t+1}=x_i)} = \frac{P_{ji}\pi_j}{\pi_i}\,. \label{eq:reversedchain}
\end{equation}\\

Note that this relationship only holds once the chains have converged. Besides, as $\pi_i$ appears in the denominator, it is only valid for distributions with $\pi_i > 0 \,\forall x_i$ \cite{seabrook2023tutorial}. \\

An important special case is one in which the Markov chain is indistinguishable from its time reversal. Any sequence $X_1,...,X_k$ would occur with the same probability as its reverse $X_k,...,X_1$, and hence the two would be statistically equivalent. The stationary dynamics of such a Markov chain has no inherent arrow of time, which is why such chains are also said to be \textit{reversible} \cite{wilmer2009markov,seabrook2023tutorial}. Since $\mathcal{X}$ and $ \mathcal{\hat X}$ are indistinguishable, they also share the same transition matrix, $\boldsymbol{\hat P} = \boldsymbol{P}$. Inserting this into Eq.~\ref{eq:reversedchain} we finally come to the concept of detailed balance.\\

\textbf{Theorem (Detailed Balance):} A recurrent Markov chain is reversible if and only if for any stationary distribution $\pi>0$ the \textit{detailed balance} condition holds \cite{seabrook2023tutorial,wilmer2009markov}:

\begin{equation}
    P_{ij}\pi_i = P_{ji}\pi_j \,\, \forall x_i,x_j\in \mathcal{S}\,.\label{eq:markovdetailedbalance}
\end{equation}\\

Detailed balance is a much stronger condition than that of global balance ($\pi_j = \sum_i P_{ij}\pi_i$, Eq.~\ref{eq:globalbalance}). For it to hold, the probability flows between any two pairs of states must balance out each other, in contrast to global balance where the probability flow must only match between one state and all others. \\

A useful tool to check for detailed balance is given by the flow matrix $\boldsymbol{F^\pi}$, defined as $\boldsymbol{F^\pi} = \boldsymbol{\Pi P}$ where $\boldsymbol{\Pi} = \text{diag}(\pi_1,...,\pi_N)$. The entries $(\boldsymbol{F^\pi})_{ij} = \pi_i P_{ij}$ indicate how much probability flows from state $x_i$ to $x_j$. For a reversible Markov chain the detailed balance condition implies that the flow matrix must be symmetric, $(\boldsymbol{F^\pi})_{ij} = (\boldsymbol{F^\pi})_{ji}$ \cite{seabrook2023tutorial}.\\

There are in general two different cases in which a Markov chain does not obey the condition of detailed balance. First, it may be non-recurrent, in which case no positive stationary distribution exists. Second, the flow matrix $\boldsymbol{F^\pi}$ may be asymmetric, in which case the Markov chain is also said to be in a \textit{non-equilibrium steady state} \cite{spinney2012fluctuation}. Overall, the concept of detailed balance in Markov chains is closely related to the concepts of thermodynamic equilibrium in statistical mechanics, where detailed balance is a prerequisite for a system to be in thermal equilibrium. \\

% Connection between reversibility and symmetric matrices, rreal eigenvalues 
\textbf{Eigenvalues of reversible transition matrices}\\

Given a Markov chain $\mathcal{X}$ described by its transition matrix $\boldsymbol{P}$, a natural question to ask is whether one can infer from the spectral properties of $\boldsymbol{P}$ directly whether $\boldsymbol{P}$ describes a reversible chain or not. Alternatively, one would have to check whether for each pair of states the detailed balance condition Eq.~\ref{eq:markovdetailedbalance} is satisfied, or check whether the flow matrix $\boldsymbol{F^\pi}$ is symmetric. In order to investigate this question, we start with some basic definitions regarding symmetric matrices \cite{seabrook2023tutorial}. \\

\textbf{Theorem (Diagonalisability of Symmetric Matrices):} A real matrix $\boldsymbol{A}$ is symmetric only if it is orthogonally diagonalisable, meaning that there exists an orthogonal matrix $\boldsymbol{Y}$ such that $\boldsymbol{\Delta} = \boldsymbol{Y}^{-1}\boldsymbol{AY} = \boldsymbol{Y}^T\boldsymbol{AY}$ with $\boldsymbol{\Delta}$ a diagonal matrix containing the eigenvalues of $\boldsymbol{A}$. The eigenvalues of $\boldsymbol{A}$ are guaranteed to be real because $\boldsymbol{Y}$ and $\boldsymbol{Y}^T$ are also real.\\

For reversible Markov chains the transition matrix $\boldsymbol{P}$ is usually not symmetric, as this would imply $P_{ij}=P_{ji}$, as opposed to the detailed balance condition Eq.~\ref{eq:markovdetailedbalance}. However, it is possible to perform scalings on the rows and columns of $\boldsymbol{P}$ that turn it into a symmetric matrix. \\

\textbf{Theorem (Symmetrised Transition Matrix):} A Markov chain $\mathcal{X}$ with transition matrix $\boldsymbol{P}$ is only reversible if for any stationary distribution $\boldsymbol{\pi}>0$ the matrix $\boldsymbol{K}$ defined by $\boldsymbol{K} = \boldsymbol{\Pi}^{1/2}\boldsymbol{P\Pi}^{-1/2}$ is symmetric.\\

Whenever two matrices $\boldsymbol{B}$ and $\boldsymbol{C}$ can be related via a transformation of the form $\boldsymbol{B} = \boldsymbol{UCU}^{-1}$, $\boldsymbol{B}$ and $\boldsymbol{C}$ are said to be \textit{similar}, in which case they share the same eigenvalues \cite{seabrook2023tutorial}. From the definition of $\boldsymbol{K}$ it follows that $\boldsymbol{K}$ and $\boldsymbol{P}$ are similar, which leads to the following theorem.  \\

\textbf{Theorem (Eigenvalues of reversible Markov Chains):} A Markov chain $\mathcal{X}$ is reversible if and only if its transition matrix $\boldsymbol{P}$ is diagonalisable with real eigenvalues and there exists a basis of right eigenvectors that are orthogonal with regard to $\langle \cdot, \cdot \rangle_\pi$, and a corresponding dual basis of left eigenvectors that are orthogonal with respect to $\langle \cdot, \cdot \rangle_{\pi^{-1}}$, where $\boldsymbol{\pi}$ is a stationary distribution of the chain and $\langle \cdot, \cdot \rangle_\pi$ is defined as $\langle \boldsymbol{a}, \boldsymbol{b} \rangle_\pi = \boldsymbol{a}^T \boldsymbol{\Pi b}$. \\

This theorem tells us that all reversible Markov chains have transition matrices that are diagonalisable and have real eigenvalues. As a result, any Markov chain with a transition matrix that has complex eigenvalues cannot satisfy detailed balance. Note that the inverse does not hold true, namely non-reversible Markov chains can also have real spectra. Thus, checking whether a transition matrix $\boldsymbol{P}$ is diagonalisable and has real eigenvalues is not a way to prove reversibility, but only a way to exclude it.  \\

\end{math_box}

After having introduced the concept of reversibility and detailed balance for Markov chains, we can apply this knowledge to our example matrices Eq.~\ref{eq:exampletransitionmatrices}. In order to check for detailed balance, one way is to see whether the detalailed balance condition Eq.~\ref{eq:markovdetailedbalance} holds for any two pairs of states. The Markov chains described by $\boldsymbol{P}_a$ and $\boldsymbol{P}_c$ have uniform stationary distributions $\boldsymbol{\pi} =(\frac{1}{5},\frac{1}{5},\frac{1}{5},\frac{1}{5},\frac{1}{5})$, which is why the detailed balance condition simplifies to $P_{ij}=P_{ji}$. Looking at the transition matrices/graphs (Fig.~\ref{fig:aem-graphrepresentation}) it is evident that this only holds for $\boldsymbol{P}_a$, where the transition probability between any two states that are $\Delta w = 1$ apart is $1/2$. Is contrast, for $\boldsymbol{P}_c$ the transition probability between states is not pair-wise matched. For example, it is possible to transition from state 0 to state 2, but not vice versa, as one has to take a detour via state 1. 

A second characteristic of reversible Markov chains is that their transition matrices are diagonalisable and have real spectra. Computing the eigenvalues of  $\boldsymbol{P}_c$ one can see that they are complex and thus the Markov chain cannot be reversible (Fig.~\ref{fig:aem-graphrepresentation}i). In contrast, for $\boldsymbol{P}_a$ the eigenvalues are purely real and one can easily verify that is diagonalisable by computing $\boldsymbol{Y}^T \boldsymbol{P Y}$ with $\boldsymbol{Y} = (\boldsymbol{r}_1 \,\, \hdots \boldsymbol{r}_N)$ the matrix containing all right eigenvectors of $\boldsymbol{P}_a$. Equivalently, computing $\boldsymbol{K} = \boldsymbol{\Pi}^{1/2}\boldsymbol{P\Pi}^{-1/2}$ reveals that $\boldsymbol{K}$ is symmetric, as required. 

We are now at a point where we can make qualitative statements about whether a Markov chain satisfies detailed balance or not, purely based on its spectrum. However, in case a Markov chain does not satisfy detailed balance, we are left with the question of quantifying the extent to which it is irreversible. The subsequent background box will answer this question by introducing the entropy production rate as a measure of irreversibility. \\

\begin{math_box}{\subsection{Markov chains 5: Entropy production as a measure for irreversibility} \label{sec:theory:markovchains5}}

This is the fifth background box about Markov chains. Here we introduce entropy production as a measure for irreversibility of stochastic processes. \\ 

\textbf{Entropy production as a measure for irreversibility}\\

The Gibbs entropy of a stochastic system with discrete states $i$ is given by \cite{zhang2012stochastic}

\begin{equation}
    S(t) = -\sum_i p_i(t) \log p_i(t)\,.
\end{equation}

The time derivative is 

\begin{equation}
    \frac{dS(t)}{dt} = -\sum_i \frac{dp_i(t)}{dt}\log p_i(t)  - \sum_i \frac{dp_i(t)}{dt}  =  -\sum_i \frac{dp_i(t)}{dt}\log p_i(t)\,,
\end{equation}

where we used $ \sum_i \frac{dp_i(t)}{dt}  = \frac{d}{dt}\sum_i p_i(t) = 0$ for the second term. \\

Inserting the master equation according to which the system evolves leads to 

\begin{equation}
    \frac{dS(t)}{dt} = - \sum_{i\neq j} (p_j(t)q_{ji} - p_i(t)q_{ij}) \log p_i(t) = - \frac{1}{2} \sum_{i\neq j} (p_j(t)q_{ji} - p_i(t)q_{ij}) \log \frac{p_i(t)}{p_j(t)}\label{eq:dSdt-entropyproduction}
\end{equation}

In steady state dynamics $\dot S =0$ must be satisfied, as the probability distribution itself does not change. However, this does not imply that no entropy is being produced in the process. The time derivative of entropy and its production rate are two different concepts \cite{zhang2012stochastic}. In particular, we can split the term in Eq.~\ref{eq:dSdt-entropyproduction} into 

\begin{equation}
    0 =-  \frac{1}{2} \sum_{i\neq j} (\pi_iq_{ij} - \pi_jq_{ji}) \log \frac{q_{ij}}{q_{ji}} + \frac{1}{2} \sum_{i\neq j} (\pi_iq_{ij} - \pi_jq_{ji}) \log \frac{\pi_i q_{ij}}{\pi_j q_{ji}} \equiv -h_p + e_p\,,\label{eq:entropyproductionrate}
\end{equation}

where $h_p$ is the heat dissipation rate and $e_p$ is the entropy production rate. Note that we used $p_i=\pi_i$ for the steady state distribution. From the two terms in Eq.~\ref{eq:entropyproductionrate} involving $\pi_iq_{ij} - \pi_jq_{ji}$, one can see that only systems that satisfy detailed balance do neither dissipate heat nor produce entropy, and are hence reversible \cite{spinney2012fluctuation,zhang2012stochastic}. The entropy production rate $e_p$ thus provides a measure of how strongly irreversible a process is, with its lower bound at $e_p=0$ when a process is completely reversible. \\

%We summarise this section by restating the most important results. Markov processes as well as thermodynamic systems are reversible only if they obey detailed balance, in which case they do not produce entropy. The entropy production rate of a system can thus be used as a measure of how irreversible the underlying dynamics are. 

\end{math_box}

After having introduced the required background on Markov chains and used 2-agent asset exchange models as examples, we are now at a point to extend the study to the original asset exchange models considered in the previous chapter. These will serve as baseline models for comparison later on, when we compute transition matrices for Ethereum data. However, asset exchange models with more than 2 agents pose a problem for the computation of the transition matrix, as will be explained in the subsequent section.

\subsection{Multi-agent case}
\label{sec:multiagentcase}

\begin{figure}
    \centering
    \includegraphics[width=6in]{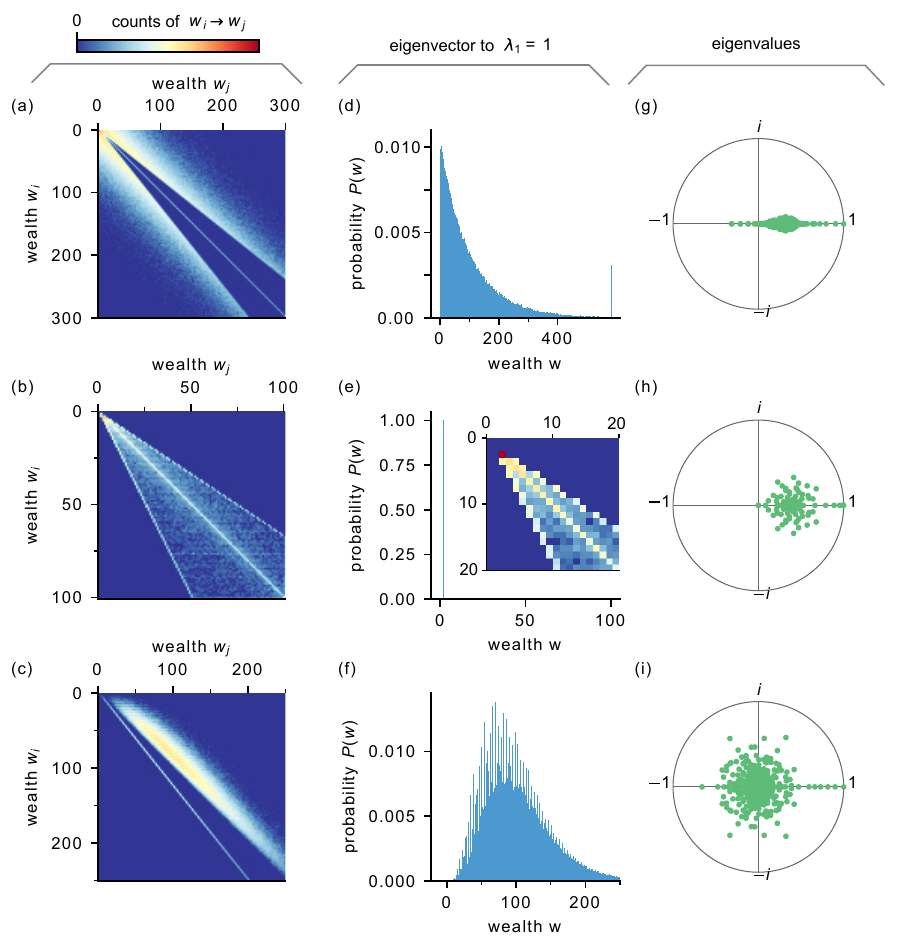}
    \caption{\textbf{Spectral analysis of transition matrices obtained from asset exchange simulations reveal qualitative differences between exchange rules.} (a)-(c) Weight matrices $\mathcal{W}$ obtained through counting the number of transitions between two wealth states in simulations of the asset exchange models. The nature of the exchange rules only allow certain transitions, leaving large areas of the weight matrices empty. (a) Dragulescu model, $\Delta w= (1-\beta)\frac{w_i+w_j}{2}$ with $\beta = 0.6$. (b) Yard-Sale model, $\Delta w = (1-\beta)\text{min}(w_i,w_j)$ with $\beta = 0.5$. As the model does not approach a steady state, the weight matrix is obtained by re-initialising the model and running it repeatedly. (c) Ispolatov model, $\Delta w = (1-\beta)w_i$ with $i$ the losing agent and $\beta=0.8$. (d)-(f) The eigenvector corresponding to the unique eigenvalue $\lambda_1=1$, representing the steady state distribution vector. (d) The steady state distribution for the Dragulescu model is a Boltzmann distribution. (e) The steady state distribution for the Yard-Sale model is the condensed state, indicated by the eigenvector $\boldsymbol{\pi} = (1,0,\hdots,0)$. Inset: Due to the discretised version of the model, the minimal attainable wealth is $w=4$. (f) The non-equilibrium steady state distribution of the Ispolatov model. (g)-(i) The eigenvalue distribution on the unit circle. (g) The Dragulescu model satisfies detailed balance, which is why the spectrum of its transition matrix should be real. Due to the stochastic way of obtaining it, complex eigenvalues persist. (h) Eigenvalues of the Yard-Sale model have purely positive real parts. (i) The Ispolatov model violates detailed balance, indicated by a complex spectrum.}
    \label{fig:transitions_linearbinning}
\end{figure}

We now turn our attention to analysing Markov transition matrices for asset exchange models involving more than two agents. However, attempting to write down the transition matrix analytically, as previously done, reveals an issue. To understand why, consider the two components that make up a transition probability. First, transitioning from state $w$ to state $w'$ necessitates either a profitable or unprofitable transaction, depending on whether $w' > w$ or $w'<w$, which occurs with probability $1/2$. Second, it depends on selecting a suitable agent such that the transaction size $\Delta w$ allows a transition from $w$ to $w'$. For a randomly chosen agent, $\Delta w$ might be either too large or too small for the desired transition. Thus, finding an appropriate agent such that $\Delta w$ matches the required transition depends on the entire probability distribution $P(w)$, namely the system's state. For a system with two agents, this issue did not arise, as the wealth of one agent determined the wealth of the other agent via $w_2 = W - w_1$. Consequently, knowing the wealth of one agent was equivalent to knowing the state of the entire system, resulting in transition probabilities of either $1/2$ or $0$.

In contrast, for three agents, knowing the wealth of one agent leaves variability in how the remaining wealth is distributed among the other two agents. For three agents, the states are $(w_1, w_2, W - w_1 - w_2)$, and for $N$ agents, the states are $(w_1, \ldots, w_{N-1}, W - \sum_{i=1}^{N-1} w_i)$. As the number of agents increases, the number of ways to distribute wealth among them grows combinatorically, making it impossible to write down the transition matrix analytically. Therefore, computing the transition matrix requires numerically simulating the system and counting the number of transitions between any two states (regardless of which agent underwent the transition). In the following, we will consider systems with a total wealth of $W$ (with an average wealth of $\bar{w}$ per agent) and a minimal unit of wealth equal to $\Delta w = 1$.

The objective is to compute a transition matrix $\boldsymbol{P}$ where the entries $P_{w_k, w_l}$ represent the transition probabilities of moving from wealth $w_k$ to $w_l$ through an asset exchange. We approximate this matrix by simulating numerous asset exchanges. Each exchange, $w_i \to w_i + \Delta w$ and $w_j \to w_j - \Delta w$, involves two transitions. These transitions occur with probabilities $\mathcal{P}(w_i + \Delta w | w_i)$ and $\mathcal{P}(w_j - \Delta w | w_j)$, which, however, are unknown. Simulating asset exchanges only allow us to count the occurrences of such transitions. This can be captured in a \textit{weight matrix} $\mathcal{W}$, with entries $\mathcal{W}_{w_k, w_l}$ that indicate the number of times we counted a transition from wealth $w_k$ to wealth $w_l$. The next step is to normalise $\mathcal{W}$ to obtain the transition matrix $\boldsymbol{P}$. Generally, $\mathcal{W}$ represents a directed weighted graph where nodes correspond to all possible wealth states and directed links represent the counts of transitions between states from the simulation. The out-degree of a node $i$, $d_i^+ = \sum_j \mathcal{W}_{ij}$, indicates the total number of transitions out of state $i$. Since the total transition probability into all other states must equal one, we can compute the normalised transition matrix as $P_{ij} = \frac{\mathcal{W}_{ij}}{d_i^+}$ \cite{seabrook2023tutorial}. 

If the weight matrix $\mathcal{W}$, obtained from counting transitions, is symmetric, it implies that transition probabilities between any two pairs of states are matched pairwise, indicating that the underlying process satisfies detailed balance. Only in the case where the stationary distribution is uniform, also the transition matrix $\boldsymbol{P}$ will be symmetric. However, according to Section~\ref{sec:theory:markovchains4}, in this case, $\boldsymbol{P}$ will be at least diagonalisable with real eigenvalues. Moreover, the entropy production rate should be zero and can be computed using the steady-state distribution, which is given by the normalised eigenvector corresponding to the eigenvalue $\lambda_1 = 1$.

In the following, we will illustrate these concepts using the Dragulescu, Yard-Sale, and Ispolatov models. We consider discretised versions with a minimal unit of wealth equal to one, a number of $N = 1000$ agents, and an average wealth per agent of $\bar{w} = 100$. We simulate each system for $T = 400$ time steps, with each time step including $N/2$ transactions, resulting in a total of $TN = \SI{400000}{}$ transitions, which are recorded in the weight matrix $\mathcal{W}$ (Fig.~\ref{fig:transitions_linearbinning}a-c). Note that for the Yard-Sale model, which does not approach a steady state, we need to perform multiple simulations to obtain the weight matrices.

Overall, the exchange rules governing asset exchanges permit only specific transitions, leading to many entries in the weight matrices being zero. For instance, in the Dragulescu model, there is a minimal transaction size $\Delta w$ for an agent with wealth $w_i$ when interacting with an agent of wealth $w_j = 0$. Given that $\Delta w = (1 - \beta) \frac{w_i + w_j}{2}$, this implies $\Delta w \geq (1 - \beta) \frac{w_i}{2}$, resulting in an empty region around the diagonal of $\mathcal{W}_a$ (Fig.\ref{fig:transitions_linearbinning}a). Similarly, in the Ispolatov model, when an agent $i$ loses wealth, the amount lost is a fraction of their own wealth, $\Delta w = (1 - \beta) w_i$. Consequently, all transitions $w_i \to w_j$ align along a straight line in the lower diagonal half of $\mathcal{W}_c$, representing the area of unprofitable transactions (Fig.\ref{fig:transitions_linearbinning}c). For the Yard-Sale model, three distinct lines emerge in $\mathcal{W}_b$. The middle line indicates transitions where wealth does not (or only minimally) change, occurring when $\Delta w$ is very small, involving interactions with poorer agents. The lower-left line represents the losing line. Since $\Delta w$ is proportional to the minimum wealth of the involved agents, for instance, at $w = 50$ with $\beta = 0.5$, the maximum loss is $\Delta w = 25$. Similarly, the maximum gain is also $\Delta w = 25$, confining all points between these two lines (Fig.~\ref{fig:transitions_linearbinning}b).

By normalising the outgoing degrees, we derive the transition matrices $\boldsymbol{P}^\mathcal{W}_a$, $\boldsymbol{P}^\mathcal{W}_b$, and $\boldsymbol{P}^\mathcal{W}_c$. We use the subscript $\mathcal{W}$ to denote that these matrices are obtained numerically from the weight matrices $\mathcal{W}$. The matrices $\boldsymbol{P}^\mathcal{W}_a$ and $\boldsymbol{P}^\mathcal{W}_c$ each exhibit one eigenvalue $\lambda_1 = 1$. The corresponding eigenvector, when properly normalised, represents the steady-state distribution for these models (Fig.\ref{fig:transitions_linearbinning}d-f). For the Dragulescu model, this distribution is the Boltzmann distribution. In the case of the Yard-Sale model, the eigenvector corresponding to $\lambda_1 = 1$ is $\boldsymbol{\pi} = (1, 0, \ldots, 0)$, indicating that the steady state distribution is one of maximal inequality with all wealth concentrated in a single agent. Note that due to the discrete nature of the model, there is a small offset from zero as the transaction size becomes zero when $\text{floor}((1 - \beta)w) = 0$ (Fig.\ref{fig:transitions_linearbinning}e, inset).

The Dragulescu model, as described by $\boldsymbol{P}^\mathcal{W}_a$, adheres to detailed balance, and therefore $\boldsymbol{P}^\mathcal{W}_a$ should be diagonalisable with real eigenvalues. Nevertheless, due to the stochastic method used to obtain $\boldsymbol{P}^\mathcal{W}_a$, a significant number of eigenvalues remain complex, although their imaginary components are relatively small (Fig.\ref{fig:transitions_linearbinning}g). In contrast, the Ispolatov model, which significantly deviates from detailed balance, has eigenvalues scattered around the origin (Fig.\ref{fig:transitions_linearbinning}i). The Yard-Sale model also has complex eigenvalues, but notably, all the real parts are positive. The entropy production rate further indicates that the Dragulescu model is 'closer' to satisfying detailed balance compared to the Ispolatov model ($e_p = 0.12$ vs $e_p = 1.42$).

In this example, we discretised the model dynamics by introducing a minimal wealth unit of one, which led to a straightforward definition of states due to their discrete nature. However, when obtaining transition matrices from data or simulating a continuous model, it is necessary to first define states between which transitions are counted. One approach is to define states by linearly binning the wealth array into intervals $[w_i, w_{i+1}]$, such as $[w_0, w_1] = [0, 1]$, $[w_1, w_2] = [1, 2]$, and so forth. The issue with this approach is that wealth distributions typically become sparse towards the tail, meaning there are few agents with very high wealth. Consequently, high-wealth bins will record fewer transitions compared to bins in the bulk of the distribution. Setting a very high bin limit may result in many bins with no transitions, leading to a number of eigenvalues becoming complex even though the model theoretically satisfies detailed balance. A natural solution is to use logarithmic binning, which efficiently captures transactions involving very high wealth, as will be demonstrated in the subsequent section.

\subsection{Logarithmic wealth bins}
\label{sec:logbinning}

\begin{figure}
    \centering
    \includegraphics[width=6in]{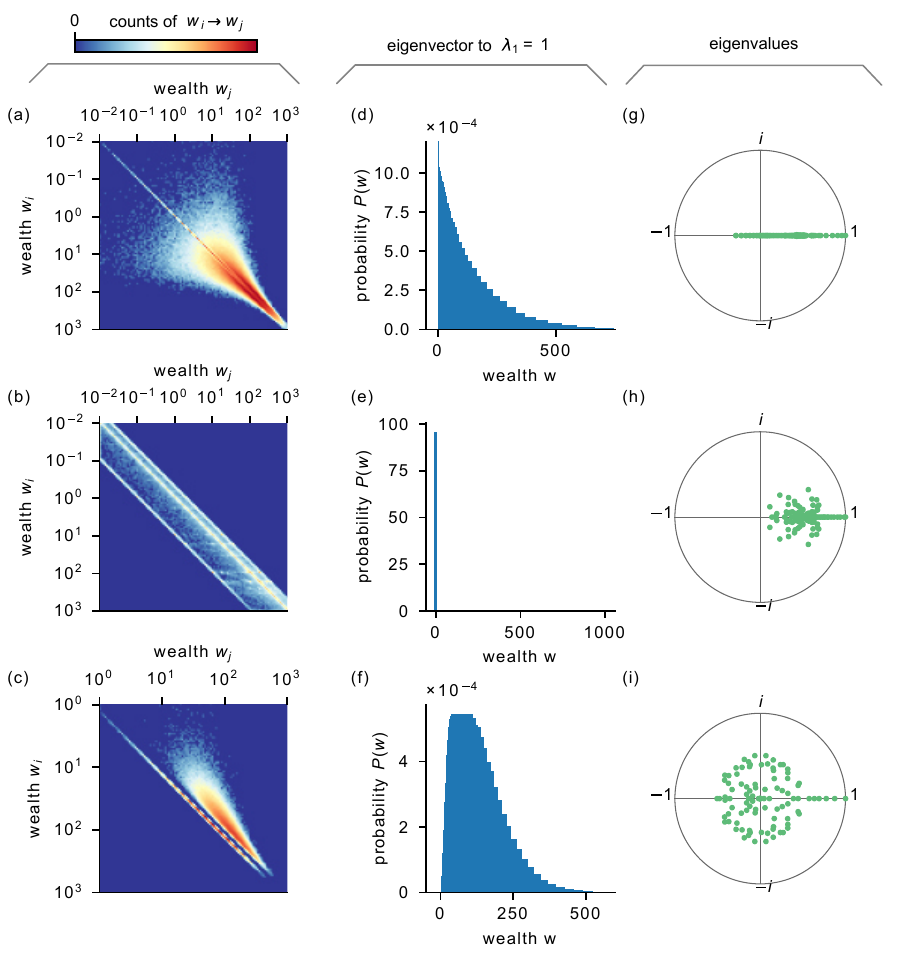}
    \caption{\textbf{Logarithmic binning of wealth captures the spectral properties of asset exchange models better than linear binning.} Wealth states are defined by logarithmically binning the wealth axis into 100 bins between an upper and a lower bound (a)-(c) Weight matrices $\mathcal{W}$ obtained through counting the number of transitions between two wealth states. (a) Dragulescu model, $\Delta w= (1-\beta)\frac{w_i+w_j}{2}$ with $\beta = 0.6$. (b) Yard-Sale model, $\Delta w = (1-\beta)\text{min}(w_i,w_j)$ with $\beta = 0.5$. (c) Ispolatov model, $\Delta w = (1-\beta)w_i$ with $i$ the losing agent and $\beta=0.8$. (d)-(f) The eigenvector corresponding to the unique eigenvalue $\lambda_1=1$. After proper rescaling, it yields the same result as for linear bins in Fig.~\ref{fig:transitions_linearbinning}. (g)-(i) The eigenvalue distribution on the unit circle. (g) Logarithmic binning captures transactions in the upper tail of the distribution better than linear binning, leading to an almost purely real spectrum for the Dragulescu model. (h) Eigenvalues of the Yard-Sale model have purely positive real parts. (i) The Ispolatov model violates detailed balance, causing a complex spectrum.}
    \label{fig:transitions_logbinning}
\end{figure}

In order to compute transition matrices using logarithmic bins we simulate our three example models in continuous version and define 100 logarithmic bins using adequate upper and lower bounds. We use again $N=\SI{1000}{}$ agents and simulate for $T=400$ time steps, collecting $\SI{400000}{}$ transitions in the weight matrices $\mathcal{W}$ (Fig.~\ref{fig:transitions_logbinning}a-c).

After normalisation, we obtain the transition matrices $\boldsymbol{P}^\mathcal{W}_{a,\rm log}, \boldsymbol{P}^\mathcal{W}_{b,\rm log}$ and $\boldsymbol{P}^\mathcal{W}_{c,\rm log}$. The steady state distributions $\pi$ again reflect a Boltzmann distribution (Fig.~\ref{fig:transitions_logbinning}d), the condensed state (Fig.~\ref{fig:transitions_logbinning}e) and a non-equilibrium steady state (Fig.~\ref{fig:transitions_logbinning}f).  The eigenvalue distributions exhibit a similar pattern to those in the linear binning examples, but it is evident that logarithmic binning provides a more suitable representation. Importantly, the spectrum of the Dragulescu model is now purely real, which aligns well with the theoretical expectation (Fig.~\ref{fig:transitions_logbinning}g). This is due to the fact that logarithmic binning allows to capture more transitions involving high wealth states, which otherwise causes complex eigenvalues. For the Yard-Sale model the eigenvalues again exhibit purely positive real parts (Fig.~\ref{fig:transitions_logbinning}h) and for the Ispolatov model the eigenvalues are found on a disc around the origin (Fig.~\ref{fig:transitions_logbinning}i). For the model with heterogeneous saving propensity (Sec.~\ref{sec:emergence-of-powerlaws}), the steady-state eigenvector shows the power-law distribution (Fig.~\ref{fig:transitionmatrix-chakrabarti}). Additionally, the entropy production underscores the difference between $\boldsymbol{P}^\mathcal{W}_{a,\rm log}$, which satisfies detailed balance, and the non-equilibrium steady state behaviour ($e_p = 0.0053$ for $\boldsymbol{P}^\mathcal{W}_{a,\rm log}$ vs $e_p = 0.82$ for $\boldsymbol{P}^\mathcal{W}_{c,\rm log}$).

% Other eigenvectors, effect of applying the transition matrix 

\begin{figure}
    \centering
    \includegraphics{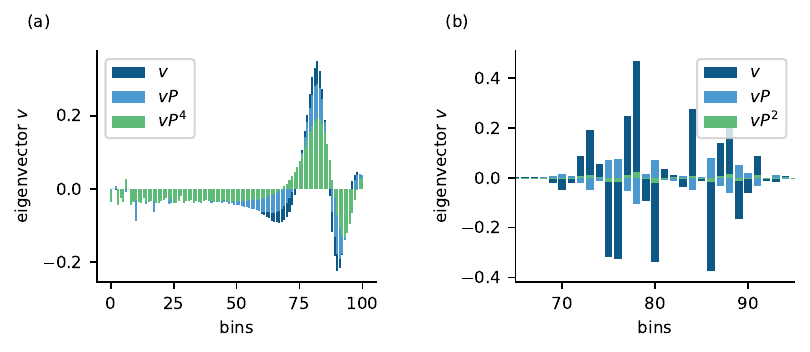}
    \caption{\textbf{Eigenvectors corresponding to $\lambda \neq 1$ indicate model behaviour while transitioning to the steady state distribution.} (a) An eigenvector $l$ to the eigenvalue $\lambda=0.9$ of the Dragulescu model. Repeated application of the transition matrix $\boldsymbol{P}$ to $l$ causes it to scale down, each time by a factor of $\lambda=0.9$. If $\lambda\approx 1$ it can require many iterations for the eigenvector to decay, indicating slow convergence of the Markov chain towards its steady state. (b) An eigenvector $l$ to the eigenvalues $\lambda = -0.2$ of the Dragulescu model. Repeated application of $\boldsymbol{P}$ flips the sign of $l$ and scales it down. For example, after applying $\boldsymbol{P}$ twice, $l$ points again in the same direction, but was scaled down by a factor of $\lambda^2 = 0.04$.}
    \label{fig:eigenvectors}
\end{figure}

After examining the largest eigenvalue $\lambda_1 = 1$ and its corresponding eigenvector, which represents the steady-state distribution, we now turn our attention to the remaining eigenvalues and their eigenvectors. What insights do they provide? To address this question, further analytical derivations are required. For the following discussion, we assume that the Markov chain in question is reversible and, consequently, that the transition matrix is diagonalisable. A first observation is that, in contrast to the steady-state distribution $\pi$, the elements of all other eigenvectors do not sum to one. Instead, the sum of their elements is zero, as will be briefly explained. Given that the transition matrix is diagonalisable, it can be expressed as \cite{seabrook2023tutorial}

\begin{equation}
    \boldsymbol{P} = \boldsymbol{Y_R}\boldsymbol{\Delta Y}_R^{-1} = (\boldsymbol{r}_1\,\,\boldsymbol{r}_2\,\,\hdots \,\,\boldsymbol{r}_N) \left(\begin{array}{cccc}
        \lambda_1 & 0 & \ldots & 0  \\
        0 & \lambda_2 & \ldots & 0  \\
        \vdots &\vdots & \ddots &  \vdots \\
        0 & 0 & \ldots & \lambda_N  \\
    \end{array}\right) \left(\begin{array}{c}
        \boldsymbol{l}_1  \\
        \boldsymbol{l}_2 \\
        \vdots \\
        \boldsymbol{l}_N \\
    \end{array}\right)\,, \label{eq:diagonalizableP}
\end{equation}

where $\boldsymbol{Y}_R$ contains the $N$ right eigenvectors $\boldsymbol{r}_i$ and $\boldsymbol{Y}_R^{-1}$ contains the $N$ left eigenvectors $\boldsymbol{l}_i$ of $\boldsymbol{P}$. Note that $\boldsymbol{r}_i$ are column vectors whereas $l_i$ are row vectors. Importantly, because $\boldsymbol{P}$ is diagonalisable, the $N$ eigenvectors are linearly independent, and thus form a basis of $\mathbf{R}^N$, called an eigenbasis \cite{seabrook2023tutorial}. Eq.~\ref{eq:diagonalizableP} can then be written as 

\begin{equation}
    \boldsymbol{P} = \sum_{i=1}^N \lambda_i \boldsymbol{r}_i \boldsymbol{l}_i\,,
\end{equation}

where each term $\boldsymbol{r}_i\boldsymbol{l}_i$ is a $N\times N$ matrix resulting from the outer product of $\boldsymbol{r}_i$ and $\boldsymbol{l}_i$. Let $\boldsymbol{l}_\gamma$ and $\boldsymbol{r}_\omega $ now be left and right eigenvectors of $\boldsymbol{P}$ with distinct eigenvalues $\lambda_\gamma \neq \lambda_\omega$. Then 

\begin{equation}
    0 = \boldsymbol{l}^T_\omega \boldsymbol{P r}_\gamma - \boldsymbol{l}^T_\omega \boldsymbol{P r}_\gamma = \lambda_\omega \boldsymbol{l}^T_\omega \boldsymbol{r}_\gamma -  \lambda_\gamma \boldsymbol{l}^T_\omega \boldsymbol{r}_\gamma  = (\lambda_\omega - \lambda_\gamma) \boldsymbol{l}^T_\omega  \boldsymbol{r}_\gamma\,. \label{eq:eigenvector_orthogonal}
\end{equation}

Because we assumed that $\lambda_\gamma \neq \lambda_\omega$ it implies that $\boldsymbol{l}_\omega \boldsymbol{r}_\gamma = 0$, namely the left and right eigenvectors corresponding to different eigenvalues are orthogonal to each other. Consider now a left eigenvector of $\boldsymbol{P}$ other than that to $\lambda_1=1$. According to Eq.~\ref{eq:eigenvector_orthogonal} it must be orthogonal to the right eigenvector with $\lambda_1=1$. The right eigenvector to $\boldsymbol{P}$ with $\lambda_1=1$ is given by the unit vector $\boldsymbol{\eta} = (1,\hdots,1)^T$ (Sec.~\ref{sec:theory:markovchains3}), as 

\begin{equation}
    (P\eta)_i = \sum_j^N P_{ij}\eta_j = \sum_j^N P_{ij} = 1\,,
\end{equation}

such that $\boldsymbol{P\eta} = \boldsymbol{\eta}$. As a result, for any left eigenvector $\boldsymbol{l}_\omega$ that is not the steady state distribution, 

\begin{equation}
    \boldsymbol{l}_\omega \boldsymbol{\eta} = \sum_i^N l_{\omega,i} = 0\,.
\end{equation}

This implies that the elements of any eigenvector other than the steady state vector must sum to zero. Consequently, these eigenvectors do not represent probabilistic vectors. So, what do they represent, and how should they be interpreted? Since the eigenvectors form a basis for $\mathbb{R}^N$, any probability vector $\boldsymbol{q}$ can be expressed as a linear combination of the eigenvectors of $\boldsymbol{P}$. By taking any such vector as an initial condition and applying the transition matrix $\boldsymbol{P}$ repeatedly, the contribution from eigenvectors associated with eigenvalues $\lambda$ of absolute value less than one will diminish over time. As a result, after many iterations, only the steady state distribution, corresponding to the eigenvalue $\lambda_1 = 1$, will persist. However, the rate at which the contributions from the other eigenvectors decays depends on the magnitude of their eigenvalues. If some eigenvalues are close to one in absolute value, it will take longer for the corresponding eigenvectors to decay, thereby prolonging the time required for the chain to converge. 

For example, the Dragulescu model exhibits both positive and negative eigenvalues. When the transition matrix is applied to eigenvectors associated with these eigenvalues, the magnitude of the eigenvectors decreases by a factor of $|\lambda|$ (Fig.~\ref{fig:eigenvectors}a). Furthermore, for negative $\lambda$, the sign of the eigenvector alternates with each application of the transition matrix, reflecting oscillatory behaviour (Fig.~\ref{fig:eigenvectors}b).

In summary, this discussion clarifies how the other eigenvectors influence the evolution of a probability vector as it approaches the steady state distribution. Eigenvalues that are purely positive, as observed in the Yard-Sale model, suggest that there are no oscillations or cycles in the probability mass while the distribution evolves towards a state of maximal inequality.

Alternative to defining transitions as occurring due to asset exchange, transitions could be defined in terms of changes in wealth over \textit{time}. For instance, by simulating asset exchange models over $t$ time steps and examining how each agent's wealth changes during this period, one could compute transition matrices for this different definition of transitions. It shows that for increasing time delay $t$, all eigenvalues tend towards zero (Fig.~\ref{fig:increases_DRA}-Fig.~\ref{fig:increases_YSM}).

Having established the necessary theoretical framework on Markov chains and illustrated these concepts with asset exchange models, we will now shift our focus to analysing transition data from the Ethereum network in the following section.

\clearpage 
\section{Study of crypto transactions in the blockchain}
\label{sec:crypto}

Having investigated the properties of transition matrices in asset exchange models, we now turn our attention to the analysis of transaction data from the digital currency Ethereum. To provide a basic understanding of this system, Sec.~\ref{sec:background:crypto} offers background information on blockchain technology. Sec.~\ref{sec:cryptodata_andmethods} outlines the data collection methodology, while Sec.~\ref{sec:crypto-balancedist} discusses the distribution of Ethereum among its users. In Sec.~\ref{sec:crypto_transitionmatrices}, we will compute and analyse transition matrices, examining their spectral properties. Recognising the highly dynamic nature of digital currency markets, Sec.~\ref{sec:crypto-dynamicmatrices} will address temporal variations in the transition matrix across different periods. All in all, this analysis will assess the extent to which the Ethereum market adheres to detailed balance and will provide insights into the degree of inequality resulting from asset exchange.

\vspace{1cm}

\begin{math_box}{\subsection{Cryptocurrencies \& the Blockchain} \label{sec:background:crypto}}

%history of block-chain (invention of bitcoin by Satoshi Nakamoto), the idea behind having a decentralized system, validation of blocks (proof of work), reward for miners.
Blockchain technology was first introduced by the pseudonymous Satoshi Nakamoto in 2008 with the invention of Bitcoin, and represents a groundbreaking innovation in decentralised financial systems \cite{nakamoto2008bitcoin}. At its core, blockchain aims to create a secure and transparent financial system that operates without the need for a central authority, such as a bank. The blockchain itself is a distributed ledger, a continuously growing list of records that is secured using cryptographic techniques. Transactions involving Bitcoin, the native currency of the Bitcoin network, are grouped into blocks and validated through a consensus mechanism known as Proof of Work (PoW). This process requires users, known as miners, to solve complex cryptographic puzzles, ensuring that each block is legitimate. Miners who successfully validate a block are rewarded with new bitcoins, providing an incentive to maintain and secure the network.\\
\
Following the invention of Bitcoin, numerous other blockchain technologies emerged, with Ethereum being the most prominent among them. Established in 2015 by Vitalik Buterin, Ethereum represents a significant evolution in blockchain technology \cite{buterin2013ethereum,buterin2014next}. Unlike Bitcoin, which focuses on peer-to-peer transactions, Ethereum supports the creation and execution of \textit{smart contracts}. These are self-executing code with predefined rules and conditions. For example, a smart contract can be programmed to collect Ether (Ethereum's cryptocurrency) until a certain threshold is reached, at which point it automatically distributes the funds or initiates a specific action. This functionality is particularly useful for applications like donation campaigns, where the smart contract can transparently manage and allocate contributions without the need for intermediaries. Overall, while Bitcoin is often regarded as a store of value, akin to digital gold, Ethereum is primarily used as a platform for decentralised applications (dApps) and financial services, making use of its robust smart contract capabilities to facilitate a wide range of applications.\\

In the following analysis, we focus on Ethereum due to two primary reasons. First, due to the existence of smart contracts, Ethereum supports a wide variety of financial instruments and services, such as lending, borrowing and trading, which closely mirrors activities in real world currency markets. Second, Ethereum employs an account-based model, where each account directly stores its balance. In Ethereum, each address, whether an externally owned account (EOA) or a contract account, has a directly associated balance. This makes it straightforward to determine the balance of any Ethereum address at any point in time. In contrast, Bitcoin uses a UTXO model that does not directly associate balances with addresses, complicating the process of linking transactions to specific users. Although clustering algorithms exist to group Bitcoin addresses that hypothetically belong to the same user, this method still lacks the direct clarity provided by Ethereum's account-based model. Therefore, Ethereum's model provides clearer and more accessible data for analysing individual financial behaviours \cite{aspembitova2019fitness,aspembitova2021behavioral}.\\

\end{math_box}

\subsection{Data and Methods}
\label{sec:cryptodata_andmethods}

\begin{figure}
    \centering
    \includegraphics{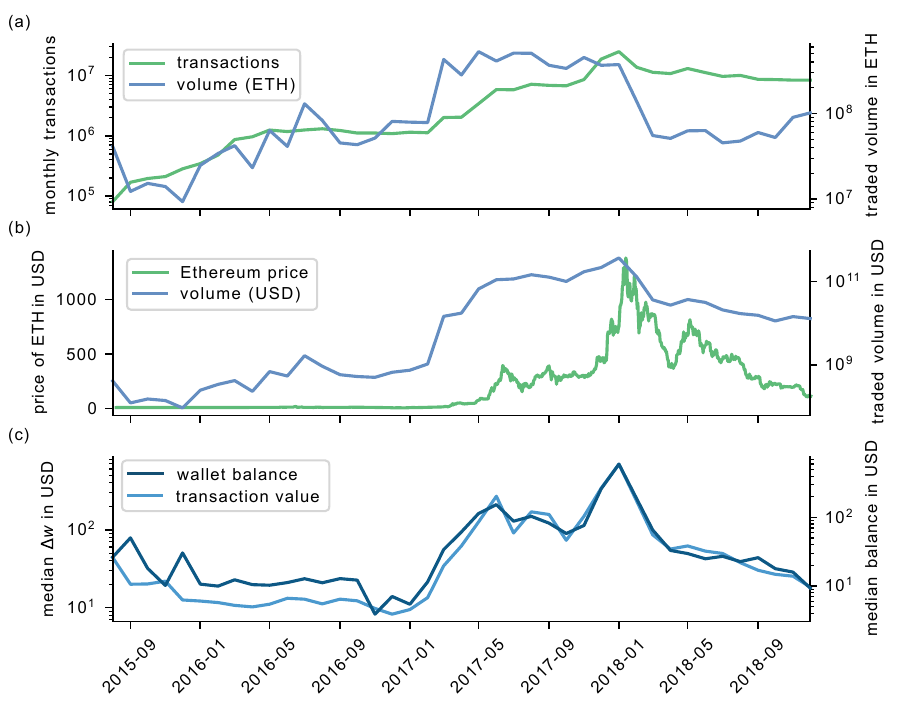}
    \caption{\textbf{Rise in popularity of Ethereum caused its price to increase.} (a) The number of transactions and traded volume of Ethereum's native currency Ether (ETH) increased 100-fold between Ethereum's inception in 2015 and the end of 2018. (b) The price of Ethereum measured in US-Dollar (USD) indicates a crypto bubble around the beginning of 2018. With Ethereum gaining popularity, also the total transacted volume in USD increased, remaining high even after the depreciation in value. (c) The median transaction value $\Delta w$ and median user wallet balance show an increase during the bubble in 2017, when measured in USD.}
    \label{fig:cryptobasics1}
\end{figure}

% Paragraph: data collection, description crypto basics 1 figure
The data used for analysis encompasses Ethereum transactions from August 2015 to December 2018. This period was selected to include transactions from Ethereum's inception in August 2015 through the cryptocurrency bubble at the beginning of 2018. We chose this time frame to examine potential changes in transaction behaviour during the bubble, which will facilitate a dynamic analysis of the transition matrices.
The dataset includes unique user addresses for both the sender and receiver of each transaction, along with the amount transacted, which we will denote as $\Delta w$. 
This data is accessible via Google's \textsc{BigQuery}. By analysing the transaction history for each user address, we can reconstruct the Ether balance in their digital wallets. To achieve this, we utilised an API from \textsc{Quicknode} in conjunction with the Python library \textsc{web3}, resulting in the collection of approximately $\SI{16000000}{transactions}$. Consequently, our dataset comprises user addresses, their wallet balances at the time of each transaction, the transaction amount, and the timestamp (Tab.~\ref{tab:dataset-overview}). The native unit of Ethereum is Ether (ETH), which we further converted to US dollar equivalent for the subsequent analysis.

\begin{table}[]
    \centering
    \begin{tabular}{c|c|c|c|c|c}
        Sender $i$ & Receiver $j$ & Value $\Delta w$ & Time $t$ & Wallet $w_i(t)$ & Wallet $w_j(t)$\\\midrule 
        $\text{0xa1e4380}\hdots$ & $\text{0x5df9b87}\hdots$ & $\SI{3.13e-14}{}$ & 2015-08-07 03:30:33 & 2000 & 0 \\
        $\text{0xbd08e0c}\hdots$ & $\text{0x5c12a8e}\hdots$ & $19.9$ & 2015-08-07 03:36:53 & 20 & 0 \\
        %$\vdots $ & $\vdots $ & $\vdots $ & $\vdots $ & $\vdots $ & $\vdots $ & \\
        $\vdots$ & $\vdots$ & $\vdots$ & $\vdots$ & $\vdots$ & $\vdots$ \\
        $\text{0xea674fd}\hdots $ & $\text{0xaf8814}\hdots $ & $0.05$ & 2018-12-31 23:59:22 & 358.184 & 2.875\\
    \end{tabular}
    \caption{\textbf{The Ethereum dataset used for analysis.} The data downloaded from Google's \textsc{Bigquery} contains unique addresses of the sending and receiving users as well as the transactions values $\Delta w$ and the time of transaction. Each transaction takes place on a \textit{block} with a certain number that keeps increasing with time, whenever a new block is created. The block number can be used to identify the wallet balance of any user at a specific time point, using the \textsc{web3} library in Python. We gathered a minimum of $\SI{10000}{}$ daily transactions between August 7th, the inception date of Ethereum, until December 31st 2028, resulting in a dataset of over $\SI{16e6}{}$ transactions. For the analysis, we further converted the transaction values and wallet balances into at the time US dollar equivalent. }
    \label{tab:dataset-overview}
\end{table}

The total number of transactions and the overall transaction volume illustrate the growing popularity of Ethereum over time, peaking with the highest monthly transaction count in early 2018 (Fig.~\ref{fig:cryptobasics1}a). Despite the high transaction frequency, the total transacted volume declined in early 2018. This decline is due to measuring volume in Ethereum's native currency, Ether (ETH), whose value increased substantially until early 2018 before experiencing a subsequent drop (Fig.~\ref{fig:cryptobasics1}b). Consequently, the volume measured in USD also decreased, though less sharply compared to the ETH-based measurement. Notably, median transaction sizes and wallet balances increased during the bubble, even when measured in USD (Fig.~\ref{fig:cryptobasics2}c). This increase likely reflects speculative behaviour, as many users entered the crypto market during this period with the expectation of financial gains.

\begin{figure}
    \centering
    \includegraphics{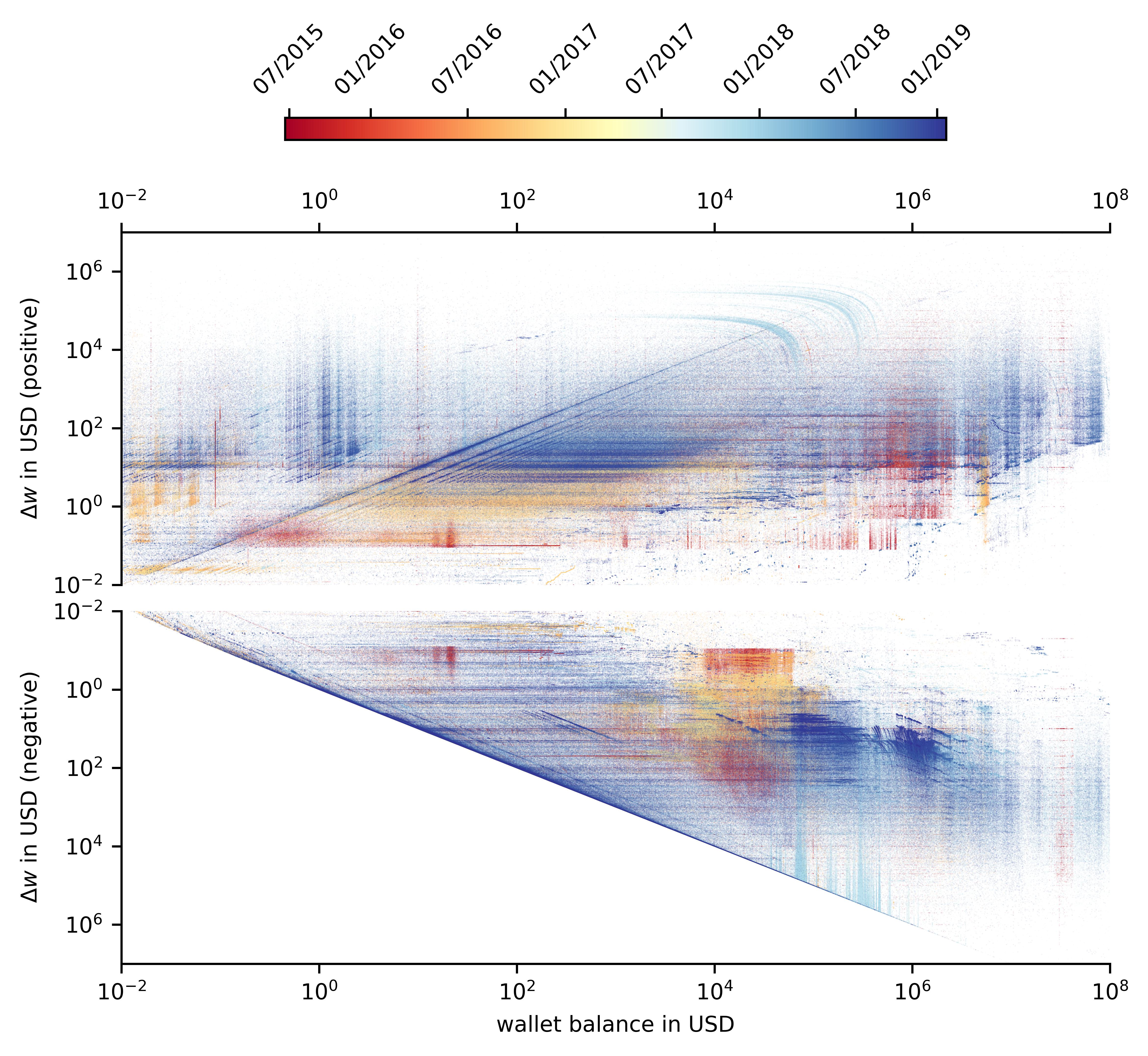}
    \caption{\textbf{Relation of transaction sizes to wallet balances.} Each transaction consists of the transaction size $\Delta w$ and the wallet balances of the sending and receiving users. Transaction sizes $\Delta w$ are scattered against the wallet balances of the users. Upper panel: for the receiving user, lower panel: for the sending user. The clear edge in the lower panel arises because wallet balances cannot drop negative.}
    \label{fig:crypto-scatterplot}
\end{figure}

In traditional currencies, it is generally expected that wealthier individuals conduct larger transactions compared to less wealthy individuals. For digital currencies to be representative of real-world markets, this relationship should similarly hold. Transaction amounts $\Delta w$ plotted against the wallet balances of corresponding users (Fig.~\ref{fig:crypto-scatterplot}) reveal indeed an increasing trend, suggesting that larger transactions are predominantly made by users with higher Ethereum balances. However, the data also shows that users with relatively small wallet balances can receive large transactions (Fig.~\ref{fig:crypto-scatterplot}, points above the 'diagonal' line in the upper panel). Such occurrences may arise when a user transfers all their Ethereum from one account to a newly created one or when an account is used for a single large transaction.

\subsection{The crypto balance distribution}
\label{sec:crypto-balancedist}

\begin{figure}
    \centering
    \includegraphics{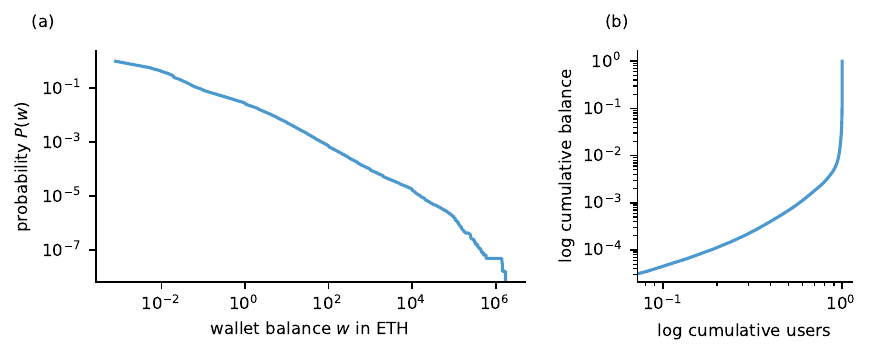}
    \caption{\textbf{The Ethereum wallet balance distribution is highly unequal.} (a) A Pareto-plot of wallet balance data in ETH taken in April 2024 shows a distribution ranging over 8 scales of magnitude. (b) The Lorenz curve with logarithmic axis, to highlight the strong increase at the very top of the user distribution. 
    %(c) The Gini index between August 2015 and December 2017 is close to $G=1$. The small drop coincides with the period of the crypto bubble (Fig.~\ref{fig:cryptobasics1}a).
    }
    \label{fig:cryptobasics2}
\end{figure}

Compared to wealth distributions in the real world, the distribution of Ethereum is even more unequal. A Pareto plot of wallet balances in 2024 demonstrates that wallet balances span across 8 orders of magnitude (Fig.~\ref{fig:cryptobasics2}a). Prior research indicates that the upper tail of this distribution is well approximated by a power-law $P(w) \sim \frac{1}{w^{1+\alpha}}$, where the exponent $\alpha$ is approximately 1. However, the remainder of the distribution is better described by a stretched exponential function \cite{kondor2014rich,aspembitova2019fitness}. The Lorenz curve illustrates that a substantial proportion of Ethereum is concentrated in the hands of a tiny fraction of users, with a steep rise near the end of the curve (Fig.~\ref{fig:cryptobasics2}b). The Gini index for 2024 reflects an almost maximal level of inequality, with a value of $G = 0.9952$. This extreme level of inequality raises questions about its origin. While digital currencies are not the primary means of asset exchange for most individuals, it is conceivable that substantial disparities arise from some individuals purchasing large amounts of Ethereum compared to others, thereby reinforcing the unequal distribution. Additionally, a considerable portion of Ethereum may still be retained by accounts linked to its initial distribution. In the following analysis, we will seek to determine the steady state distribution of Ethereum resulting from transition probabilities between different states of the distribution. Instead of reflecting a highly unequal state as just described, this method should help to understand the degree of inequality that results purely from the statistics of the exchange process.

\subsection{Static Transition matrix for Ethereum transactions}
\label{sec:crypto_transitionmatrices}

\begin{figure}
    \centering
    \includegraphics{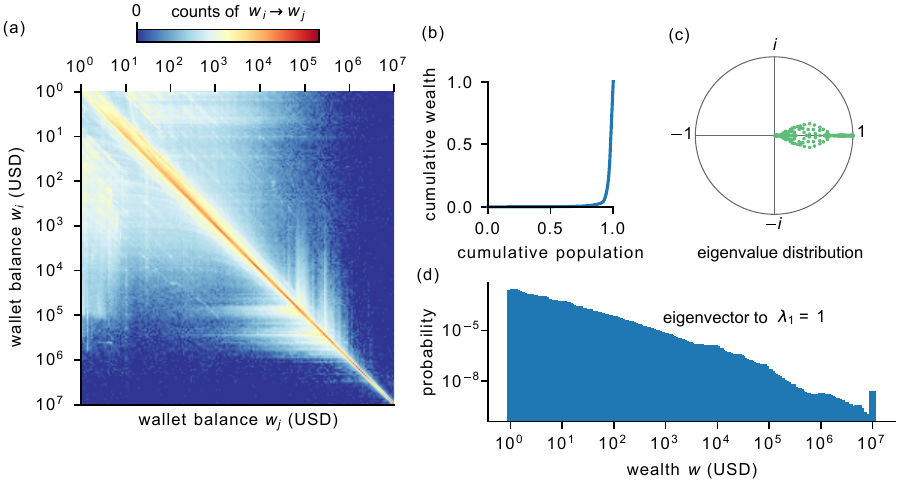}
    \caption{\textbf{Ethereum transaction dynamics lead to unequal steady state distribution.} (a) The weight matrix $\mathcal{W}^{\rm ETH}$ for $\SI{16e6}{}$ Ethereum transactions collected between August 2015 and December 2018. Each transaction is converted to the US dollar (USD) equivalent at the time of taking place. Normalising $\mathcal{W}^{\rm ETH}$ yields the transition matrix $\mathbf{P}^{\rm ETH}$. (b) The Lorenz curve constructed using the eigenvector corresponding to the unique eigenvalue $\lambda_1=1$ of $\mathbf{P}^{\rm ETH}$. The corresponding Gini index is $G=0.94$. (c) The eigenvalue distribution. All eigenvalues have a positive real part. (d) Log-log histogram of the steady state eigenvector. The increase at the tail of the distribution is due to collecting all transitions above the threshold $w=\SI{1e7}{}$ USD in the last bin.}
    \label{fig:crypto_transitionmatrix_all}
\end{figure}

Using our dataset comprising $\SI{16e6}{}$ transactions we computed the weight matrix $\mathcal{W}^{\rm ETH}$ by counting transitions between any two pairs of states (Fig.~\ref{fig:crypto_transitionmatrix_all}a). In order to do so, we first converted the transacted values $\Delta w$ as well as the user's wallet balances $w_i$ and $w_j$ into USD equivalent. We then defined 200 logarithmically spaced states ranging from $w_{\rm min} = \SI{1}{USD}$ to $w_{\rm max} = \SI{10000000}{USD}$. Finally, the transition matrix $\mathbf{P}^{\rm ETH}$ was computed by normalising $\mathcal{W}^{\rm ETH}$ such that the rows sum to one. 

The matrix $\mathbf{P}^{\rm ETH}$ has a unique eigenvector $\lambda_1=1$ and is thus guaranteed to end up in a steady-state distribution. However, the second largest eigenvalue is $\lambda = 0.999905$, the third largest is $\lambda = 0.99939$, suggesting a slow convergence to the steady state. The Lorenz curve reveals a highly unequal distribution of Ethereum (Fig.~\ref{fig:crypto_transitionmatrix_all}b). The corresponding Gini index, computed by sampling from the steady state probability distribution, is $G=0.94$. All eigenvalues of $\mathbf{P}^{\rm ETH}$ have positive real parts, indicating the absence of oscillatory behaviour as the distribution approaches its steady state (Fig.~\ref{fig:crypto_transitionmatrix_all}c). This closely resembles the eigenvalue distribution of the Yard-Sale model, that similarly lacks eigenvalues with negative real parts (Fig.~\ref{fig:transitions_logbinning}h). A Pareto plot of the steady state eigenvector shows a distribution over 6 orders of magnitude that decreases approximately as a power-law between $10^0$ and $\SI{1e4}{USD}$, after which it is truncated (Fig.~\ref{fig:crypto_transitionmatrix_all}d). 

Note that the choice of binning has an effect on the resulting steady-state vector. In particular, reducing the upper bin limit from $\SI{1e7}{}$ to $\SI{1e6}{}$ reduces the equilibrium Gini index from $G=0.94$ to $G=0.89$ (Fig.~\ref{fig:appendix-transitionmatrix_all-1e6-200bins}). In contrast, reducing the number of bins from 200 to 100 does not affect the Gini index (Fig.~\ref{fig:appendix-transitionmatrix_all-1e6-100bins}).  

The Ethereum market is undeniably dynamic, as evidenced by the volatile price movements (Fig.~\ref{fig:cryptobasics1}b). The transition matrix $\mathbf{P}^{\rm ETH}$ encompasses all transactions in our dataset regardless of their timing. However, transitions from different time periods are expected to differ. For instance, transitions from the early phases predominantly occupy the upper left corner of $\mathbf{P}^{\rm ETH}$ due to their smaller typical sizes. These temporal variations are expected to influence the eigenvalue distribution and the steady-state distribution, which will be analysed in the subsequent section.

\subsection{Dynamic Analysis of Transition Matrices}
\label{sec:crypto-dynamicmatrices}

\begin{figure}
    \centering
    \includegraphics{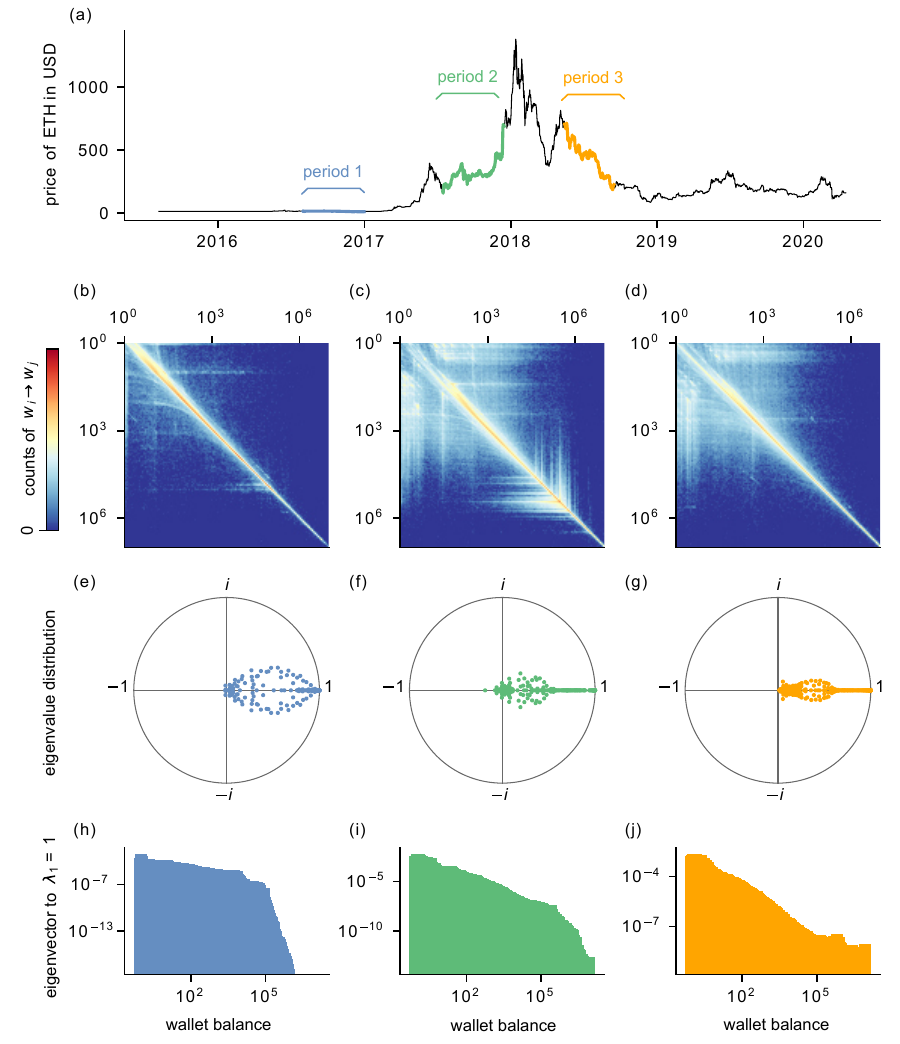}
    \caption{\textbf{Changes in the eigenvalue spectrum for different time periods.} (a) The Ethereum price in USD over time. We select three periods of 4 months each to compute the transition matrices. Period 1 covers the low price period before the first Ethereum bubble (blue). Period 2 covers the first price increase period (green). Period 3 covers the period when Ethereum dropped in price (orange). (b)-(d) The weight matrices showcasing the number of transitions between states. States are defined as 200 logarithmic bins between $w=\SI{1}{USD}$ and $w=\SI{1e7}{USD}$. During period 2, more transactions involving large amounts took place (lower right corner of transition matrices). (e)-(g) The eigenvalue spectra of the transition matrices. (h)-(j) The steady state eigenvectors to the eigenvalue $\lambda_1=1$. The Gini indices and entropy production are $G_1=0.89, e_{p,1}=0.24$, $G_2=0.92, e_{p,2}=0.52$ and $G_3=0.77, e_{p,3}=0.19$ for periods 1-3, respectively.}
    \label{fig:transitionmatrix-comparison}
\end{figure}

To compare transition matrices across different time periods, we selected three distinct 4-month intervals characterised by different Ethereum market behaviours. Period 1 spans from August 2016 to December 2016, when Ethereum's price was relatively low (Fig.~\ref{fig:transitionmatrix-comparison}a, blue). Period 2 covers the bubble phase from mid-July 2017 to mid-December 2017, during which Ethereum experienced a steep price increase (Fig.~\ref{fig:transitionmatrix-comparison}a, green). Period 3 includes mid-May 2018 to mid-September 2018, a time when the cryptocurrency bubble burst and the price of Ethereum declined (Fig.~\ref{fig:transitionmatrix-comparison}a, orange).  

The three different weight matrices look qualitatively different. 
Notably, only during Period 2, when prices were at their peak, did transitions involving users with balances of $\SI{1e6}{USD}$ or more occur. Consequently, the weight matrices for Periods 1 and 3 exhibit sparse entries in the lower right corner (Fig.~\ref{fig:transitionmatrix-comparison}b-d). 
The eigenvalue distribution appears to shift leftward during the price surge in period 2, and eigenvalues decreasing in absolute value (Fig.~\ref{fig:transitionmatrix-comparison}e,f). In Period 3, following the price drop, the distribution reverts to the right, but all eigenvalues remain with positive real parts (Fig.~\ref{fig:transitionmatrix-comparison}f,g). The entropy production rate is highest during Period 2, with $e_{p,2}=0.52$, compared to Periods 1 and 3 with $e_{p,1}=0.24$ and $e_{p,3}=0.19$, respectively. This increased entropy in Period 2 may result from a greater number of non-zero entries in the transition matrix, as zeros do not contribute to entropy. To account for this, we computed an adjusted entropy production rate, normalising $e_p$ by the fraction of zero-valued entries. This adjustment reveals that the highest entropy production rate occurred in Period 1, followed by Periods 2 and 3 ($e^{\rm adj.}{p,1}=2.32$, $e^{\rm adj.}{p,2}=1.30$, $e^{\rm adj.}_{p,3}=0.57$). The steady-state distribution vectors indicate highly unequal distributions for Periods 1 and 2, with Gini indices of $G_1 = 0.89$ and $G_2 = 0.92$, and a more equitable distribution for Period 3, with $G_3=0.77$. This shift may be due to the influx of new users during and after the bubble. Log-log histograms further illustrate qualitatively different truncations at the tails of the distributions (Fig.~\ref{fig:transitionmatrix-comparison}h-j).

\begin{figure}[!ht]
    \centering
    \includegraphics{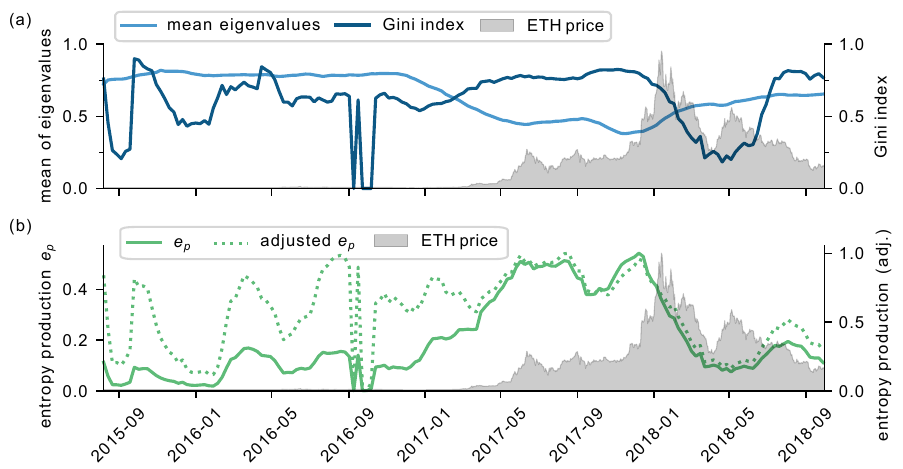}
    \caption{\textbf{The eigenvalue spectrum, Gini index and entropy production over time.} Transition matrices are computed as moving aggregates starting in August 2015 and ending in September 2018, with a time window of 90 days and increments of 7 days. States are defined as 100 logarithmic bins between $\SI{10}{USD}$ and $\SI{1e6}{USD}$. (a) The mean of the eigenvalues $\langle \lambda \rangle $ of the aggregated transition matrices over time, as well as the Gini index computed by sampling from the steady state eigenvector. During the bubble beginning in 2017, the eigenvalue spectrum shifts to the left (in negative direction) as indicated by a decrease in the mean of the eigenvalues $\langle \lambda \rangle $. The Gini index drops after the bubble burst but increases thereafter. (b) The entropy production rate $e_p$ increases during the crypto bubble (solid green line). However, this could be due to the presence of higher transactions taking place during that period. The adjusted entropy production rate does not exhibit the increase, but does exhibit the decrease after the Ethereum price drop (dotted green line). Note that the computation of the Gini index and entropy production rate requires a sensible steady state eigenvector, which is not always guaranteed and leads to issues around September 2016. 
    }
    \label{fig:movingaggregate}
\end{figure}

To better understand the temporal evolution of the eigenvalue spectrum in relation to Ethereum's price changes, we computed aggregated transition matrices using a sliding window approach with a 3-month window and a 7-day step size. This approach allows us to track spectral properties over time by shifting the 3-month period in 7-day increments (animations can be found in supplementary Sec.~\ref{sec:s-animations-gif}). During the crypto bubble, the eigenvalue spectrum notably shifted to the left, as evidenced by the mean real part of the eigenvalues, $\langle \lambda \rangle$ (Fig.~\ref{fig:movingaggregate}a). Since eigenvalues occur in complex conjugate pairs, the mean of the imaginary parts is zero, making the mean real part equivalent to the 'centre of mass' of the eigenvalues, when thinking of them as a point cloud. The Gini index remained high throughout, with a significant decrease following Ethereum's price drop at the beginning of 2018 (Fig.~\ref{fig:movingaggregate}a). It is important to note that the lower Gini values in this analysis, compared to Fig.~\ref{fig:transitionmatrix-comparison}, are due to the different upper bin limits of $\SI{1e7}{USD}$ compared to $\SI{1e6}{USD}$. The entropy production rate increased during the bubble (Fig.~\ref{fig:movingaggregate}b), which can be largely attributed to the higher number of non-zero entries in the transition matrix. However, the adjusted entropy rate, which accounts for these non-zero entries, did not increase during the bubble but similarly experienced a substantial decrease afterwards.

Overall, the dynamic analysis of Ethereum's transition matrices reveals significant variations in the eigenvalue spectrum across different time periods. Most notably, the eigenvalues shifted leftward towards the origin during price increases. Eigenvalues equal to zero indicate that the matrix has a determinant equal to zero and is thus noninvertible. One explanation for that could be that times of volatile price movement lead to speculative behaviour where many users trade similar transaction amounts, which could result in linear dependencies between the rows or columns of the transition matrix, causing zero-valued eigenvalues. Further analysis is required to understand these qualitative changes of the eigenvalue spectrum and whether they possibly precede Ethereum's market behaviour.

\clearpage 
\section{External influences}
\label{sec:bistability}

In addition to the natural exchange dynamics among individuals, a real-world economic system may also be influenced by external measures instituted by the government, such as taxes. Unlike asset exchanges, these interventions can be deterministic, meaning they apply consistently each time they are imposed. In this section, we will conceptually formalise such interventions within a Markov framework. First, we will describe how to formulate the transition matrix for a wealth tax, as discussed in Sec.~\ref{sec:yardsale-wealthcondensatoin}. Next, we will introduce and examine a phenomenon we term bistability, which refers to the existence of two qualitatively different stable states in the wealth distribution.

\subsection{Deterministic transition matrix}

Here, we focus on mechanisms that directly impact the wealth of one or all agents as an external influence. An example is the wealth tax used to address wealth inequality in the Yard-Sale model. After each asset exchange step, the wealth of all agents is adjusted according to

\begin{equation}
w_i \to w_i' = w_i + \chi_w(\bar{w} - w_i),,\label{eq:wealthtax}
\end{equation}

where $\chi_w$ is a parameter that controls the strength of the tax intervention. This mechanism is deterministic; given an initial wealth $w_i$, an agent will always transition to $w_i'$ after the tax is applied. Consequently, we can define a transition matrix for this process, $\mathbf{P}^{\rm det}$, which is deterministic rather than stochastic. Each row of this matrix contains only one nonzero entry, corresponding to the state to which the transition occurs with probability 1. The combined effect of the stochastic asset exchange and the deterministic mechanism on a vector $\boldsymbol{\mu}$ is given by the matrix product

\begin{equation}
\boldsymbol{\mu'} = \boldsymbol{\mu} \, \mathbf{P}^{\rm sto} \, \mathbf{P}^{\rm det},.
\end{equation}

In this context, the deterministic matrix alone would result in a state of maximal equality, where every agent has wealth $w = \bar{w}$. When combined with the stochastic matrix, which typically leads to more unequal states, the resulting state depends on the balance between the stochastic and deterministic part. 

From the perspective of Dynamical Systems, the wealth tax can also be expressed as a function $w_i(t+1) = f(w_i(t)) = w_i(t) + \chi_w(\bar{w} - w_i(t))$, which has a fixed point at $w = \bar{w}$ (Fig.~\ref{fig:bistability-2agent}a). In principle, it is possible that such a deterministic mechanism exhibits not just one but two fixed points. The following section will explore such a scenario in the context of asset exchange models and discuss its implications.

\subsection{Bistable wealth distribution}

In this section, we examine a deterministic mechanism that has two fixed points. Unlike the wealth tax, which drives all agents toward equal wealth, this mechanism increases the disparity between agents. It does so by exacerbating inequality: poorer agents become poorer, while wealthier agents become richer. One can achieve this by modifying the wealth tax Eq.~\ref{eq:wealthtax}. Here, a total amount of $\chi \sum_i^N w_i$ is collected from all agents and then equally redistributed, with each agent receiving $\chi \frac{1}{N}\sum_i^N w_i = \chi \bar{w}$. In contrast, we will explore a scenario where the collected amount is redistributed unequally. For instance, wealthier agents might invest a larger fraction of their wealth compared to poorer agents, who need to use most of their wealth to meet basic living expenses. Thus, the proportion of wealth invested by wealthy agents is higher than that of poorer agents, giving them another advantage.

\begin{figure}
    \centering
    \includegraphics{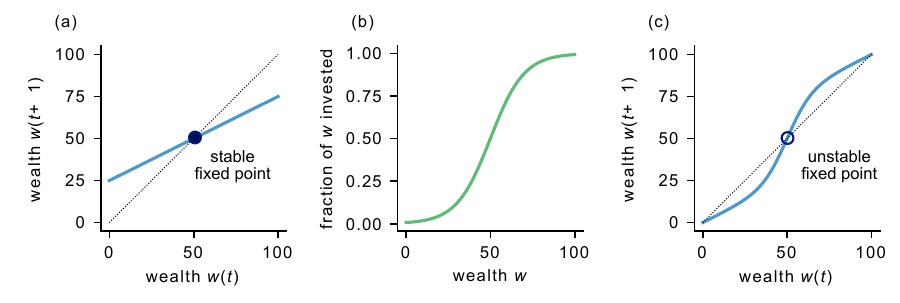}
    \caption{\textbf{Deterministic redistribution mechanisms can have one or more fixed points.} (a) A wealth tax applied to $w(t)$ changes it to $w(t+1) = w(t) + \chi (\bar w-w(t))$ (blue line). The fixed point is the intersection with the diagonal (dotted grey line), which is stable. (b) The fraction of wealth agents invest for an unequal redistribution can be given by a Sigmoid. (c) The unequal redistribution causes the stable fixed point of the wealth tax to lose stability. The mechanism increases inequality by reducing wealth of the agent with $w<\bar w$ and increasing wealth of the agent with $w>\bar w$.}
    \label{fig:bistability-2agent}
\end{figure}

Consider the redistribution mechanism defined by

\begin{equation}
    w_i' = w_i - \chi w_i +  \frac{w_i f_i}{\sum_i w_i f_i}\sum_i \chi w_i\, \hspace{0.5cm}\text{with}\hspace{0.5cm} f_i = \frac{1}{1+e^{-\alpha(w_i-b)}}\, ,\label{eq:unequalredistribution}
\end{equation}

where $f_i$ is a Sigmoid function that regulates the fraction of wealth an agent can invest (Fig.~\ref{fig:bistability-2agent}b).

This mechanism preserves the total wealth but gives an advantage to richer agents, allowing them to accumulate even more wealth if they start with a high amount. In a model with only two agents, the wealth of one agent determines the state of the other and hence $w_i' = f(w_i)$ is clearly defined. This function has an unstable fixed point at $w=\bar w$ and stable fixed points at $w=0$ and $w=W$, respectively (Fig.~\ref{fig:bistability-2agent}c). Depending on whether $w_i \lessgtr \bar w$, agent $i$'s wealth will either end up at $w=0$ or $w=W$, resulting in a maximally unequal state.

\begin{figure}
    \centering
    \includegraphics{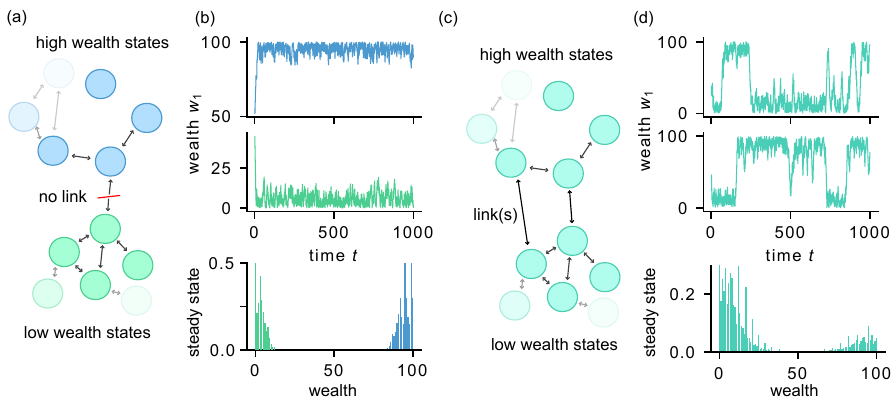}
    \caption{\textbf{Deterministic, unequal redistribution can lead to a disconnected Markov chain.} An asset exchange of $\Delta w$ takes place between two agents with a total amount of wealth $W=100$. After each asset exchange, wealth is redistributed unequally according to Eq.~\ref{eq:unequalredistribution}. (a) For $\Delta w = 4$ the Markov chain is disconnected, it is not possible to transition between the high wealth states to the low wealth states. (b) Top: Temporal simulations indicate that agents either end up solely in the high wealth states, or the low wealth states, but do not transition between them. Bottom: The two eigenvectors corresponding to the degenerate eigenvalues $\lambda=1$, coloured distinctly. (c) For $\Delta w = 10$ the Markov chain is connected and it is possible to transition between the high wealth states and the low wealth states, albeit with low probability. (d) Top: Temporal simulations indicate that agents can transition between the high and low wealth states, but spend the majority of time close to one of either state. Bottom: The eigenvector corresponding to the unique eigenvalue $\lambda_1=1$ has two peaks, around the high wealth and low wealth states.}
    \label{fig:bistability-temporal}
\end{figure}

We now examine the impact of combining an unequal redistribution mechanism with an asset exchange step. We consider a discretised version of the Dragulescu model with $N = 2$ agents, where an amount $\Delta w\in \mathbf{N}$ is exchanged and wealth states range from $0, 1, 2, \ldots, W$, similar to $\mathbf{P}_a$ in Eq.\ref{eq:exampletransitionmatrices}. The overall transition matrix for the stochastic asset exchange combined with the unequal redistribution mechanism is the product of the two respective transition matrices. For a small value of $\Delta w$, the resulting matrix exhibits two disconnected sub-components, meaning transitions between high and low wealth states are not possible  (Fig.~\ref{fig:bistability-temporal}a).
As a result, the transition matrix has two eigenvalues equal to 1, each corresponding to one of the communicating classes. Numerical simulations indicate that, after initialisation, the agents' wealth consistently fluctuates around either $w = 0$ or $w = W$ (Fig.~\ref{fig:bistability-temporal}b, top). The eigenvectors corresponding to the two eigenvalues $\lambda = 1$ show distributions concentrated around $w=0$ and $w=W$ (Fig.~\ref{fig:bistability-temporal}b, bottom). In contrast, when $\Delta w$ is larger, transitions between lower and higher states of the wealth spectrum become possible, resulting in a fully connected transition graph (Fig.~\ref{fig:bistability-temporal}c). Numerical simulations of this scenario demonstrate that agents switch between the two wealth states, spending extended periods near $w = 0$ and $w = W$, but occasionally transitioning between these states (Fig.~\ref{fig:bistability-temporal}d, top). The transition matrix has only one eigenvalue equal to one, which corresponds to the unique stationary distribution of the chain, showing peaks near $w = 0$ and $w = W$ (Fig.~\ref{fig:bistability-temporal}d, bottom).

\begin{figure}
    \centering
    \includegraphics{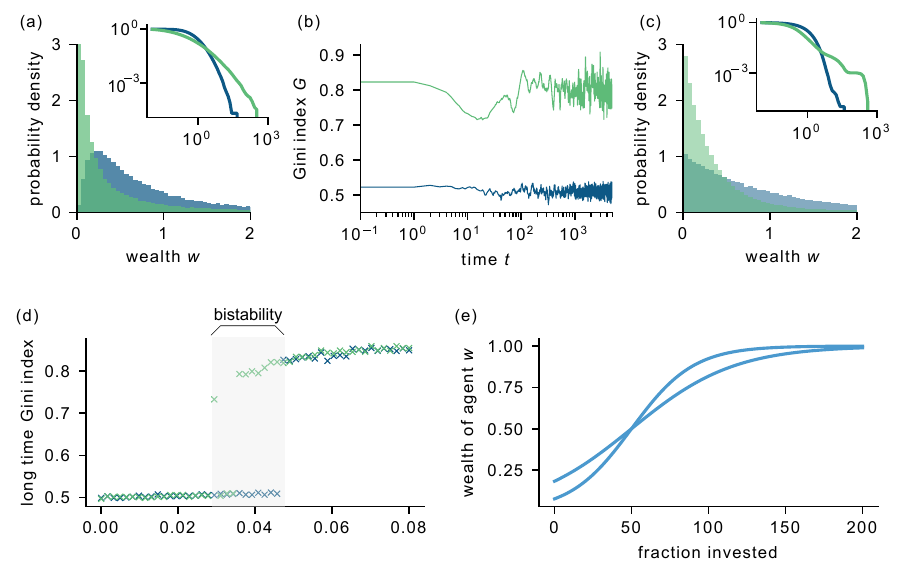}
    \caption{\textbf{Bistability in asset exchange models through unequal redistribution.} (a) The model is initialised with lognormal distributions with mean $\mu=1$ and variance $\sigma=1$ (blue, Gini index $G=0.52$) and $\sigma=1.9$ (green, Gini index $G=0.84$), respectively. The Pareto plot shows the presence of high wealth individuals for $\sigma=1.9$ (inset, green). (b) The model is evolved using the Dragulescu asset exchange ($\Delta w \propto w_i+w_j$) and the unequal redistribution mechanism Eq.~\ref{eq:unequalredistribution}. The different initial distributions do not converge to the same steady state distribution. (c) The distributions after time $t=10000$. The blue distribution is a Boltzmann distribution with $G=0.5$. The green distribution, although its histogram resembles a Boltzmann distribution, contains high wealth individuals as apparent by the Pareto plot (inset, green). The Gini index is $G=0.76$. (d) Gini indices after $t=10000$ for different values of $a$ in the Sigmoid Eq.~\ref{eq:unequalredistribution}. Initial distributions are again logormal with mean $\mu=1$ and variances $\sigma=1$ (blue) and $\sigma=1.9$ (green). In the range $a\approx 0.03$ to $a\approx 0.05$ the initial distributions converge to different final distributions. (e) Unequal redistribution is introduced by a Sigmoid function that dictates the fraction of wealth agents can invest. The two curves show $a=0.03$ and $a=0.05$, in between which bistability arises. The mean wealth of agents is $\bar w = 1$, allowing only the very rich to benefit substantially through the redistribution.}
    \label{fig:bistability-dragulescu}
\end{figure}

The effects of unequal redistribution can also be examined in a system with more than two agents. We initialise two distinct wealth distributions, each with varying levels of inequality, and evolve them according to the Dragulescu asset exchange rule and the unequal redistribution mechanism (Eq.~\ref{eq:unequalredistribution}). Specifically, we use lognormal initial wealth distributions with different variances to control the degree of inequality (Fig.~\ref{fig:bistability-dragulescu}a). Throughout the simulation, the initial distribution with a lower Gini index ($G(t=0)=0.52$) remains more equal than the initial distribution with a higher Gini index ($G(t=0)=0.84$) (Fig.~\ref{fig:bistability-dragulescu}b). When starting with an unequal initial distribution, a few agents possess a level of wealth that allows them to benefit substantially from the redistribution mechanism and amass a large fraction of the total available wealth. When these wealthiest agents engage in transactions, substantial amounts of $\Delta w \propto w_i + w_j$ can be exchanged, causing significant fluctuations in the Gini index (Fig.~\ref{fig:bistability-dragulescu}b, green line). The final wealth distribution resembles a Boltzmann distribution in the bulk but includes several agents with wealth levels significantly exceeding the standard deviation (Fig.~\ref{fig:bistability-dragulescu}c). This is highly unlikely for a true Boltzmann distribution --- in an exponential distribution with mean wealth $\bar{w}$, the standard deviation is $\sigma = \bar{w}$. The probability of an agent having wealth $w = \bar{w} + 10\sigma$ or more is given by

\begin{equation}
    \text{Pr}[w\geq \bar w + 10\sigma | P(w)] = \int_{11\bar w}^\infty \frac{1}{\bar w}e^{-\frac{w}{\bar w}}\,\text{d}w = \frac{1}{e^{11}} \approx \SI{1.67e-5}{}\,. 
\end{equation}

For $N = 1000$ agents with $\bar{w} = 1$, there is roughly a $1.67\%$ chance of having an agent with wealth $w \geq 10$. In a test simulation with unequal redistribution, however, there are 8 agents with wealth $w > 10$. If the distribution were truly exponential, this would occur with a probability of approximately $ 0.0167^8 \approx \SI{6e-15}{}$.

Bistability arises only for specific parameters in the unequal redistribution mechanism. The Sigmoid function that determines the fraction of wealth that agents can invest must increase steeply, but not excessively so; otherwise, all starting distributions converge to the same final distribution (Fig.~\ref{fig:bistability-dragulescu}d,e). 

The emergence of bistability necessitates an asset exchange mechanism that has a stable steady-state distribution with $G<1$. In the Yard-Sale model, the system converges to the maximally unequal state $G=1$, and introducing an additional unequal redistribution mechanism alone will not alter this outcome. Instead, bistability in the Yard-Sale model occurs only if an equal redistribution mechanism is also implemented. We can formalise this combined mechanism as:

\begin{equation}
    w_i' = w_i - \chi w_i +  \epsilon \frac{w_i f_i}{\sum_i w_i f_i}\sum_i \chi w_i + (1-\epsilon)\frac{1}{N}\sum_i \chi w_i \, \hspace{0.5cm}\text{with}\hspace{0.5cm} f_i = \frac{1}{1+e^{-\alpha(w_i-b)}}\, ,\label{eq:unequalredistribution2}
\end{equation}

where the parameter $\epsilon$ controls the fraction of the collected amount $\sum_i \chi w_i$ that is redistributed equally versus unequally. For $\epsilon = 0$, the mechanism corresponds to the wealth tax Eq.~\ref{eq:wealthtax}, while for $\epsilon = 1$, it corresponds to the unequal redistribution mechanism Eq.~\ref{eq:unequalredistribution}.

\begin{figure}
    \centering
    \includegraphics{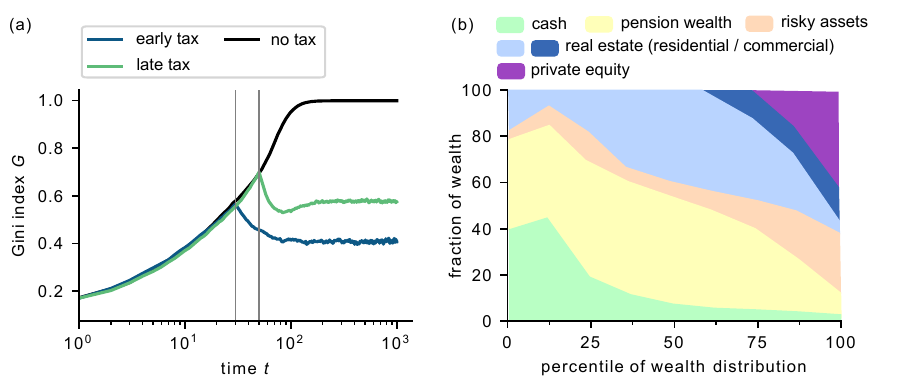}
    \caption{\textbf{The timing of tax intervention can determine the long term evolution of the system.} (a) A Yard-Sale model with tax that is redistributed partly equally, and partly unequally (Eq.~\ref{eq:unequalredistribution}). Without any tax, the Gini index approaches $G=1$ (black line). Timely introduction of the tax ($\epsilon=0.45$) at $t=30$, when the Gini index has reached $G=0.55$, stabilises the system around $G=0.4$ (blue line). In contrast, late introduction at $t=50$, when $G$ has reached $G=0.69$, only reverts the system to $G=0.57$ (green line). (b) Individual wealth composition across the wealth distribution, qualitatively reproduced from data of Swedish households \cite{bach2020rich}. In the upper percentiles, wealth is primarily held in risky financial assets, real estate and private equity, whereas it is primarily held in cash and pension wealth in the lower percentiles. }
    \label{fig:taxintervention_wealthcomposition}
\end{figure}

Changing $\epsilon$ from 1 to 0 can thus be interpreted as introducing a wealth tax intended to restore a more equal state within the system. However, the timing of implementing such a wealth tax is critical. If introduced too late, it may only revert the system to the more unequal state. For instance, in a temporal simulation starting with $\epsilon = 1$, the system tends toward its maximally unequal state with a Gini index of $G = 1$. Introducing the tax $\epsilon = 0.45$ at $t = 30$, when the system's Gini index is $G = 0.55$, successfully guides the system to a path with a Gini index of $G = 0.4$. Conversely, introducing the tax at a later time, $t = 50$, when the Gini index has risen to $G = 0.69$, only reduces the Gini index to $G = 0.57$ (Fig.~\ref{fig:taxintervention_wealthcomposition}a).

We defined the unequal redistribution mechanism described in Eq.~\ref{eq:unequalredistribution} based on the concept of unequal investment shares among agents. To further illustrate this, we take wealth composition data from Sweden as an example (Fig.\ref{fig:taxintervention_wealthcomposition}b, qualitatively reproduced) \cite{bach2020rich}. It shows that wealthy individuals predominantly hold their assets in real estate, risky financial assets, and private equity-investment vehicles that have the potential to further increase wealth. In contrast, individuals at the lower end of the wealth distribution tend to keep most of their assets in cash, which does not contribute to wealth accumulation. Cash is required to meet basic living expenses, which is why we assume that once these needs are met, there is a sharp increase in the proportion of wealth available for investment, incorporated by the Sigmoid in Eq.~\ref{eq:unequalredistribution}. By incorporating a redistribution mechanism that amplifies inequality based on the share of investments agents undertake, we demonstrate that this approach leads to distinctive dynamic properties of the wealth distribution. When combined with stochastic asset exchange, the model exhibits path dependence, introducing a novel phenomenon in the study of asset exchange models.

% ADD: Implication of bistability: Transition probabilities depend on the entire state of the system, one cannot even numerically compute a transition matrix 

\clearpage
\chapter{Concluding Remarks}
\label{ch:conclusion_outlook}

\section{Conclusion}
\label{sec:conclusion}

In this thesis, we pursued four primary objectives. First, in Chapter~\ref{ch:theoretical_background}, we provided an overview on the history of wealth distribution modelling, the Gini index and asset exchange models. Our analysis revealed that Pareto's observed power-law holds only for the upper tail of most empirical wealth distributions, while the majority of the distribution is better approximated by an exponential distribution. The Gini index, a measure of inequality, exceeds $G = 0.7$ when measuring wealth, and is generally lower when measuring income. In an effort to explain the formation of wealth distributions from microscopic interactions between individuals, physicists have developed asset exchange models. These models describe wealth formation through a pairwise exchange process among economic agents, governed by specific exchange rules. 
The Master equation of such processes is a simplified version of the Boltzmann equation. 
It shows that if the exchange process adheres to detailed balance, the Boltzmann distribution emerges as the equilibrium distribution of wealth. In the absence of detailed balance, non-equilibrium steady-state distributions result. Additionally, some models exhibit wealth condensation, where a single individual accumulates the entire wealth. This outcome, counterintuitive at first glance, can be understood by considering the asset exchange as a random walk with state-dependent step sizes. As agents descend the wealth distribution, their step size diminishes, making it increasingly difficult to ascend. Wealth condensation can be proved by demonstrating that the Gini index is an $H$-function of the Fokker-Planck approximation of the dynamics, similar to Boltzmann's $H$-Theorem. Finally, we analysed modifications to the Yard-Sale model, including external tax interventions and adjusted winning probabilities among agents, and demonstrated that a wealth tax can prevent wealth condensation.

Chapter ~\ref{ch:results} presented the results of this thesis. 
In Section~\ref{sec:markovchains}, we aimed to gain a more detailed understanding of exchange dynamics of asset exchange models by computing Markov transition matrices. We began with simplified 2-agent asset exchange models, for which the transition probabilities between states can be written down analytically, as the state of one agent defines the state of the whole system. Transition matrices can be represented as weighted directed graphs and are called \textit{irreducible} when all states are mutually reachable. 
In that case, the transition matrix possesses a unique left eigenvalue $\lambda_1 = 1$, with the corresponding eigenvector representing the steady-state distribution. The other eigenvalues describe the system's convergence to the steady state and may be complex, always appearing in conjugate pairs. When transition probabilities between states are balanced, the Markov chain satisfies \textit{detailed balance} and is \textit{reversible}, meaning that processes occur with equal probability in both forward and backward directions. In that case, the transition matrix is orthogonally diagonalisable and its spectrum is purely real. The reversibility of Markov chains is analogous to the reversibility of thermodynamic systems, which must also satisfy detailed balance to be reversible. According to the second law of thermodynamics, non-reversible systems produce entropy, and the entropy production rate quantifies the degree of irreversibility.
Simulations of the asset exchange models discussed in Chapter~\ref{ch:theoretical_background} allowed us to numerically estimate the transition matrices. Results confirmed that the model satisfying detailed balance exhibits a purely real spectrum and lower entropy production rates compared to the model that does not satisfy detailed balance. Besides, the model leading to wealth condensation shows a purely positive spectrum, the implications of which are to be further understood. 

In Section~\ref{sec:crypto}, we analysed transition matrices computed for a dataset of Ethereum transactions. We collected $\SI{16e6}{}$ transactions between Ethereum's inception and early 2019, including a period of significant price volatility. We hypothesised that qualitative changes in user behaviour during this period would be reflected in the transition matrix spectra. The distribution of Ethereum differs from real world wealth distribution as it is highly unequal, with a Gini index $G \approx 1$. The steady-state distribution derived from the transition matrix using all transactions resulted in a Gini index of $G = 0.94$. Analysis of transition matrices across different time periods revealed that during the price bubble, the eigenvalue spectrum shifted downward, while most eigenvalues retained a positive real part, similar to the model leading to wealth condensation. Following the bubble, during the price crash, the entropy production rate decreased substantially.

In Section~\ref{sec:bistability}, we examined the impact of deterministic external influences on asset exchange models. Specifically, we investigated the effects of a wealth tax, which collects a proportional amount from each agent’s wealth and redistributes it. We analysed scenarios where the collected tax amount is redistributed unequally, favouring wealthier agents. This unequal advantage can be interpreted as investment benefits, as wealthier agents can invest larger shares of their wealth compared to poorer agents, who must first cover basic living costs. This deterministic investment process leads to wealth condensation, while a stochastic asset exchange results in a stable wealth distribution. Together, these mechanisms achieve a state of balance that, importantly, depends on the initial wealth distribution.
In cases where the initial distribution is already highly unequal, rich agents further benefit from unequal investments, exacerbating inequality. In contrast, for a more equal initial distribution, rich agents cannot take advantage through investments, which leads to a more equal wealth distribution. By introducing a parameter $\epsilon$ to adjust between equal and unequal redistribution, one can interpret a decrease in $\epsilon$ as the imposition of a wealth tax. The timing of this tax is crucial: a timely introduction during rising inequality can guide the system to a lower Gini index compared to a late introduction. To our knowledge, this is the first demonstration of how path dependence can emerge in asset exchange models.

\section{Discussion}
\label{sec:discussion}

\textbf{Asset exchange models and their assumptions}

Asset exchange models, inspired by energy exchange in fluids, are far from capturing the complexity of real-world wealth dynamics within human societies. Real economic systems involve various mechanisms that create, destroy and transfer wealth among individuals, that extend beyond simple pairwise transactions in a closed system. Consequently, various extensions have been incorporated into asset exchange models, such as inflation, production, and taxes. These modifications generally affect the resulting wealth distribution, and several studies have examined these effects \cite{banzhaf_effects_2021,boghosian_fokkerplanck_2014}. Notably, in many cases, additional factors do not alter the Gini index, especially when the economy is uniformly affected. For example, in models with a Boltzmann equilibrium distribution, the mean wealth $\bar{w}$ in $P(w)\frac{1}{\bar{w}}e^{-\frac{w}{\bar{w}}}$ cancels out when computing the Lorenz curve, indicating that it is indifferent to the total wealth within a society. Thus, additional effects such as inflation do not impact the degree of inequality in models satisfying detailed balance. In contrast, an influx of wealth at, e.g., the bottom of the distribution would affect the wealth of agents differently across the distribution, and thus no longer lead to a Boltzmann distribution . 

The approach in asset exchange models contrasts sharply with economic approaches to studying wealth accumulation \cite{boghosian_fokkerplanck_2014, boghosian_economically_nodate}. Asset exchange models depict wealth formation as a purely stochastic process where each interaction results in a clear winner and loser. In contrast, economic theory suggests that agents would avoid transactions where they might incur losses. In most microeconomic frameworks, the value of a commodity is not intrinsic but is determined by the price that a buyer is willing to pay \cite{boghosian_economically_nodate}. When valuing commodities differently, it is possible that one individual will overvalue a commodity and another will undervalue it. Hence, two individuals may be perfectly satisfied with the price they agree upon to exchange the commodity for. In contrast, asset exchange models, assume that one agent gains intrinsic wealth while the other loses, implying a 'mistake' by the losing agent.

The closest economic counterpart to asset exchange models is the \textit{search theory of money}, which examines the costs associated with finding suitable trading partners \cite{yakovenko_colloquium_2009}. In this framework, money facilitates transactions by reducing the friction associated with finding trading partners \cite{kiyotaki1993search}. Unlike asset exchange models, which analyse the statistical properties of wealth distributions, search theory focuses on market efficiency, such as why individuals hold onto money rather than spending it immediately, and how inflation affects the use of money. Thus, asset exchange models remain unique in their focus on the statistical properties of wealth distributions and their emergence from microscopic interactions.

Two fundamental concepts in economics are general equilibrium frameworks and the utility maximisation principle. Efforts have been made to reconcile asset exchange models with these economic principles \cite{boghosian_economically_nodate}. It has been demonstrated that, within both frameworks, at least two tradable goods are necessary for trade, unlike asset exchange models that consider only a single quantity. 
Moreover, general equilibrium models incorporate an intrinsic advantage for poorer agents, which stabilises the system but is unlikely to hold in the real world. \cite{boghosian_oligarchy_2017,boghosian_economically_nodate}. Instead, asset exchange models align with certain aspects of prospect theory, which explains decision-making under uncertainty, particularly asymmetric losses and gains. Empirical findings indicate that utility functions differ based on the potential benefit or loss: for gains, utility is concave, meaning satisfaction diminishes with increasing gains; for losses, utility is convex, meaning the pain of losing intensifies with increasing losses. It is possible to construct asset exchange models that incorporate prospect theory principles, providing them with a solid microeconomic foundation \cite{boghosian_economically_nodate}.

Despite ongoing debates within economics, asset exchange models offer clear advantages over traditional economic approaches. They provide insight into how wealth outcomes can result from stochastic processes, namely pure chance, contrasting with economic frameworks where agents are assumed to be perfectly rational and fully informed. In reality, individuals are neither perfectly rational nor possess complete information. The idealised economic models attribute outcomes solely to rational decision-making, implying that success and failure are solely the result of individual choices. However, the role of unequal opportunities and luck in real-life economic outcomes is undeniable. Thus, the perspective that rational decisions alone determine success or failure is both inaccurate and potentially misleading. Asset exchange models highlight that even in seemingly fair economies, significant inequality can arise, challenging the traditional economic viewpoints.

\textbf{Markov analysis of transition matrices}
% issues with Markov assumption
% what approximation do we take by simulating? 

One major criticism of asset exchange models is the implicit Markov assumption. Specifically, if asset exchange is interpreted as monetary transactions occurring against a flow of goods, the Markov assumption becomes unrealistic. For instance, if an agent has recently experienced a long period of spending, they would likely need to sell goods in the near future to restore their wealth, violating the Markov property. In contrast, the interpretation of asset exchange as a 'mistake' by one of the agents does not face this issue and can indeed adhere to the Markov property.
However, as mistakes can be assumed to be minor relative to an agent’s wealth, this interpretation would require transaction amounts to be very small.

Another challenge related to the Markov property is the assumption that agents are chosen randomly for transactions. In reality, trading partners often remain consistent over extended periods. For example, when visiting a supermarket, one typically does not select a random store but rather frequents the same as always. More realistic models would incorporate a transaction network with directed weights, reflecting the agents’ preferences for interaction partners. Additionally, specifying interaction rates based on agents' wealth could make the models more realistic. For example, poorer agents might predominantly engage with wealthier agents, as they produce different types of goods. The subsequent section, Sec.~\ref{sec:otherapproaches-models}, briefly outlines how to incorporate such interaction rates into asset exchange models.

In Sec.\ref{sec:markovchains}, we made extensive use of the Markov property to analyse spectra of transition matrices. While we could analytically derive transition probabilities in the case of two agents, this approach becomes impractical for systems with more agents. In asset exchange models, the transition probability from a wealth state $w$ to another state depends not solely on $w$ but also on the availability of a suitable interacting agent such that $\Delta w$ allows the transition $w \to w'$. Thus, wealth changes depend on interactions with other agents, which is why the joint probability distribution $P(w_i, w_j)$ appears in the Master equation discussed in Sec.\ref{sec:wealth_modelling_overview}. By simulating asset exchange models and counting transitions, we implicitly assumed that $P(w)$ can change on its own, namely that $\frac{dP(w,t)}{dt}$ depends only on $P(w)$ itself.

The eigenvalue analysis of continuous systems requires the definition of suitable wealth bins. 
Initially, we employed linear bins for asset exchange models, which proved impractical. The results are highly sensitive to the upper bound of the bins, and when wealth spans several orders of magnitude, this approach results in excessively large transition matrices. Additionally, higher wealth bins become sparse, making transitions rare and potentially inducing complex-valued eigenvalues even when they should be real. Logarithmic bins address the issue of upper bounds by accommodating high-wealth transactions within a manageable number of bins. Nonetheless, the choice of upper bounds still has an effect on the spectrum, although it is much smaller than in the linearly binned case (Fig.~\ref{fig:appendix-transitionmatrix_all-1e6-200bins}, Fig.~\ref{fig:appendix-transitionmatrix_all-1e6-100bins}).

\textbf{Ethereum data analysis }

One downside of asset exchange models is that there is no straightforward way of comparing their mechanisms to data. For an effective comparison, this would require information on the wealth of individuals engaging in transactions, which is difficult, if not impossible, to access.
In this thesis, we circumvented this issue by using data on cryptocurrency transactions, made accessible by blockchain technology. This allowed us to compute transition matrices from millions of transactions. However, the question remains of how closely these transactions reflect real-world transactions. Significant differences are evident: many transactions are comparable in magnitude to the wallet balances of the users involved (Fig.~\ref{fig:crypto-scatterplot}), which would rarely occur in the real world. Furthermore, the distribution of wallet balances in the cryptocurrency space is much more unequal than real-world wealth distributions. While it is likely that users with substantial Ethereum holdings are also wealthy in reality, the converse is less certain --- users with low Ethereum balances could be anything from rich to poor in the real world.

To potentially enhance the quality of the data, one approach could be to filter transactions based on specific criteria. For instance, transactions could be restricted to those made by users who are active regularly or who use cryptocurrencies in ways that mirror real-world behaviours. Another method could involve filtering transactions according to 'realistic' sizes that are small relative to the users' wallet balances. Additionally, analysing network structures might provide insights; for example, identifying subnetworks where interactions are predominantly internal could better reflect a closed system similar to asset exchange models. However, applying these filters could introduce biases, which is why we chose to select transactions randomly for this analysis.

\textbf{Bistability} 

Generally, the steady-state distribution of basic asset exchange models is independent of the initial distribution. This is expected; in a container of gas particles the Boltzmann distribution also does not depend on the initialisation of the particles. Yet, path dependence in economic systems is rather intuitive ---
historical conditions and past events undeniably influence a country’s economic trajectory. For instance, certain countries exhibit significantly higher levels of inequality compared to others. These countries often have less democratic governance (Fig.~\ref{fig:appendix-incomegini-vs-democracy}, Fig.~\ref{fig:appendix-wealthgini-vs-democracy}) and lower Human Development Indexes compared to countries with more equitable distributions (Fig.~\ref{fig:appendix-incomegini-vs-hdi}, Fig.~\ref{fig:appendix-wealthgini-vs-hdi}). It is plausible that high levels of inequality in these countries obstruct efforts to reduce inequality. For example, policies intended to address inequality may be ineffective if corruption is too pronounced. Conversely, in countries with more equitable distributions, similar measures could result in a further reduction of inequality.

In this thesis, we made an initial attempt to incorporate path dependence into asset exchange models by introducing an additional mechanism that drives inequality. Specifically, we implemented a mechanism that collects a fraction of each individual’s wealth, akin to a wealth tax, but redistributes it unequally rather than equally. We interpreted it as investments and assumed that wealthier agents can invest a larger portion of their wealth compared to poorer agents, reflecting different investment opportunities across the wealth distribution. In reality, however, investment returns are stochastic rather than deterministic, as assumed in our model. Our scenario is thus somewhat artificial, as it presumes all agents invest in shares of the same company within a closed economy. It remains an open question for future research to explore which other mechanisms, when integrated with asset exchange processes, might lead to path dependence.

\section{Other approaches to wealth distribution modelling}
\label{sec:otherapproaches-models}

Asset exchange models are not the only physics-inspired models used to study wealth distribution. Here, we outline several noteworthy models that pursue a different approach.

\textbf{Random growth models}

Random growth models examine the impact of compounding returns on investments. In these models, wealth is neither conserved nor exchanged between agents; instead, it is assumed to grow or shrink independently for each agent who invests in the stock market. This framework demonstrates that here too, chance alone can lead to a maximally unequal state with $G=1$ \cite{fargione_entrepreneurs_2011,cardoso_equal_2022}.  

The analysis involves an individual-based stochastic model that incorporates compounding returns. All individuals are assumed to have equal talent and start with identical amounts of capital. Success in one year is independent of success in previous or subsequent years, and labour income (additive income) does not contribute to capital wealth growth.

In each time period $k$, each agent $i$ invests their capital and earns a return rate $r_{i,k}$, which is randomly drawn from a common normal distribution with mean $\mu$ and variance $\sigma^2$. The factor by which an agent's capital increases in time period $k$ is $e^{r_{i,k}}$, so the total capital accumulated at time $t$ is given by $e^{r_{i,1}} \cdot e^{r_{i,2}} \cdot \ldots \cdot e^{r_{i,t}} = e^{r_{i,1} + r_{i,2} + \cdots + r_{i,t}} = e^{x_i}$. Here, $x_i = \sum_{k=1}^t r_{i,k}$ is, according to the central limit theorem, normally distributed with mean $\mu t$ and variance $\sigma^2 t$.

The total wealth of the society is then the integral of individual wealth over this probability density,

\begin{equation*}
    \int_{-\infty}^\infty dx \frac{1}{\sqrt{2\pi\sigma^2}} e^{\frac{(x-\mu t)^2}{2 t\sigma^2}}e^x\,.
\end{equation*}

The wealth held by a small fraction of the population is given by the ratio

\begin{equation*}
    \frac{\int_{\mu t + h\sigma \sqrt{t}}^\infty \frac{1}{\sqrt{2\pi\sigma^2}} e^{\frac{(x-\mu t)^2}{2 t\sigma^2}}e^x\,dx}{\int_{-\infty}^\infty  \frac{1}{\sqrt{2\pi\sigma^2}} e^{\frac{(x-\mu t)^2}{2 t\sigma^2}}e^x \,dx} = \frac{1}{2}\left(1+\text{erf}\frac{\sigma\sqrt{t}-h}{\sqrt{2}}\right)\,,
\end{equation*}

where $h$ denotes the number of standard deviations from the mean. As $\text{erf}(x)$ approaches 1 for $x \to \infty$, this integral approaches 1 as $t \to \infty$, for any fixed $h$. Thus, over time, the portion of wealth held by an arbitrarily small fraction of individuals converges to 1 \cite{fargione_entrepreneurs_2011}. Importantly, this result is independent of the mean $\mu$, implying that it applies to shrinking, stagnant, and growing economies alike.

Wealth concentration in this model arises because some individuals, by chance, receive a series of high growth rates and thus accumulate substantial capital. Note that this model predicts a log-normal distribution of wealth, which contrasts with the exponential - Pareto distribution observed empirically.

All in all, random growth models consider compounding returns as the sole mechanism of wealth change. A straightforward extension of this model type is to incorporate money exchange, leading to the well-known Bouchaud-Mezard Model.

\textbf{Bouchaud-Mezard Model}

In contrast to random growth models, this approach is based on a stochastic differential equation and includes interactions between agents. The evolution of wealth $w_i$ is described by \cite{bouchaud_wealth_2000}

\begin{equation*}
    \frac{dw_i}{dt} = \eta_i(t) w_i + \sum_{j\neq i} J_{ji}w_j - \sum_{j\neq i} J_{ij}w_i\,,
\end{equation*}

where $\eta_i$ is a Gaussian random variable with common mean $\mu$ and variance $2\sigma^2$, representing the growth of wealth due to investments in the stock market. The terms $\sum_{j \neq i} J_{ji} w_j$ and $\sum_{j \neq i} J_{ij} w_i$ describe interactions between agents, with $J_{ij}$ denoting the amount of wealth transferred from agent $i$ to agent $j$ when agent $j$ purchases goods from agent $i$.

In the simplest case, it is assumed that $J_{ij} = J/N$ for all $i \neq j$. This represents a mean-field approximation, where all agents experience the same external environment, leading to the equation

\begin{equation}
    \frac{dw_i}{dt} = \eta_i(t)w_i + J(\bar w - w_i)\,. \label{eq:bouchaudmezard-meanfield}
\end{equation}

Using the corresponding Fokker-Planck equation, it can be shown that wealth follows a power-law distribution in the long-term limit with a Pareto exponent $\alpha = 1 + \frac{J}{\sigma^2}$. Increased connectivity between agents, represented by $J$, raises the Pareto exponent, thereby reducing inequality. This is expected since Eq.~\ref{eq:bouchaudmezard-meanfield} includes a competition between two terms: the investment term, which mirrors the random growth model and leads to inequality, and the trading term $J$, which tends toward equality. As connectivity among agents increases, the relative impact of money exchange grows, resulting in a more equitable distribution.

By integrating the effects of investments and trading, the Bouchaud-Mezard Model provided an initial framework to analyse how both types of economic mechanisms influence inequality, and has led to similar work studying the joint effects \cite{nirei_two_2007}.

\textbf{Growth rate model for income distribution}

An approach to modelling income distribution is using a Master equation with growth and reset terms \cite{neda_scaling_2020}. Income of individuals is assumed to grow in increments, e.g., due to career progression, and decrease due to the death of individuals, leading to a reset term. Both growth and reset occur at a certain rate that depends on the income.

Assume that the income distribution is binned into $n$ bins and $P_k(t)$ is the probability that an individual has income in bin $k$ at time $t$. Normalisation requires that $\sum_k P_k(t) = 1$. The Master equation of the described process is given by

\begin{equation*}
    \frac{dP_k}{dt} = \mu_{k-1}P_{k-1} - \mu_kP_k -\gamma_k P_k + \delta_{k,0}\langle\gamma\rangle\,.
\end{equation*}

Here, $\mu_k$ denotes the growth rate of income and $\gamma_k$ denotes the reset rate. The last term ensures normalisation and is given by $\langle\gamma\rangle(t) = \sum_i \gamma_i(t)P_i(t)$.

In the continuous limit, the Master equation becomes

\begin{equation*}
    \frac{dP(w)}{dt} = \frac{d}{dw}(\mu(w)P(w)) - \gamma(w)P(w) + \delta(w)\langle\gamma(w)\rangle(t)\,,
\end{equation*}

where $\mu(w)$ and $\gamma(w)$ are the growth and reset kernels. By inferring the kernels $\mu$ and $\gamma$ from individual income data, it was shown that the Master equation can be solved analytically \cite{neda_scaling_2020}. The resulting steady-state distribution corresponds to a Pearson Type I distribution, which is notable because it can model both the exponential bulk and the power-law tail of the empirical income distribution. This represents a significant advance in attempts to capture the full range of the income distribution with a single function, as opposed to modelling the bulk and tail separately .

\textbf{Aggregation models}

A wide range of phenomena in natural sciences can be described by aggregation processes. For example, aerosol formation in clouds, the formation of clusters in the rings of Saturn, but also social phenomena such as opinion formation or herding behaviour in financial markets \cite{leyvraz_scaling_2003,pruppacher1998microphysics,bohorquez_common_2009,brilliantov_size_2015}. In all these processes, clusters of size $i$ collide or interact with clusters of size $j$ to produce clusters of size $k = i + j$. For example, two groups of individuals with different opinions on some subject (size $i$ and $j$) might interact and form a single joint opinion on that subject (size $i+j$).

In asset exchange models, in the limit where agents do not save anything during an exchange, interactions can also be viewed as an aggregation process. Upon interaction, agents $i$ and $j$ form a new agent that combines both agents' wealth from before, $w_i + w_j$, and an empty cluster representing an agent with zero wealth, $w = 0$. The Master equation for such an aggregation process is generally given by

\begin{equation*}
    \frac{dP_k}{dt} =  \frac{1}{2}\sum_{i+j=k} P_iP_jK_{ij} - P_k\sum_i P_i K_{ik}\,,
\end{equation*}

where $P_k$ denotes the concentration of clusters of size $k$, i.e., individuals with wealth $k$. The first term accounts for the creation of a cluster of size $k$ by aggregating clusters of sizes $i$ and $j$, and the second term accounts for losses of clusters of size $k$ due to aggregation with clusters of other sizes. The kernel $K_{ij}$ specifies the rate at which clusters of size $i$ and $j$ interact. 
In some systems for example, large clusters are approximately stationary due to their large mass, and only smaller clusters are free to move. In such a scenario, the reaction is strongly dominated by the interaction of small clusters with large clusters, which can be specified by properly choosing $K_{ij}$.

While this process is just one limiting case and may not be very realistic for wealth exchange, studying such aggregation models can help in extending asset exchange models. In contrast to asset exchange models, where the focus has primarily been on the details of the exchange rules, aggregation-fragmentation systems have mainly focused on analysing the kernels $K_{ij}$. For instance, it has been found that systems behave qualitatively differently depending on whether interactions are dominated by large-large, small-small, or large-small interactions. Aggregation models also exhibit a phenomenon known as 'gelation,' where a single cluster emerges that encompasses all the system's mass, corresponding to wealth condensation in asset exchange models. Importantly, the gelation threshold in aggregation models depends on the kernels; for example, multiplicative kernels of the form $K_{ij} = ij$ tend to favour gelation more than constant ($K_{ij} = a$) or additive kernels ($K_{ij} = i + j$).

In asset exchange models, introducing kernels that govern the rate at which rich and poor agents interact could influence, e.g., the transition for which wealth condensation occurs. 
However, analytical treatment of such extended models is difficult. The Master equations for aggregation models and exchange models have one important difference. In aggregation models, the Master equation can be analysed by examining how the moments of the distribution behave. Similarly, in asset exchange models, we utilised this method to show that the second moment of the Yard-Sale model diverges (Sec.~\ref{sec:yardsale-wealthcondensatoin}), indicating wealth condensation. The challenge arises when introducing a kernel of the form $K(w, w')$ in asset exchange models to denote the rate of interactions between agents with wealth $w$ and $w'$. In this case, the moment change $\frac{dm_k}{dt}$ depends on the next moment $m_{k+1}$, leading to a system of equations that does not yield a closed-form solution. Therefore, apart from a few existing results \cite{ispolatov_wealth_1998,krapivsky_kinetic_2010}, analysing rate kernels in asset exchange models would require numerical simulations.

\textbf{Model free explanation of exponential - power-law distributions}

The fact that empirical wealth distributions often show both exponential and power-law behaviour can be qualitatively explained without using a specific model \cite{silva406385thermal}. Consider the general Fokker-Planck equation that governs the change in the wealth distribution,

\begin{equation*}
    \frac{\partial P(w,t)}{\partial t} = -\frac{\partial }{\partial w}[\langle \Delta w\rangle P(w)] + \frac{\partial^2}{\partial w^2}\left[\frac{\langle \Delta w\rangle ^2}{2}P(w)\right]\,.
\end{equation*}

For the lower part of the distribution, it is reasonable to assume that the change in wealth $\Delta w$ is independent of $w$, as they are driven by wages, which are additive. In this case, both $\langle \Delta w \rangle$ and $\langle \Delta w \rangle^2$ are constant, and the stationary distribution of the Fokker-Planck equation is an exponential distribution $P(w) \propto \exp(-w/T)$ with $T = - \frac{\langle \Delta w \rangle^2}{\langle \Delta w \rangle}$. Note that this requires $\langle \Delta w \rangle < 0$ for a steady-state solution.

On the other hand, changes in the upper tail of the wealth distribution can be assumed to be proportional to wealth, $\Delta w \propto w$, so $\langle \Delta w \rangle = aw$ and $\langle \Delta w \rangle^2 = bw^2$. The stationary solution in this case is a power-law distribution \cite{silva406385thermal} $P(w) \propto \frac{1}{w^{1+\alpha}}$ with Pareto exponent $\alpha = 1 + \frac{a}{b}$. 

All in all, this presents a model free approach to explain the different scaling regimes in the upper and lower part of the distribution. Wealth diffusion is assumed to be independent of wealth in the lower part, but proportional to wealth in the upper part, resulting in exponential and power-law solutions, respectively.

\section{Outlook}
\label{sec:outlook}

Our analysis of Markov transition matrices has revealed some unresolved questions. Among these, is a more detailed understanding of the full spectrum, in particular the position of the eigenvalues.
Most importantly, we observed that the Yard-Sale model only has eigenvalues with positive real parts, which raises the question of whether the position of eigenvalues can generally be linked to the degree of inequality of the steady state eigenvector.

A significant challenge in relating Ethereum transactions to asset exchange models is the interpretation of a transaction. In asset exchange models, transactions are viewed as shifts in wealth due to imperfectly fair exchanges, which implies that transactions should be relatively small. However, our analysis of Ethereum data considered the full amounts of transactions, which can be very large. Future research could explore the effects of scaling down transaction sizes and introducing a random element to determine which agent benefits or loses from an exchange.

The transition matrix analysis used in this study has potential applications beyond systems where a quantity such as wealth is exchanged. The approach requires only that a given system is stationary, with constant transition probabilities between states. For instance, one could apply this approach to income distributions. By constructing a transition matrix from individual income data over time, collected for many individuals, the steady-state eigenvector should represent the actual income distribution observed in the society. If the spectrum of this matrix is purely real, it would suggest detailed balance, indicating equal probabilities of moving up and down the income distribution.

Spectral properties could also provide a measure of mobility or 'dynamic inequality'. While the Gini index effectively measures the degree of inequality of a wealth distribution, it considers the wealth distribution as an aggregated quantity. For example, consider two identical wealth distributions that differ in terms of the mobility of individuals: in one, individuals are free to move up and down the distribution; whereas in the other, they have their fixed place in the distribution. We might be tempted to say that the former case is more 'fair' than the latter, as it allows individuals to change their position. While this difference does not become apparent in the Gini index, it does so in the eigenvalue spectrum. The dynamic system would show a distribution of eigenvalues in the unit circle, while the static system would only have eigenvalues equal to one, as the transition matrix would be diagonal. Thus, the spectrum could offer a new way to quantify mobility / dynamic inequality.

The study of wealth distributions using asset exchange models has become a major part of Econophysics. Since their inception more than 20 years ago, various exchange rules and possible extensions to AEMs have been explored. 
In this thesis, we have pursued three novel approaches: examining spectral properties of transition matrices, analysing blockchain data, and exploring path dependence. These contributions provide a foundation for future research to build upon and expand.

\bibliography{library}

\clearpage 
\appendix 

\chapter{Supplementary Material}

\section{Ethereum eigenvalue animations}
\label{sec:s-animations-gif}
Animations of temporal changes of the spectrum of transition matrices computed for Ethereum (Sec.~\ref{sec:crypto-dynamicmatrices}) can be found under \href{https://owncloud.gwdg.de/index.php/s/2uziWR7sSIda7dt}{this link} or in the .zip folder that was submitted with this thesis. Results are computed for different binning and window sizes that are taken to compute the transition matrix. The top panel shows the price of Ethereum and the window used to compute the transition matrix (coloured in red). The middle panel shows the mean of the eigenvalues, which is always real. The lower left panel shows the weight matrix $\mathcal{W}$ and the lower right panel the eigenvalues in the complex plane. 

\section{Mathematical background}

\begin{math_box}{\subsection{Thin vs fat tailed distributions} \label{sec:theory:thin_vs_fat_tails}}

    Consider $N$ identically and independently distributed (iid) random variables $x_1,...,x_N$. They are distributed according to the common probability density function $P(x)$, which is defined such that the probability of a variable falling between $x$ and $x+\text{d}x$ is $P(x)\text{d}x$ \cite{walck1996hand}. 
    $P(x)$ is said to be \textit{thin-tailed} if all moments of $P(x)$ exist \cite{rolski2009stochastic, bouchaud2003theory}, i.e., 
    \begin{equation*}
        m_n = \int_{-\infty}^\infty \text{d}x\,x^n\,P(x) < \infty \, \forall n \in \mathbf{N}_0\,.
    \end{equation*}
    Contrarily, $P(x)$ is said to be \textit{fat-tailed} if some of the moments diverge \cite{rolski2009stochastic}. Clearly, for all moments to exist, $P(x)$ must decrease sufficiently quick for large $x$ to counter the increase of $x^n$. 

    Two famous thin-tailed distributions are the Gaussian distribution $P_G(x)$ and the Laplace distribution $P_L(x)$ \cite{bouchaud2003theory}, given by
    \begin{equation*}
        P_G(x) = \frac{1}{\sqrt{2\pi \sigma^2}} e^{-\frac{(x-\langle x\rangle)^2}{2\sigma^2}} \hspace{0.5cm}\text{and}\hspace{0.5cm} P_L(x) = \lambda e^{-\lambda x}\,,
    \end{equation*}
    respectively \cite{walck1996hand}. For the Gaussian distribution, $\langle x\rangle = m_1$ is the mean and $\sigma^2 = m_2 - m_1^2$ is the variance. For both distributions all moments exist as $x^n$ is suppressed exponentially. 
    In contrast, two examples of fat-tailed distributions are the Pareto distribution $P_P(x)$ and the Student's-t distribution $P_T(x)$ given by 
    \begin{equation*}
        P_P(x) = \begin{cases} 
      0 & \text{if } x < x_{\rm min} \\
      \frac{\alpha x_{\rm min}^\mu}{x^{\mu + 1}} & \text{otherwise}
      \end{cases} \hspace{0.5cm} \text{and}  \hspace{0.5cm} 
      P_T(x) = \frac{N}{(a^2+x^2)^\frac{1+\mu}{2}}\,,
    \end{equation*}
    respectively \cite{walck1996hand}. $N$ is a normalisation factor in the Student's-t distribution. Note that the Pareto-distribution is a full power-law distribution, while the Student's-t distribution is only a power-law distribution $P_T\sim x^{-(1+\mu)}$ asymptotically. Nonetheless, for both distributions only moments $m_n$ with $n<\mu$ exist \cite{walck1996hand}. This implies that for $\mu <2$ both distributions have a diverging variance and for $\mu < 1$ even the mean is diverging. 

    Intuitively, the implications of fat tails can be illustrated if asking for the probability of a $10$-$\sigma$ event, i.e., the probability of an extreme event larger than 10 times the standard deviation:
    \begin{equation*}
        \text{Pr}[x>\langle x\rangle + 10\sigma|P(x)] = \int_{\langle x \rangle + 10\sigma}^\infty P(x')\,\text{d}x'\,.
    \end{equation*}
    For the Gaussian distribution this probability is extremely small, $\text{Prob}[x>\langle x\rangle + 10\sigma|P_G(x)] \approx 10^{-22}$. If an event took place once per second, the probability that a $10$-$\sigma$-event would have occurred in the lifetime of the universe ($10^{16}$ seconds) would be 1 in one million. However, in reality, many $10$-$\sigma$ events occur for various processes, implying that the underlying distribution is far from Gaussian and exhibits fat tails instead \cite{taleb2010black}. \\
    
    Another important insight can be gained by comparing how the maximum of a set of random variables $M_N=\text{max}\,(x_1,...,x_N)$ grows with $N$ compared to the sum of the random variables $S_N=\sum_i^N x_i$. \\%The former is given using the Fisher-Tippett-Gnedenko-theorem, also known as the extreme value theorem and the latter is given by the central limit theorem.\\ 
    
    \textbf{The maximum of $N$ random variables.}\\
    The behaviour of the maximum of a set of random variables can be evaluated using the Fisher-Tippett-Gnedenko-theorem, also known as the extreme value theorem \cite{fisher1928limiting}. 
    It states that  
    \begin{equation*}
        M_{N|G} \sim \sigma\sqrt{2\log N} \hspace{0.5cm}\text{and}\hspace{0.5cm} M_{N|L} \sim \frac{1}{\lambda} \log N,\label{eq:m_n_mu_larger1}
    \end{equation*}
    for the Gaussian and Laplace distribution \cite{zarfaty2021accurately}. In contrast, for the Pareto and Student's-t distribution the maximum grows considerably faster (if $\mu$ is small) \cite{zarfaty2021accurately,bouchaud2003theory}:
    \begin{equation}
        M_{N|P,T} \sim N^{\frac{1}{\mu}}\,.\label{eq:m_n_mu_smaller2}
    \end{equation}
    Note that if $\mu<1$ the maximum of the set of random variables $M_N$ grows faster than the number of its members $N$.\\
    
    \textbf{The sum of $N$ random variables.}\\
    The behaviour of the sum of random variables is governed by the central limit theorem. Its well-known statement is that when adding random variables $x_i$, their sum $S_N$ follows a Gaussian distribution \cite{walck1996hand}. However, this only holds if the second moment of the underlying distribution of $x_i$ exists, i.e., for $\mu>2$. The precise statement in that case is that the probability of finding the rescaled variable $\frac{S_N-\langle x\rangle N}{\sigma \sqrt{N}}$ within the interval $[a,b]$ is given by 
\begin{equation*}
    \text{Pr}[a<\frac{S_N-\langle x\rangle N}{\sigma \sqrt{N}}<b] \xrightarrow{N \to \infty} \int_a^b \frac{\text{d}x}{2\pi}e^{-\frac{x^2}{2}}\,.
\end{equation*}
    This statement is surprisingly universal as it does not depend on the details of the underlying distribution, only on the existence of the second moment \cite{montgomery2010applied,walck1996hand}. 
    If that is not the case, i.e., for $\mu<2$, the generalised version of the central limit theorem states that 
    \begin{equation*}
        S_N = \langle x\rangle N + uN^{\frac{1}{\mu}} \hspace{0.5cm}\text{with}\hspace{0.5cm} P(u)\xrightarrow{N \to \infty}L_{\mu,\beta}(u)\,,
    \end{equation*}
    where $L_{\mu,\beta}(u)$ denotes the Levi stable distribution with shape parameter $\beta$ \cite{gnedenko1968limit,stanley2000introduction}. Note that if $\mu<1$ the term $\langle x\rangle N$ does not exist. 
    Consider now the case for $\langle x \rangle =0$. Depending on whether $\mu>2$ or $\mu<2$, the sum $S_N$ grows as 
\begin{equation*}
    S_N \sim \sqrt{N} \,\,(\mu>2) \hspace{0.5cm}\text{and}\hspace{0.5cm} S_N\sim N^\frac{1}{\mu} \,\,(\mu<2)\,.
\end{equation*}

    Comparing this to how the maximum $M_N$ scales with $N$ (Eqs.~\ref{eq:m_n_mu_larger1},\ref{eq:m_n_mu_smaller2}), it becomes evident that for the case of $\mu>2$ the sum of random variables is way larger than any of its members, $S_N \gg M_N$ because $\sqrt{N}\gg \sqrt{\log N}$. 
    
    However, if $\mu<2$, both grow as $N^\frac{1}{\mu}$, hence $S_N \sim M_N$. This implies that one single member of the set of random variables can 'dominate' the set and be of similar magnitude than all remaining members combined.

\end{math_box}

\clearpage

\begin{math_box}{\subsection{Stochastic dynamics} \label{sec:theory:fokkerplanck}}

\textbf{Derivation of the Fokker-Planck equation}\\

A classic example of the use of the Fokker-Planck equation is the description of the dynamics of a particle, such as a pollen grain, suspended in a fluid \cite{risken1996fokker,zwanzig2001nonequilibrium}. It experiences friction according to the Stokes' law, $F_{\rm Stokes} = av$. Classically, the equation of motion is given by \cite{risken1996fokker}

\begin{equation}
    m\dot v + av = 0\,.
\end{equation}

Equivalently, one can write

\begin{equation}
    \dot v + \gamma v = 0\,,
\end{equation}

with $\gamma = a/m = 1/\tau$. $\tau$ is called the relaxation time --- an initial velocity $v(0)$ decreases exponentially to zero with relaxation time $\tau$, $v(t) = v(0)\exp(-t/\tau)$. \\

The friction term  $\gamma v$ is motivated by the interaction between the particle and the fluid molecules. As the particle collides with these molecules, its momentum is gradually transferred to the fluid, resulting in a reduction of its velocity to zero. In this context, this process is treated as a deterministic effect.\\

This deterministic description is a good approximation only when the particle’s mass is sufficiently large, such that the velocity fluctuations due to thermal motion are negligible. However, if the particle is sufficiently light, so that individual collisions with fluid molecules significantly impact its velocity, the deterministic approximation is no longer valid. In this scenario, the particle's trajectory is influenced not only by macroscopic forces but also by microscopic interactions with the fluid molecules.\\

If we wanted to treat the problem exactly, we would need to solve the coupled equations of motion for both the particle and the surrounding fluid molecules. As the number of molecules is very large ($\sim 10^{23}$), this poses an impossible task. Instead, the collision with the molecules can be approximated by introducing a stochastic noise term $F_{\rm noise}$, such that $F(t) = F_{\rm Stokes}(t) + F_{\rm noise}(t)$.\\

Dividing by the mass, we obtain $\dot v + \gamma v = \Gamma(t)$, where $\Gamma$ denotes the Langevin force. It satisfies $\langle \Gamma(t)\rangle = 0$ and $\langle \Gamma(t)\Gamma(t')\rangle = q\delta(t-t')$, which arises from the limit that collision times are much shorter than the relaxation time $\tau$ \cite{risken1996fokker}. \\

Because $\Gamma$ is a stochastic quantity, so will be the velocity $v$. Hence, we can ask about the probability $P(v)$ of finding $v$ in an interval $v$ and $v+dv$. The evolution of this probability is given by 

\begin{equation}
    \frac{\partial P(v)}{\partial t} = \gamma \frac{\partial }{\partial v}(vP(v)) + \frac{q}{2}\frac{\partial ^2}{\partial v^2}(P(v))\,,
\end{equation}

which is one of the simplest Fokker Planck equations. The general, the Fokker-Planck equation has the form

\begin{equation}
\frac{\partial P(v)}{\partial t} = \sum_n \left(-\frac{\partial}{\partial v}\right)^n \left[D^{(n)}(v)P(v)\right]\,,    
\end{equation}

also known as Kramers-Moyal expansion. The Kramers-Moyal coefficients $D^{(n)}$ are given by 

\begin{equation}
    D^{(n)}(v) = \frac{1}{n!}\text{lim}_{\Delta t\to 0}\frac{\langle [\Delta v(t)]^n \rangle}{\Delta t}\,.
\end{equation}

It can be shown that if $v$ obeys a Langevin equation with Gaussian $\delta$-correlated noise, all coefficients $D^{(n)}$ for $n\geq 3$ vanish. This leads to the Fokker-Planck equation

\begin{equation}
    \frac{\partial P}{\partial t} = -\frac{\partial }{\partial v}(D^{(1)}(v)P(v) ) + \frac{\partial^2 }{\partial v^2}(D^{(2)}(v)P(v) )\,.
\end{equation}

$D^{(1)}(v)$ is called the drift coefficient and $D^{(2)}(v)$ the diffusion coefficient. \\

The Fokker-Planck equation usually appears for variables describing a macroscopic but small subsystem, under the influence of external influences, e.g., collisions. If the subsystem was larger, the fluctuations can usually be neglected and a deterministic equation is a good approximation.\\

The Fokker-Planck equation is not the only equation of motion for the distribution function of a variable. Another common description is given using the Master equation, which is a generalised version of the Fokker-Planck equation. \\

\textbf{The Master equation}\\

A general linear equation for the time evolution of a probability density is the Master equation. Consider the case in which a variable $x$ only takes integer values $m$. Then, 

\begin{equation}
    \frac{\partial P_n}{\partial t} = \sum_m [ P_m w(m\to n) - P_n w(n\to m) ]\,,
\end{equation}

where $w(m\to n)$ denotes the transition rate from $m$ to $n$. For a continuous variable, this can be generalised to 

\begin{equation}
    \frac{\partial P(x)}{\partial t} = \int [P(x')w(x'\to x) - P(x)w(x\to x')]\,dx'\,.
\end{equation}

The Fokker-Planck equation is just a special case of the Master equation where the transition probabilities $w(x'\to x)$ are given by 

\begin{equation}
    w(x'\to x) = \left[ -\frac{\partial}{\partial x}D^{(1)}(x) + \frac{\partial^2}{\partial x^2}D^{(2)}(x) \right]\delta(x-x')\,.
\end{equation}

Note that here the probability at a later time is completely determined by the probability at the current time, i.e., the system is Markovian. A more general equation is given when taking memory effects into account:

\begin{equation}
    \frac{\partial P(x)}{\partial t} = \int_{-\infty}^t  K(x,t-\tau) P(x,\tau)\,d\tau\,.
\end{equation}

In this case, one can derive a memory-dependent Fokker-Planck equation,

\begin{equation}
     \frac{\partial P(x)}{\partial t} = \int_{-\infty}^t \left[-\frac{\partial }{\partial x}(D^{(1)}(x,t-\tau)P(x,\tau) ) + \frac{\partial^2 }{\partial x^2}(D^{(2)}(x,t-\tau)P(x,\tau) )\right]\,d\tau\,.
\end{equation}\\

\textbf{Fokker-Planck equation for wealth diffusion}\\

Assuming that wealth exchanges $\Delta w$ are small, the evolution of the wealth distribution can also be described by a Fokker-Planck equation. For the Yard-Sale model (Sec.~\ref{sec:yardsale-wealthcondensatoin}), the change $\Delta w$ can be written as \cite{boghosian_fokkerplanck_2014}

\begin{equation}
    \Delta w \equiv \Delta (w,w',\eta) = (1-\beta)\eta [w\theta(w'-w) + w'\theta(w-w')]\,,
\end{equation}

with $\eta \in {-1,1}$ and $\theta$ the Heaviside step function. \\

In order to compute the drift and diffusion coefficients $D^{(1)}(v)$ and $D^{(2)}(v)$, we have to consider two sources of uncertainty. First, choosing an interaction partner is sampled from the probability distribution $P(w')$. Second, the random variable $\eta$ decides whether the interaction is profitable or unprofitable. To account for these two sources of uncertainty, one can compute the average of a quantity $f(w',\eta)$ as

\begin{equation}
    \langle f(w',\eta)\rangle  = \int_0^\infty \dw' P(w')\frac{f(w',1)+f(w',-1)}{2}\,.
\end{equation}

The average $\langle \Delta(w) \rangle$ is then given by

\begin{equation*}
    \langle \Delta(w)\rangle = \int_0^\infty \dw'\, P(w') \frac{\Delta(w,w',+1) + \Delta(w,w',-1)}{2}= 0\,,
\end{equation*}

which equals zero as the profiting agent is selected with even odds. Hence, there is no deterministic drift in the Fokker-Planck equation for the Yard-Sale model. Instead, all the change in $P(w)$ results from the stochastic diffusion of wealth:

\begin{align*}
    \langle \Delta(w)^2\rangle &= \int_0^\infty \dw' P(w') \frac{\Delta(w,w',+1)^2 + \Delta(w,w',-1)^2}{2}\,\\
    &= (1-\beta)^2 \int_0^\infty \dw'\, P(w')[w\theta(w'-w) + w'\theta(w-w')]^2\\
    &=  (1-\beta)^2 \int_0^\infty \dw'\, P(w')[w^2\theta(w'-w) + w'^2\theta(w-w')]\\
    &= (1-\beta)^2 w^2 \underbrace{\int_w^\infty \dw'\, P(w')}_{=A(w)} + (1-\beta)^2 \underbrace{\int_0^w \dw'\,P(w')w'^2}_{\equiv 2B(w)}\,,\\
\end{align*}

where $A(w)$ is the Pareto function and $B(w)$ is the incomplete second moment of $P(w)$. All in all, the Fokker-Planck description of the Yard-Sale dynamics is thus given by the nonlinear integrodifferential Fokker-Planck equation \cite{boghosian_fokkerplanck_2014,boghosian_kinetics_2014}

\begin{equation}
     \frac{\partial P(w,t)}{\partial t} =  (1-\beta)^2 \frac{\partial^2}{\partial w^2}\left[\left(\frac{w^2}{2}\int_w^\infty \dw'\, P(w')+\int_0^w \dw'\,P(w')w'^2\right)P(w)\right]\,.
\end{equation}

\end{math_box}

\clearpage
\begin{math_box}{\subsection{Boltzmann's $H$-Theorem} \label{sec:theory:boltzmann-htheorem}}

\textbf{$H$-Theorem with Boltzmann equation for a dilute gas}\\

A fundamental problem in statistical mechanics is understanding the irreversibility observed in macroscopic systems, despite the fact that the underlying microscopic equations of motion are time-reversible. In macroscopic phenomena, such as the diffusion of gases or the flow of heat, we consistently observe a clear direction of time --- entropy increases, and systems evolve toward equilibrium. This observation stands in contrast to the time-reversible nature of classical mechanics, where reversing the velocities of all particles in a system would theoretically allow the system to retrace its path back to its initial state. Reconciling this apparent contradiction between microscopic reversibility and macroscopic irreversibility has been a central challenge in the development of statistical mechanics \cite{kardar2007statistical,krapivsky_kinetic_2010}.\\

Ludwig Boltzmann addressed this problem through his formulation of the $H$-Theorem. It is derived from the Boltzmann equation, which describes the time evolution of the distribution function $p(x,v,t)$ of particles in a dilute gas. It is defined such that $p(x,v,t) d^3xd^3v$ is the number of gas molecules in the volume element $d^3x d^3v$ of the position and velocity space. \\

For particles moving in an external field, the Boltzmann equation takes the form

\begin{equation}
    \left( \frac{\partial}{\partial t} + v\cdot\nabla_r + \frac{F(x)}{m}\nabla_v \right)p(x,v,t) = \left( \frac{\partial p}{\partial t}\right)_{\rm collision}\,,
\end{equation}

with the collision term

\begin{equation}
    \left( \frac{\partial p}{\partial t}\right)_{\rm collision} = \int d^3v_1 \int d\Omega\,|v-v_1| \sigma(v,v_1|v',v_1') \cdot [p(x,v',t)p(x,v_1',t) - p(x,v,t)p(x,v_1,t)]\,.
\end{equation}

$\sigma$ describes the cross section of the collision process of particles with velocity $v$ and $v_1$ before the collision and $v'$ and $v_1'$ after the collision.\\

The $H$-Theorem states that if $p(x,v,t)$ satisfies the Boltzmann equation, then

\begin{equation}
   \frac{dH}{dt}\leq 0 \hspace{0.5cm}\text{with}\hspace{0.5cm} H(t) = \int d^3x\, d^3v\, p(x,v,t)\log p(x,v,t)\,.
\end{equation}

This result suggests that the distribution function $p(x,v,t)$ evolves towards a Maxwell-Boltzmann distribution, corresponding to thermodynamic equilibrium, where $H(t)$ reaches its minimum value.\\

Proving the $H$-Theorem involves computing the time derivative of $H$,

\begin{equation}
    \frac{dH(t)}{dt} = \int d^3x \,d^3v \,\frac{\partial p}{\partial t}(\log p + 1) = \int d^3x\, d^3v\, \frac{\partial p}{\partial t}\log p\,.
\end{equation}

Inserting the Boltzmann equation for $\frac{\partial p}{\partial t}$ (details in \cite{kardar2007statistical}) leads to an expression of the form 

\begin{equation}
    -(p(v)p(v_1) - p(v')p(v_1'))(\log \frac{p(v)p(v_1)}{p(v')p(v_1')}) \geq 0\,.
\end{equation}

If the first term $p(v)p(v_1)$ is larger than the second term $p(v')p(v_1')$, both parts are positive. If not, both parts are negative. In either case, the product is positive, and due to the minus sign, the expression becomes negative, proving that $dH/dt \leq 0$. \\

The validity of the $H$-Theorem depends on specific assumptions, most importantly the assumption of \textit{molecular chaos}. This assumption states that the velocities of colliding particles are uncorrelated before the collision, meaning that the distribution of particle velocities is statistically independent for each pair of particles about to collide. It allows to approximate the joint two-particle distribution function by a product of one-particle distribution functions. 
While this assumption is reasonable for a gas in a non-equilibrium state, it implicitly introduces a preferred direction of time, thereby breaking the time-reversal symmetry that is characteristic of the underlying equations of motion. Boltzmann’s approach thus provides a statistical explanation for the observed macroscopic irreversibility, even though the microscopic dynamics are reversible.\\

Despite its success, the $H$-Theorem faced criticism, particularly the Loschmidt paradox. Josef Loschmidt pointed out that if the microscopic dynamics are reversible, then reversing the velocities of all particles should lead to a decrease in entropy, contradicting the second law of thermodynamics. Boltzmann responded to this paradox by emphasising the statistical nature of the second law: while it is theoretically possible for all particles to reverse their velocities and decrease entropy, the probability of such an event occurring in a system with a large number of particles is extremely small. As a result, entropy increase is overwhelmingly likely in practice, which accounts for the observed irreversibility in macroscopic systems.\\

\textbf{$H$-Theorem for the Gini index in the Yard-Sale model}\\

Similar to Boltzmann's proof of the $H$-Theorem, one can prove that the Gini index is an $H$-function of the Master equation of the Yard-Sale model, namely that it strictly increases \cite{boghosian_h_2015}.\\

For the derivation it is useful to work with the Fokker-Planck equation of the Yard-Sale model, given by 

\begin{equation}
     \frac{\partial P(w,t)}{\partial t} =  \frac{\partial^2}{\partial w^2}\left[\left(\gamma \frac{w^2}{2}\int_w^\infty \dw'\, P(w')+\int_0^w \dw'\,P(w')w'^2\right)P(w)\right]\,,\label{eq:yardsale-htheorem}
\end{equation}

with $\gamma$ being a constant. One can rewrite Eq.~\ref{eq:yardsale-htheorem} to

\begin{equation}
     \frac{\partial P(w,t)}{\partial t} =  \frac{\partial^2}{\partial w^2}\left(\gamma \frac{w^2}{2}C(w)P(w)\right)\,\label{eq:fp_for_proof}
\end{equation}

with 

\begin{equation}
    C(w) = 1 - \int_0^w dx \,P(x)\left(1-\frac{x^2}{w^2}\right)\,,
\end{equation}

which is bounded between $0\leq C(w) \leq 1$. \\

The time derivative of the Gini index is (Eq.~\ref{eq:dGdt_frechet})

\begin{equation}
    \frac{dG}{dt} = \int_0^\infty dw\, \frac{\delta G}{\delta P(w)}\frac{\partial P(w,t)}{\partial t}\,,\label{eq:dGdt-appendix}
\end{equation}

where one can compute the term $ \frac{\delta G}{\delta P(w)}$ by rewriting the Gini index as 

\begin{equation}
    G = 1 - \frac{2}{\bar w} \int_0^\infty dw\,P(w)A(w)\,,
\end{equation}

with $A(w)$ the Pareto function $A(w) = \int_w^\infty P(x)dx$. This leads to 

\begin{equation}
    \frac{\delta G}{\delta P(w)} = \frac{2}{\bar w}\left[-w + \int_0^wdx\,P(x)(w-x)\right]\,.
\end{equation}

Inserting this as well as Eq.~\ref{eq:fp_for_proof} into Eq.~\ref{eq:dGdt-appendix} results in 

\begin{equation}
    \frac{dG}{dt} = \frac{2}{\bar w} \int_0^\infty dw \left[-w + \int_0^wdx\,P(x)(w-x)\right] \frac{\partial^2}{\partial w^2}\left(\gamma \frac{w^2}{2}C(w)P(w)\right)\,.
\end{equation}

Integrating by parts twice then leads to

\begin{align*}
    \frac{dG}{dt} &= -\frac{2}{\bar w} \int_0^\infty dw\,\left[-1 + \int_0^w dx\,P(x)\right]\frac{\partial}{\partial w}\left(\gamma \frac{w^2}{2}C(w)P(w)\right)\\
    &= \frac{2}{\bar w} \int_0^\infty dw\, P(w) \left(\gamma \frac{w^2}{2}C(w)P(w)\right)\\
    &= \frac{\gamma}{\bar w} \int_0^\infty dw\, w^2C(w)P(w)^2 \geq 0\,,
\end{align*}

which proves that the Gini index strictly increases under the Yard-sale dynamics. 

\end{math_box}
\clearpage

\section{Supplementary figures}

\subsection{Heterogeneous saving propensity for power-law distribution}

\begin{figure}
    \centering
    \includegraphics{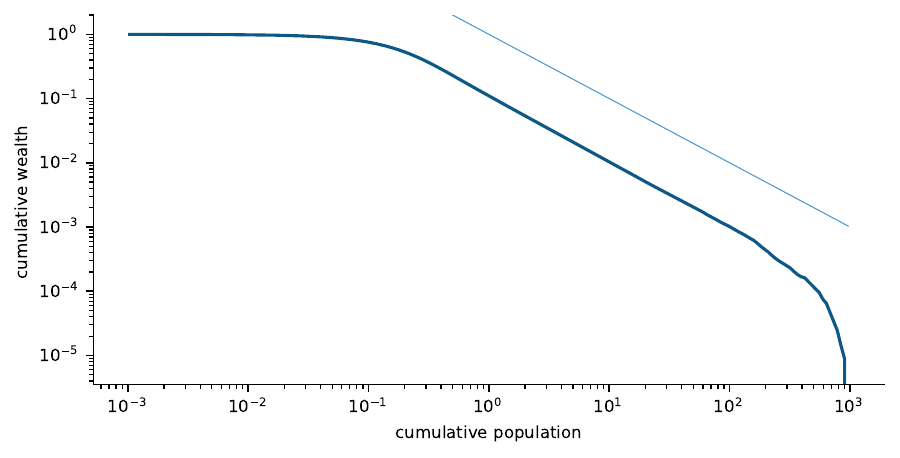}
    \caption{\textbf{Heterogeneous saving propensity results in power-law tails.} The steady state wealth distribution for uniform saving propensity $\lambda$ among agents (Eq.~\ref{eq:chakrabarti-saving}). It follows a power-law with exponent $\alpha = 1$ (light blue line). The simulation is performed using $N=1000$ agents and a mean wealth of $\bar w = 1$. The system is equilibrated for $t^{\rm eq} = \SI{100000}{}$ (one time step constituting 500 transactions) before averaging the distribution over $t^{\rm meas} = \SI{1000}{}$, and taking 450 configurational averages for different realisations of the distribution of $\lambda$.}
    \label{fig:chakrabarti-powerlaw}
\end{figure}

\begin{figure}
    \centering
    \includegraphics{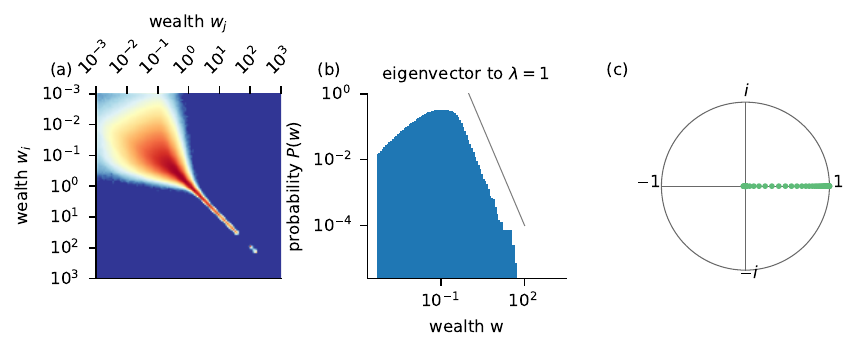}
    \caption{\textbf{Transition matrix and eigenvalue distribution for heterogeneous saving propensity.} (a) Weight matrix. (b) The steady-state eigenvector follows a power-law with Pareto exponent $\alpha=1$ (grey line). (c) The eigenvalue distribution.}
    \label{fig:transitionmatrix-chakrabarti}
\end{figure} 

\clearpage

\subsection{Transition matrices over longer time windows}

\begin{figure}
    \centering
    \includegraphics{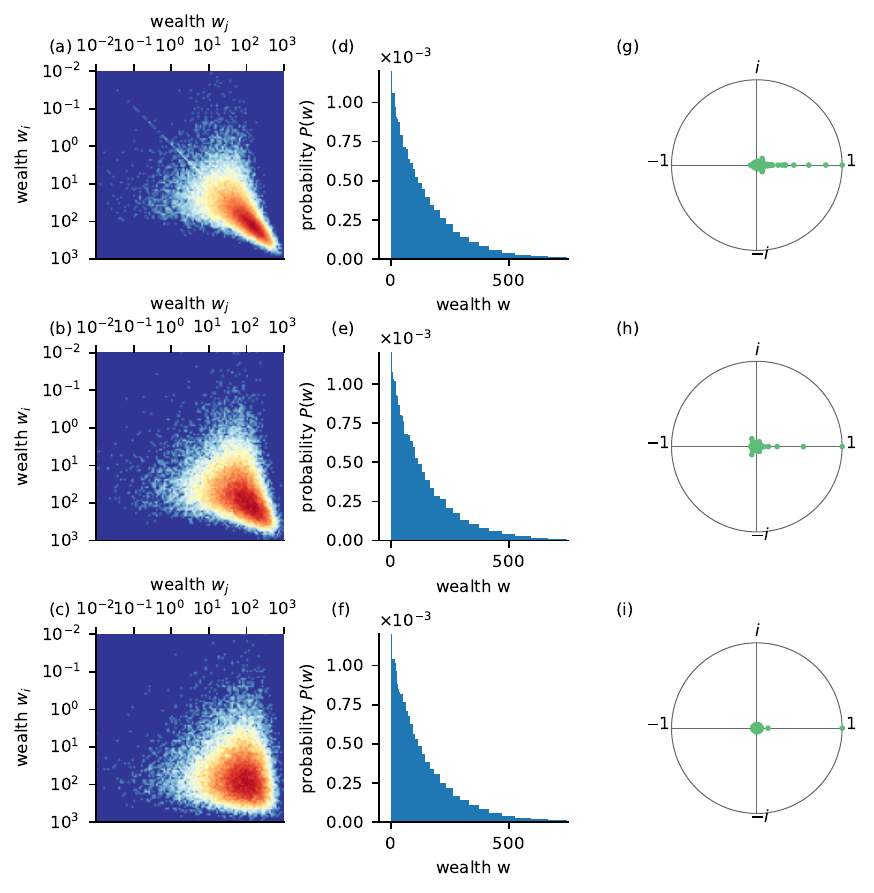}
    \caption{\textbf{Dragulescu model} }
    \label{fig:increases_DRA}
\end{figure}

\begin{figure}
    \centering
    \includegraphics{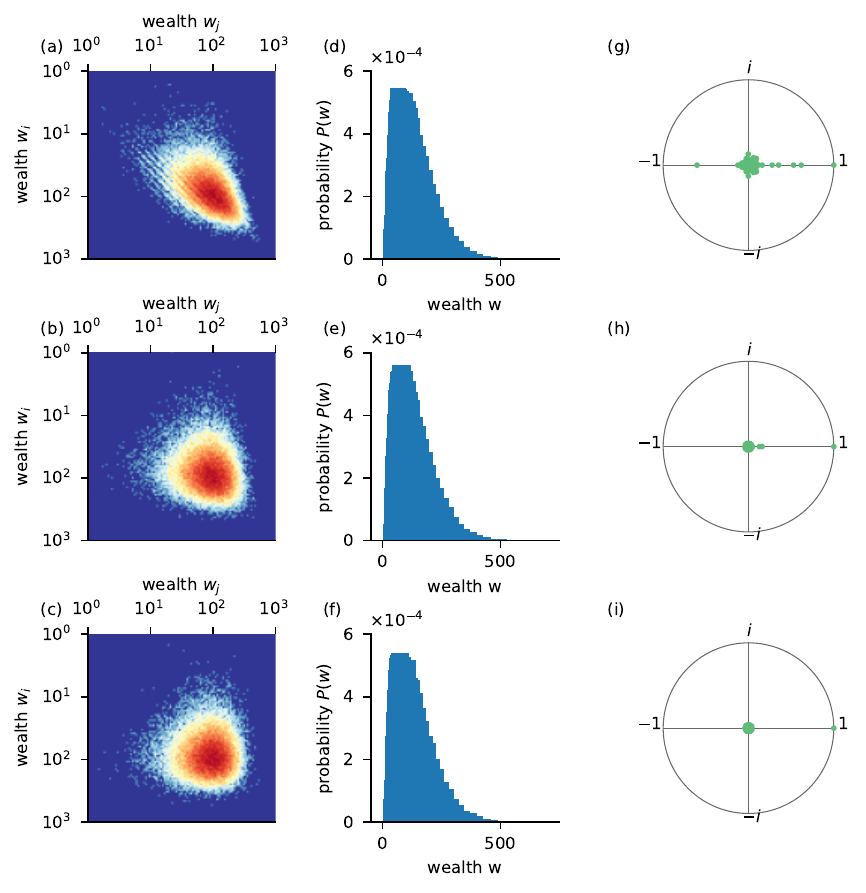}
    \caption{\textbf{Ispolatov model}}
    \label{fig:increases_ISP}
\end{figure}

\begin{figure}
    \centering
    \includegraphics{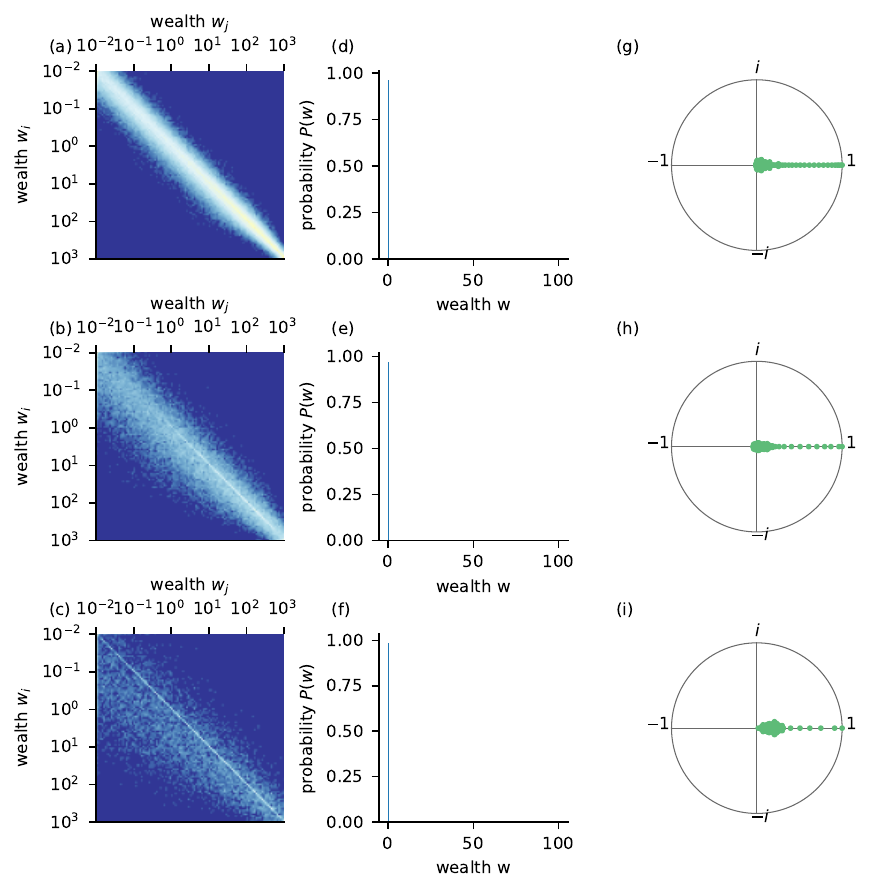}
    \caption{\textbf{Yard-Sale model}}
    \label{fig:increases_YSM}
\end{figure}

Fig.~\ref{fig:increases_DRA}, Fig.~\ref{fig:increases_ISP} and Fig.~\ref{fig:increases_YSM} show transition matrices for the different definition of transitions, as described at the end of Sec.~\ref{sec:logbinning}.
Transition matrices are computed by collecting transitions not from individual transactions, but changes in wealth of agents over time. Time delays $\delta t$ are chosen as $\delta t=5$, $\delta t = 15$ and $\delta t=50$ time units (top to bottom). One time unit is defined as $N/2$ transactions. (a)-(c) Weight matrices, for longer time delays $\delta t$ it gets more and more likely to transition into states far away. (d)-(f) The steady state eigenvector corresponding to the eigenvalue $\lambda_1=1$. (g)-(i) All eigenvalues except $\lambda_1=1$ tend towards zero with increasing time delay.

\clearpage

\subsection{Influence of binning}
\label{sec:appendix-influence-binnumber}

\begin{figure}
    \centering
    \includegraphics{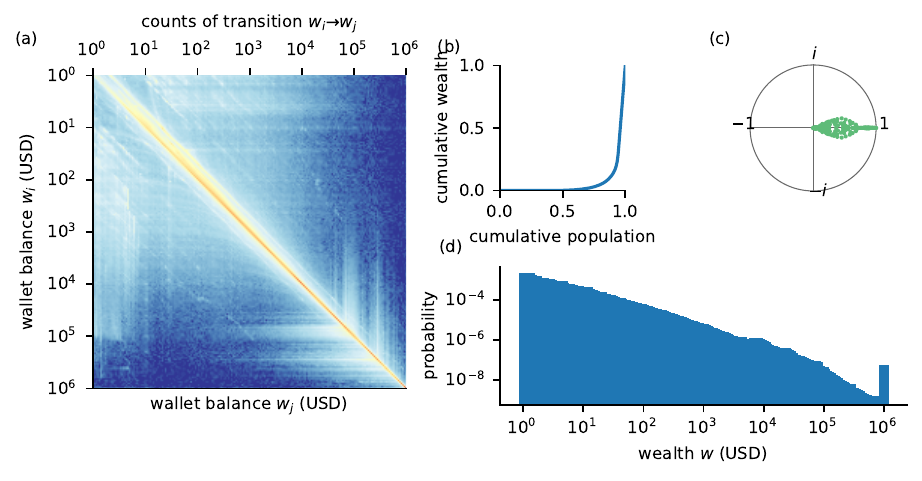}
    \caption{\textbf{Lower upper bin decreases equilibrium Gini index.} Same panels as Fig.~\ref{fig:crypto_transitionmatrix_all} but with $\SI{1e6}{}$ as upper bin limit instead of $\SI{1e7}{}$. This reduces the equilibrium Gini index from $G=0.94$ in Fig.~\ref{fig:crypto_transitionmatrix_all} to $G=0.89$.}
    \label{fig:appendix-transitionmatrix_all-1e6-200bins}
\end{figure}

\begin{figure}
    \centering
    \includegraphics{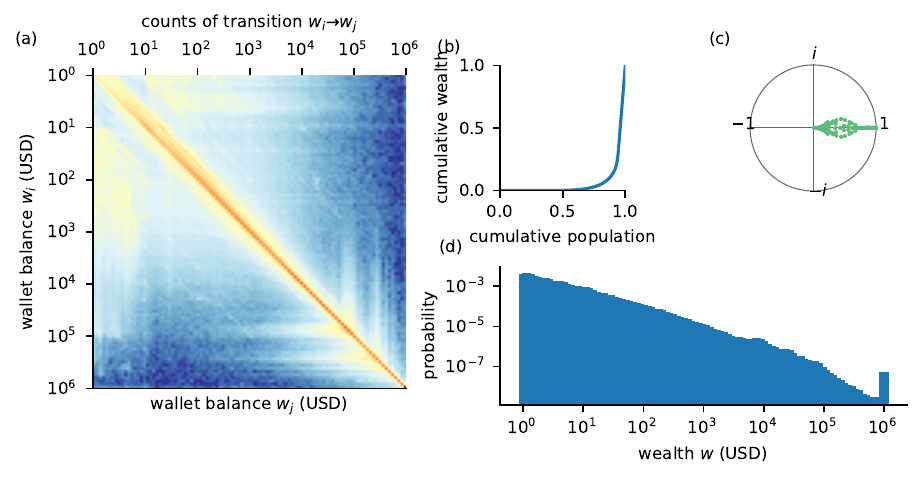}
    \caption{\textbf{.} Same panels as Fig.~\ref{fig:appendix-transitionmatrix_all-1e6-200bins} but with 100 bins instead of 200. The equilibrium Gini index remains at $G=0.89$.}
    \label{fig:appendix-transitionmatrix_all-1e6-100bins}
\end{figure}

In continuous systems, the choice of how to define the Markov states influences the transition matrix. Fig.~\ref{fig:appendix-transitionmatrix_all-1e6-200bins} and Fig.~\ref{fig:appendix-transitionmatrix_all-1e6-100bins} show how a change in upper bin limit and reduced number of bins has an effect on the equilibrium Gini index.

\clearpage

\subsection{Relation between Gini index, Democracy index and HDI}
\label{sec:appendix-giniindex_vsdemocracy_vshdi}

The Gini index correlates with socio-economic variables such as the Democracy Index and Human Development Index. The following figures show that the income Gini index reflects a higher correlation with those variables than the wealth Gini index. Data is taken from \textit{Our World in Data} \cite{owid_democracy,owid-human-development-index} and the \textit{World Inequality Database} \cite{wid}.

\begin{figure}
    \centering
    \includegraphics{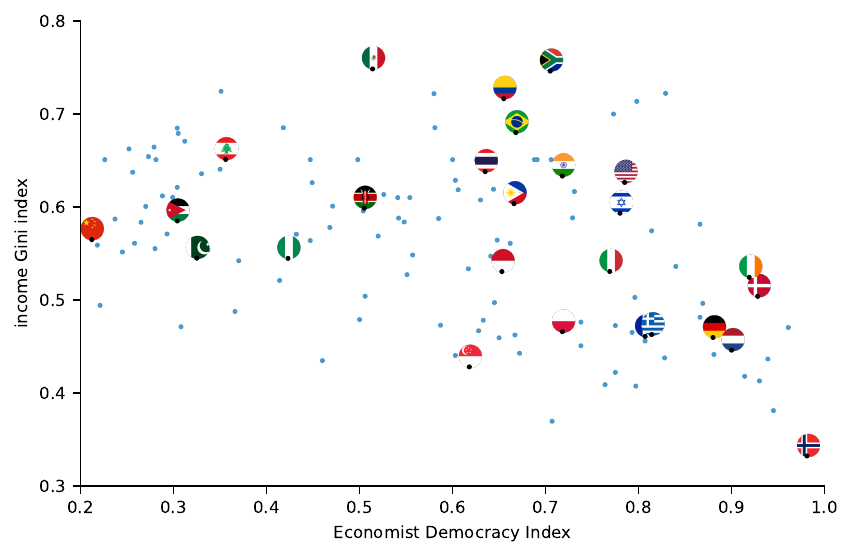}
    \caption{\textbf{Income Gini index vs Economist Democracy Index.} Pearson correlation: $r=-0.43$, $p$-value: $p=\SI{3.87e-7}{}$. Economist Democracy Index is scaled down by factor of 10.}
    \label{fig:appendix-incomegini-vs-democracy}
\end{figure}

\begin{figure}
    \centering
    \includegraphics{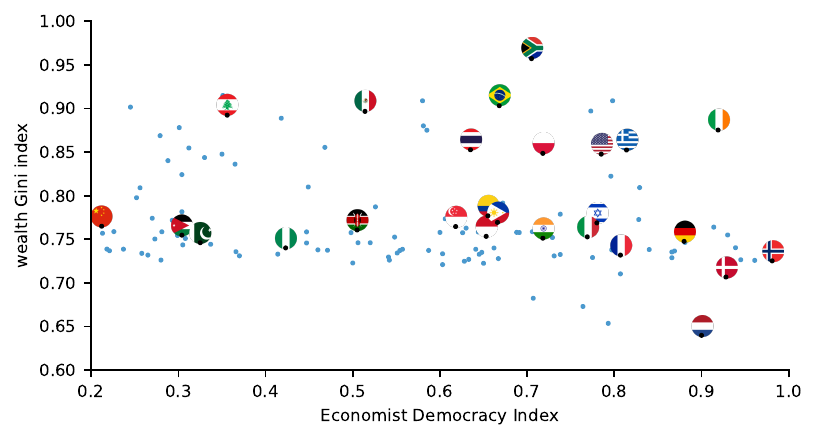}
    \caption{\textbf{Wealth Gini index vs Economist Democracy Index.} Pearson correlation: $r=-0.18$, $p$-value: $p=\SI{0.041}{}$. Economist Democracy Index is scaled down by factor of 10.}
    \label{fig:appendix-wealthgini-vs-democracy}
\end{figure}

\begin{figure}
    \centering
    \includegraphics{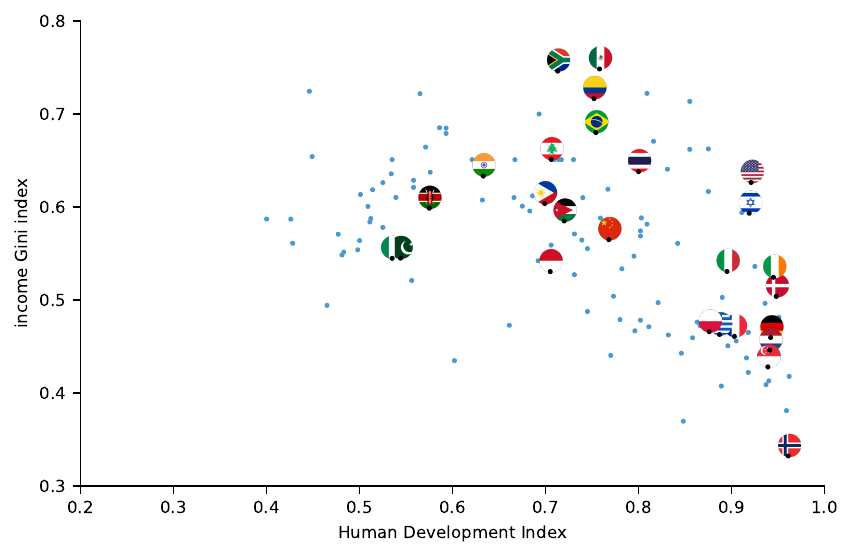}
    \caption{\textbf{Income Gini index vs Human Development Index.} Pearson correlation: $r=-0.48$, $p$-value: $p=\SI{5.39e-9}{}$.}
    \label{fig:appendix-incomegini-vs-hdi}
\end{figure}

\begin{figure}
    \centering
    \includegraphics{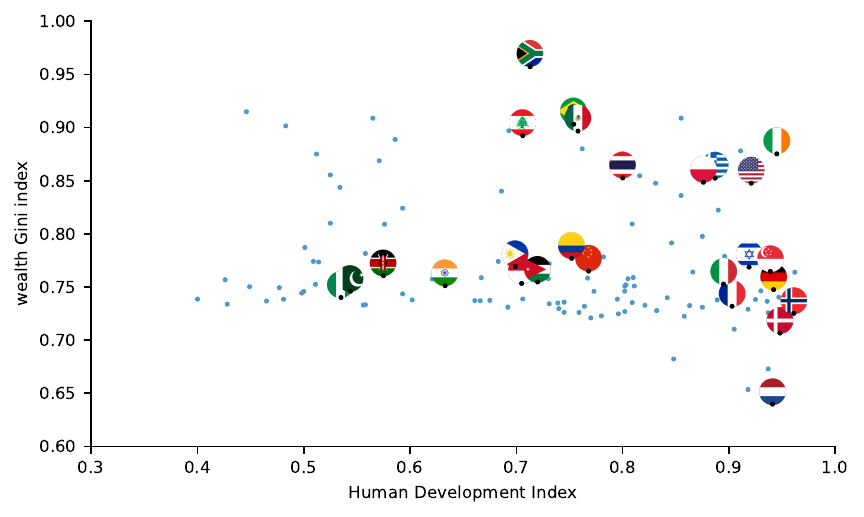}
    \caption{\textbf{Wealth Gini index vs Human Development Index.} Pearson correlation: $r=-0.18$, $p$-value: $p=\SI{0.045}{}$. Economist Democracy Index is scaled down by factor of 10.}
    \label{fig:appendix-wealthgini-vs-hdi}
\end{figure}

\end{document}